\documentclass{aastex62}

\usepackage{colortbl}
\usepackage{xcolor}
\newcommand{\get}{\sc Getsources}
\newcommand{\fw}{\sc FellWalker}
\newcommand{\nhp}{N$_2$H$^+$}
\newcommand{\ndp}{N$_2$D$^+$}

\definecolor{forestgreen}{rgb}{0.013,0.55,0.13}

\newcommand{\x}[1]{\text{#1}}
\accepted{March 1, 2019}
\submitjournal{ApJ}

\shorttitle{GAS: A Virial Analysis of Gould Belt Clouds in Data Release 1}
\shortauthors{Kerr et al}
\graphicspath{./}

\begin{document}
\turnoffeditone
\turnoffedittwo
\title{The Green Bank Ammonia Survey: A Virial Analysis of Gould Belt Clouds in Data Release 1}

\correspondingauthor{Ronan Kerr}
\email{rmp.kerr@gmail.com}

%\author[0000-0002-0786-7307]{Greg J. Schwarz}
%\affil{American Astronomical Society \\
%2000 Florida Ave., NW, Suite 300 \\
%Washington, DC 20009-1231, USA}

\author{Ronan Kerr}
\affiliation{Department of Physics and Astronomy, University of British Columbia\\
2329 West Mall, Vancouver, BC, V6T 1Z1, Canada}

\author{Helen Kirk}
\affiliation{Herzberg Astronomy \& Astrophysics Program, National Research Council of Canada \\
5071 West Saanich Road, Victoria, Bc, V9E 2E7, Canada}
\affiliation{Department of Physics and Astronomy, University of Victoria \\ 3800 Finnerty Rd., Victoria, BC, V8P 5C2, Canada}
\author{James Di Francesco}
\affiliation{Herzberg Astronomy \& Astrophysics Program, National Research Council of Canada \\
5071 West Saanich Road, Victoria, Bc, V9E 2E7, Canada}
\affiliation{Department of Physics and Astronomy, University of Victoria \\ 3800 Finnerty Rd., Victoria, BC, V8P 5C2, Canada}
\author{Jared Keown}
\affiliation{Department of Physics and Astronomy, University of Victoria \\ 3800 Finnerty Rd., Victoria, BC, V8P 5C2, Canada}
\author{Mike Chen}
\affiliation{Department of Physics and Astronomy, University of Victoria \\ 3800 Finnerty Rd., Victoria, BC, V8P 5C2, Canada}
\author{Erik Rosolowsky}
\affil{Department of Physics, University of Alberta, Edmonton, AB, Canada}
\author{Stella S. R. Offner}
\affil{The University of Texas at Austin
Department of Astronomy, College of Natural Sciences
2515 Speedway, Stop C1400
Austin, TX 78712}
\author{Rachel Friesen}
\affil{National Radio Astronomy Observatory, 520 Edgemont Rd., Charlottesville, VA 22901, USA}
\author{Jaime E. Pineda}
\affil{Max-Planck-Institut für extraterrestrische Physik, Giessenbachstrasse 1, 85748 Garching, Germany}
\author{Yancy Shirley}
\affil{Steward Observatory, 933 North Cherry Avenue, Tucson, AZ 85721, USA}
\author{Elena Redaelli} 	
\affil{Max-Planck-Institut für extraterrestrische Physik, Giessenbachstrasse 1, 85748 Garching, Germany}
\author{Paola Caselli}
\affil{Max-Planck-Institut für extraterrestrische Physik, Giessenbachstrasse 1, 85748 Garching, Germany}
\author{Anna Punanova}
\affil{Ural Federal University, 620002 Mira st. 19, Yekaterinburg, Russia }
\author{Youngmin Seo}
\affil{Jet Propulsion Laboratory, California Institute of Technology, 4800 Oak Grove Drive, Pasadena, CA, 91109, USA}
\author{Felipe Alves}
\affil{Max-Planck-Institut für extraterrestrische Physik, Giessenbachstrasse 1, 85748 Garching, Germany}
\author{Ana Chac\'{o}n-Tanarro}
\affil{Max-Planck-Institut für extraterrestrische Physik, Giessenbachstrasse 1, 85748 Garching, Germany}
\author{Hope How-Huan Chen}
\affil{Harvard-Smithsonian Center for Astrophysics, 60 Garden St., Cambridge, MA 02138, USA}

\begin{abstract}
We perform a virial analysis of \edit2{starless} dense cores in three nearby star-forming regions : L1688 in Ophiuchus, NGC 1333 in Perseus, and B18 in Taurus. Our analysis takes advantage of comprehensive kinematic information for the dense gas in all of these regions \edit2{made publicly available through the Green Bank Ammonia Survey Data Release 1, which used to estimate internal support against collapse}. We combine this information \edit2{with ancillary data used to estimate other important properties of the cores, including} continuum data from the James Clerk Maxwell Telescope Gould Belt Survey \edit2{for core identification, core masses, and core sizes.  Additionally, we used \textit{Planck} and \textit{Herschel}-based column density maps for external cloud weight pressure, and Five College Radio Astronomy Observatory $^{13}$CO observations for external turbulent pressure. }Our self-consistent analysis suggests that many dense cores in all three star-forming regions are not bound by gravity alone, but rather require additional pressure confinement to remain bound. Unlike a recent, similar study in Orion~A, we find that turbulent pressure represents a significant portion of the external pressure budget. Our broad conclusion emphasizing the importance of pressure confinement in dense core evolution, however, agrees with earlier work.

\end{abstract}

\keywords{stars:formation, ISM:kinematics and dynamics}
%HK notes to Ronan: allowed keywords are listed here:
%https://journals.aas.org/authors/keywords2013.html

\section{Introduction} \label{sec:intro}

Dense cores are the birthplaces of stars. Throughout our Galaxy, these relatively small but crucially important structures can be found inside much larger molecular clouds. Current telescope instrumentation allows us to perform cloud-wide observations of dense cores, providing an opportunity to characterize unbiased and complete dense core populations at a range of evolutionary stages.  Our focus here is on simple kinematic and (column) density observations, which provide insight into the virial states of the cores. Determining which dense cores are likely unstable to collapse, and the relative strength of the various forces acting on the cores helps to shed light on future star formation within the clouds.

\edit1{Dense cores can be studied through a variety of complementary observations.  Dust continuum observations in the millimeter and submillimeter regime (a good tracer for the total dust column density) have long been a popular and efficient method for characterizing the locations, sizes, and approximate masses of dense cores \citep[e.g.,][]{Motte98, Kirk16}.  
At the same time, spectral observations of emission lines of molecular species such as NH$_3$ \citep[e.g.,][]{Benson89, Friesen16,Sokolov18} offer a powerful way to study the material close to or above the transition critical density, and directly derive the gas temperatures and non-thermal motions associated with dense core gas.}

This paper draws on the extensive core kinematic data produced by the Green Bank Telescope's Green Bank Ammonia Survey (GAS) \citep{Friesen16}, combined with dust continuum emission maps acquired by the James Clerk Maxwell Telescope's Gould Belt Survey (GBS) \citep{WardThompson07}. \edit1{Observations of dust continuum and NH$_3$ produce highly complementary sets of information, collectively providing a mechanism to estimate the main core properties needed for a virial analysis.}
\citet{Kirk17b} previously presented a virial analysis of the dense cores in the Orion A molecular cloud, in which most cores were found to be bound, with external cloud weight pressure dominating the forces binding cores. 
Here, we consider the three additional star-forming regions that make up GAS's Data Release 1 (DR1): NGC 1333 in Perseus, L1688 in Ophiuchus, and B18 in Taurus.  These three regions span a wide range of star formation rates and molecular cloud properties \citep[e.g., see][]{Kenyon08, Walawender08, Wilking08}. 

%\rknotes{[The following could use a secondary check from someone more familiar with the material]}

The combination of \edit1{the} GBS and GAS data is \edit1{powerful.} GBS traces the cold and compact dust emission structures associated with dense cores, allowing for the identification of cores, and estimation of their masses and sizes.  GAS meanwhile \edit1{examines the emission from cold and dense gas in these same structures.  Specifically, GAS uses emission from the lowest NH$_3$ rotation-inversion transitions to measure the gas temperature and non-thermal velocity dispersion. % of the gas. \soedit
Radiative transitions between $K$-ladders (states with the same angular momentum projected on the molecule's axis of symmetry) are generally forbidden through radiative processes, so their excitation is therefore controlled by collisions \citep{Ho83}.  Since the level populations in these states are mediated by collisions, their distribution reflects the kinetic temperature of the gas. The lowest two inversion transitions of para-NH$_3$, the (1,1) and (2,2) transitions at 1.3~cm, are widely enough spaced in energy ($\Delta E/k = 41.2$~K) that they provide a good gas kinetic temperature probe in cold gas ($T_k < 30$~K).  Furthermore, densities of $> 10^3 - 10^4$ cm$^{-3}$ are required to excite the (1,1) and (2,2) transitions effectively \citep{Shirley15}.}

This paper is organized as follows: in \S~\ref{sec:methods}, we introduce the data used, and \edit1{extract} basic core properties needed for a virial analysis, such as the \edit1{masses, radii, temperatures, and velocity dispersions} of the cores. In \S~\ref{sec:Jeans}, we run a \edit1{standard} Jeans analysis on the cores, while in
\S~\ref{sec:fullvirial}, we perform a more complete virial analysis, including the effects of gravity, pressure from the ambient cloud weight, and turbulent pressure as binding forces in opposition to internal kinetic support. In \S~\ref{sec:discussion}, we discuss our results, and the implications with respect to the role of turbulent pressure in local star formation. Finally, we summarize our results and conclude in \S~\ref{sec:conclusion}.

\section{Data} \label{sec:methods}
\subsection{JCMT GBS Observations} \label{sec:gbs}

\begin{figure}[htbp]
\includegraphics[width=7.0in]{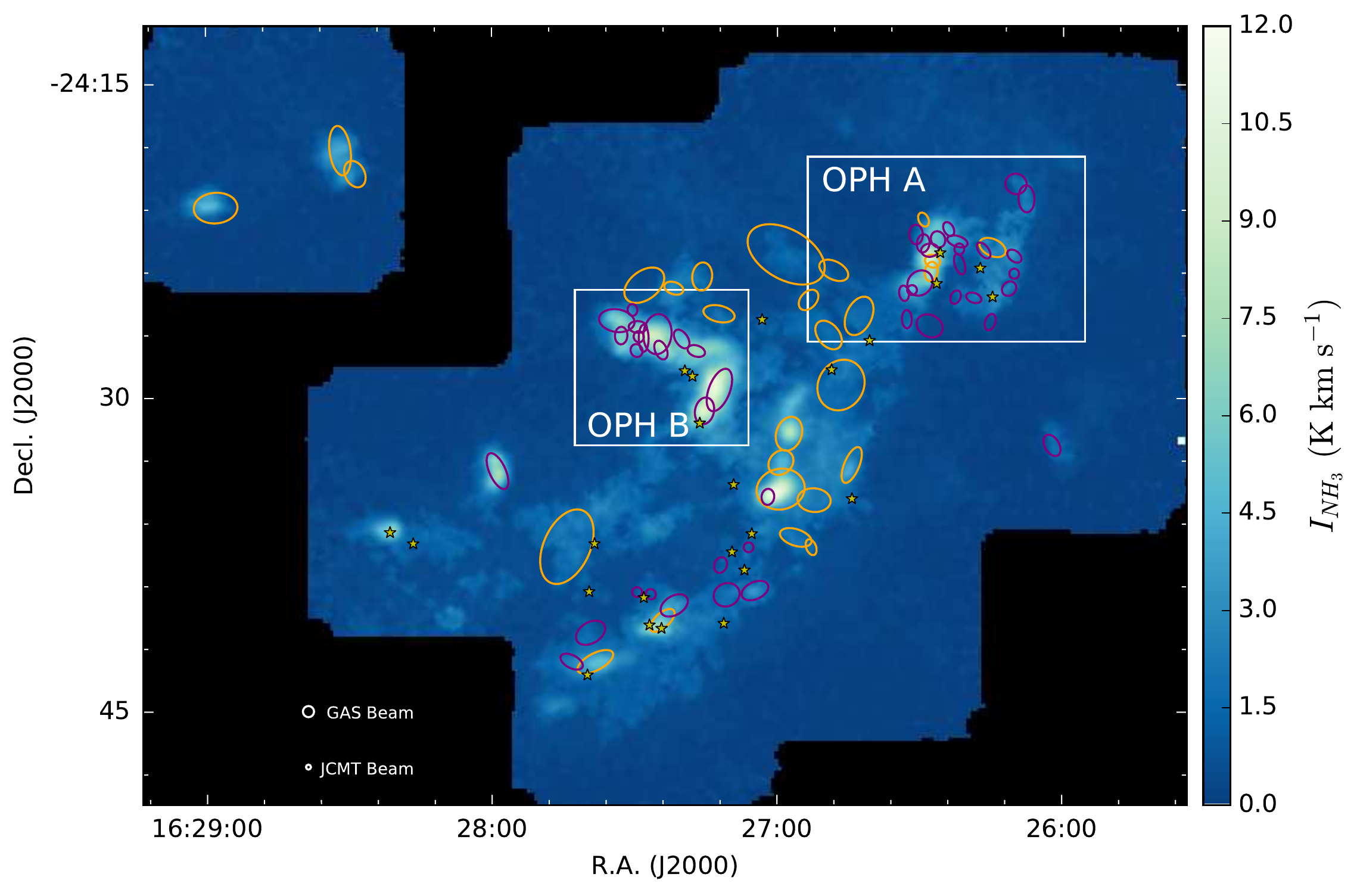}
\caption{Starless dense cores identified in L1688 in Ophiuchus using the {\get} algorithm. The background bluescale image shows integrated intensity emission from NH$_3$ (1,1) emission as observed from the Green Bank Ammonia Survey \citep{Friesen16}. The ellipses correspond to the full extent of each core (2$\times$FWHM),\edit1{ i.e., the {\get}-derived values used to compute the effective radius (see Section \ref{sec:gbs})}. Orange contours indicate cores that appear to be bound by a combination of gravity and pressure, while purple contours indicate cores that appear unbound in our analysis (see Section~\ref{sec:fullvirial} for more details). Protostellar cores identified by {\get} are marked with yellow stars, and their extents are omitted from the figure. These objects are excluded from the virial analysis. Note that the sources found by {\get}, which use 450 $\mu$m and 850 $\mu$m continuum data, do not always align directly with NH$_3$ integrated intensity peaks, hence the slight mismatch between some {\get} extents and NH$_3$ integrated intensity peaks evident in the figure. The \edit2{regions} assigned to Oph A \edit2{and Oph B are} shown in white, and the "JCMT beam" label refers to the 14\farcs1 beam for the 850 $\mu$m continuum observations. }
\label{fig_oph_gs}
\end{figure}

\begin{figure}[htb]
\centering
\includegraphics[width=6.4in]{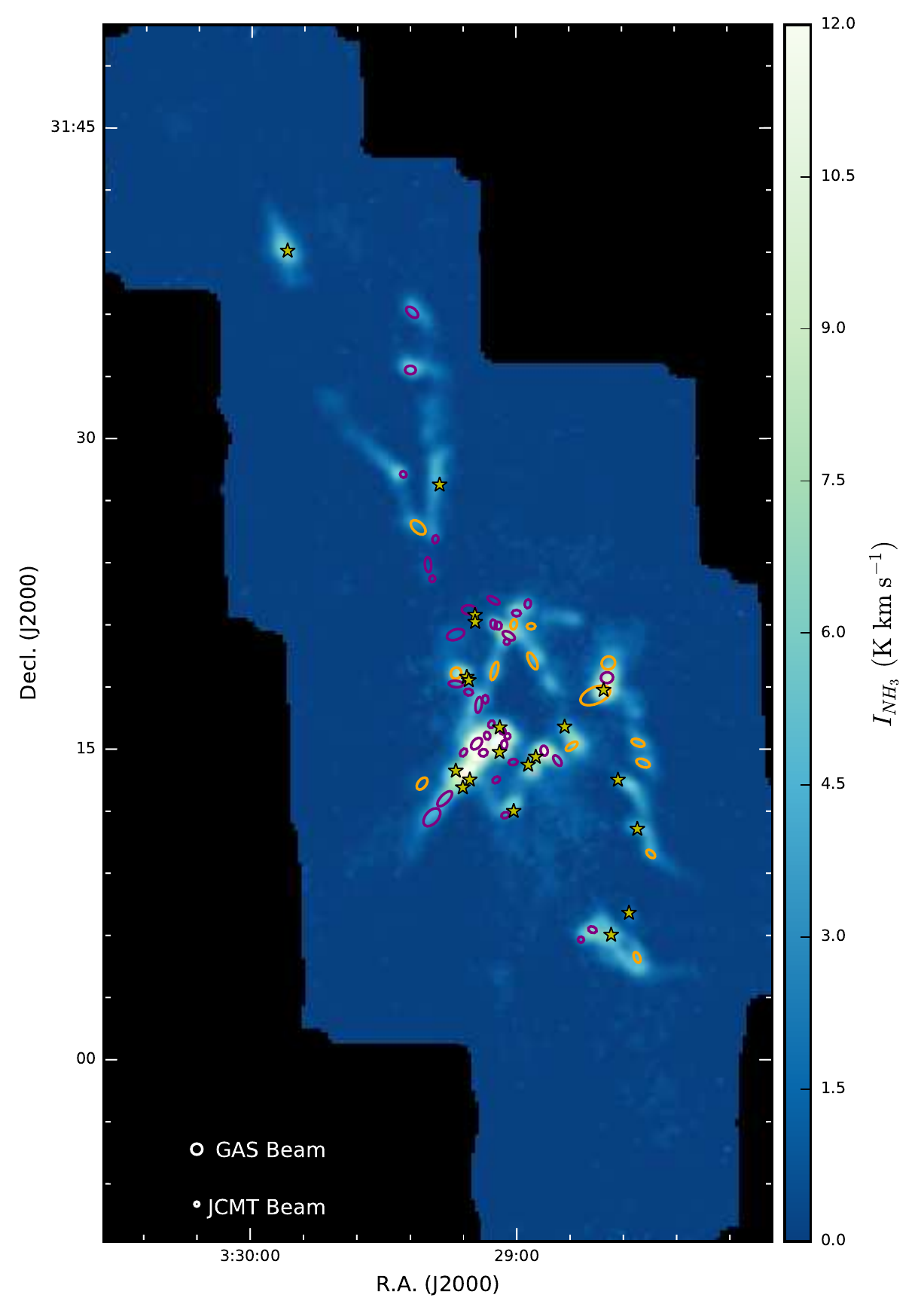}
\caption{Dense cores identified in NGC1333 in Perseus using the {\get} algorithm.  See Figure~\ref{fig_oph_gs} for details.}
\label{fig_pers_gs}
\end{figure}

\begin{figure}[htb]
\centering
\includegraphics[width=8.8in, angle=270]{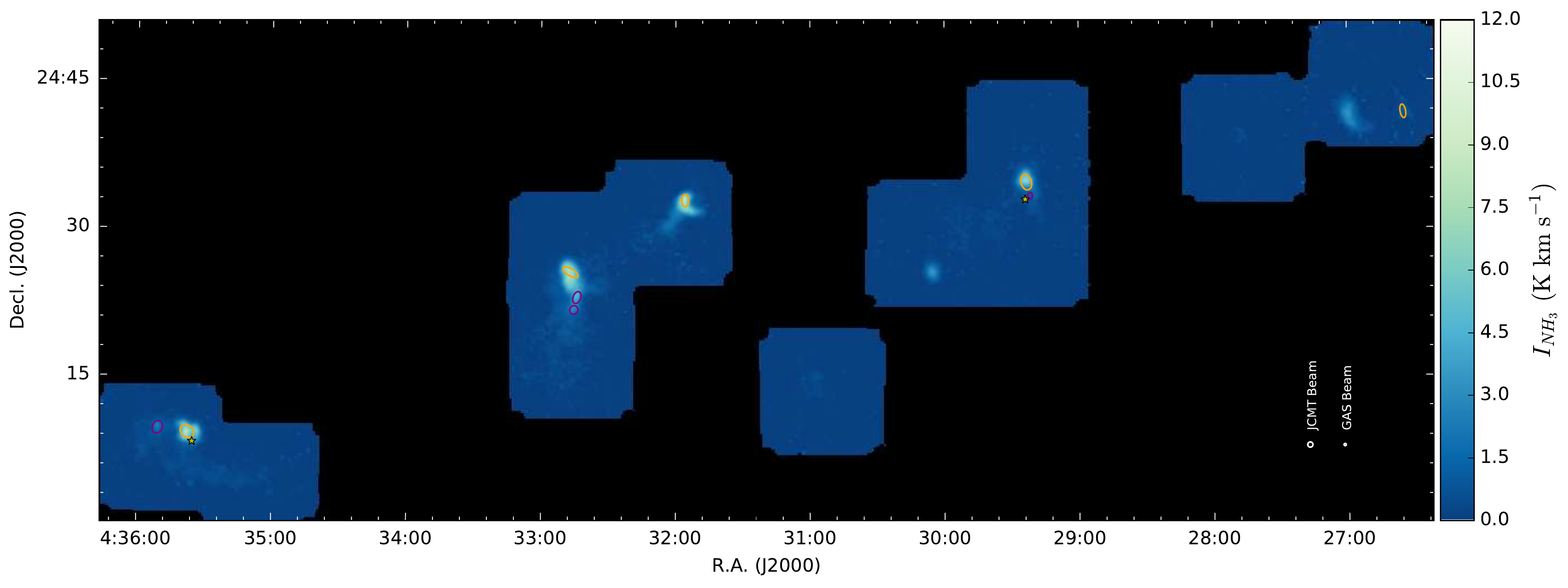}
\caption{Dense cores identified in B18 in Taurus using the {\get} algorithm.  See Figure~\ref{fig_oph_gs} for details.}
\label{fig_tau_gs}
\end{figure}

We use dust emission maps from the James Clerk Maxwell Telescope (JCMT) Gould Belt Survey \citep[GBS;][]{WardThompson07} to identify the dense cores studied in our virial analysis. The JCMT GBS obtained large-area maps of dust continuum emission at 850~$\mu$m and 450~$\mu$m across all nearby visible star-forming regions, including the subset for which we have GAS kinematic data.  Here, we used the most recent maps available at the time of our analysis, from Data Release 2 \citep{Kirk18}. For our main analysis, we use the {\get} algorithm \citep{Menshchikov12} to identify dense cores, which forms part of the JCMT GBS full core catalog to be presented in \citet{Pattle19}. {\get} allows users to identify compact structures using information available at multiple wavelengths, and we find it does a good job of identifying visually apparent dense cores in the three regions we analyze.  We discuss the details of the {\get} algorithm and its application in Appendix \ref{app:getsources}, and compare this to the results obtained with a second core-identification algorithm, {\fw}, in Appendix \ref{app:fellwalker}. Including all types of cores (starless cores and protostellar objects), we identify 134, 78, and 17 cores in L1688, NGC 1333, and B18, respectively, using {\get}. Figures~\ref{fig_oph_gs} to \ref{fig_tau_gs} show the dense cores identified in each of the three regions overlaid on the GAS-based NH$_3$ emission. A sample source identification using both {\get} and {\fw} on the GBS 850~$\mu$m maps is shown in Figure \ref{fig:gsvfw}, focused on Oph~B in L1688.

Measurements of the core size and total flux from {\get} are obtained for our virial analysis. We define the effective \textit{radius} as the geometric mean of the full diameter of the major and minor axes of the elliptical {\get} cores, which we take to represent the full extent of the cores in our analysis.
To account for the JCMT beam size, we furthermore deconvolve the core diameters by subtracting in quadrature the beam size of 14.1\arcsec\ at 850~$\mu$m \citep{Dempsey13}.
Deconvolution becomes highly uncertain for sources of sizes similar to the beam, so for cores with calculated deconvolved diameters of less than the JCMT beam size, we instead use a value of half of the beam as an upper limit to the true radius. These unresolved cores represent 19\% of those detected.

We convert the total flux measured at 850~$\mu$m to a mass assuming the mean kinetic temperature for each core (see Section~\ref{sec:gasObs}), using the equation

\begin{equation}\label{eqn:mass}
M=1.06\,S_{850}\,\exp\left(\frac{17}{T}-1\right) \left(\frac{\kappa_{850}}{0.0125}\right)^{-1} \left(\frac{D}{415}\right)^2
\end{equation}

\edit1{\citep{Kirk17b,Hildebrand83}}, where $M$ is the \edit1{core} mass in $M_{\odot}$, $S_{850}$ is the total flux at 850~$\mu$m in Jy, $T$ is the dust temperature in K, $\kappa_{850}$ is the dust opacity in cm$^2$g$^{-1}$, and $D$ is the distance to the cloud in pc.  Following previous JCMT GBS studies \citep[e.g.,][]{Kirk17b}, we assume $\kappa_{850} = 0.0125$~cm$^{2}$g$^{-1}$.  Furthermore, we assume that the dust temperature is equal to the gas temperature estimated from the GAS NH$_3$ observations (see Section~2.3).
Finally, we adopt new distance estimates where available.  For L1688, we use the VLBI-based GOBELINS distance of 137~pc \citep{OrtizLeon17}, noting that this is nearly identical to previous values \citep[e.g.,][]{Mamajek08}.  For NGC~1333, we adopt a distance based on recent Gaia observations, which suggest a distance of 271~pc \citep{Herczeg17}. This distance is somewhat larger than the standard 235~pc adopted by \citet{Hirota08}, but also smaller than in another recent study combining Gaia and GOBELINS data \citep{OrtizLeon18}. 
For B18, we adopt a distance of 140~pc \citep{Loinard08}.  The GOBELINS team recently published results for the Taurus molecular cloud \citep{Galli18}, however, none of the targets with measured distances appear to be associated with B18. We note that the GOBELINS results show that the Taurus molecular cloud has a large inclination relative to the plane of the sky, with line-of-sight distances ranging from 127~pc to 163~pc along its extent. Our distance estimate of 140~pc for B18 lies comfortably in the middle of the GOBELINS range.

%\edit1{In light of new distance measurements and the low flux uncertainties in the JCMT observations, we expect the mass uncertainties to be dominated by systematic effects. The choice of source extraction method likely has the most significant impact, and the implications of changing the selected algorithm are shown in Appendix \ref{app:sfa}. The assumption of a constant opacity may also introduce systematic error. }
\edit1{The uncertainty in the core masses estimated using Eqn~\ref{eqn:mass} is expected to be dominated by the assumed dust grain opacity: the distances and temperatures are typically accurate to 10\% to 20\% or better, and the flux calibration factor at 850~$\mu$m is also accurate to about 10\% \citep{Dempsey13}.  The dust opacity at 850~$\mu$m could vary by a larger amount. For example, \citet{Chen16} \edit1{found} that the slope of the dust emissivity spectral index ($\beta$) varies between 1.0 and 2.7 within the Perseus molecular cloud, while a constant value of 2 was used for our estimation of the dust opacity at 850~$\mu$m. Similarly, \citet{ChaconTanarro19} found variations of roughly a factor of two in the opacity for the dense core L1544.
Opacity values under a variety of conditions are provided by \citet{Ossenkopf94}. Within star-forming environments, these values can vary by up to an order of magnitude, but normally they do not vary by more than a factor of two.  In addition to the formal errors associated with the mass estimation, the choice of the source identification method can also have a noticeable effect on both the total fluxes and sizes of objects identified.  We explore the implications of adopting different source identification algorithms on our virial analysis in Appendix~\ref{app:sfa}.}

\subsection{{\it Spitzer} Observations}

We use data from the {\it Spitzer Space Telescope} to search for protostars associated with each of the dense cores. For L1688 and NGC~1333, we use the young stellar object (YSO) catalogue from \citet{Dunham15}. The \citet{Dunham15} catalogue, however, does not include Taurus, so for B18, we use the catalogue of \citet{Rebull10}. The observations used for the \citet{Dunham15} catalogue are discussed extensively in terms of completeness by \citet{Evans09}, who estimate 90\% completeness down to 0.05 L$_{\odot}$, and 50\% completeness down to 0.01 L$_{\odot}$ for the highly crowded Serpens region, then assumed to be at a distance of 260 pc. Due to a combination of distance and crowdedness, \citet{Evans09} presented the Serpens completeness as a lower limit for the remainder of the data set, including L1688 and NGC~1333. Therefore, the vast majority of protostars should be detected in both  L1688 and NGC~1333.
While a full completeness analysis of the Taurus protostars does not appear to be published, \citet{Luhman09} used a comparison of {\it Spitzer} \edit1{data and x-ray observations from the XMM-Newton Extended Survey of the Taurus molecular cloud \citep{Gudel07}} to suggest that the YSO census is complete down to $\sim$0.02~$M_{\odot}$ and therefore also covers nearly all of the protostars present.
%HK notes: Luhman09 paper is at: http://adsabs.harvard.edu/abs/2009ApJ...703..399L
%Completeness discussion for Spitzer at the bottom of page 413 in the pdf version.
%Therefore, some stars may be missed, particularly pre-main sequence stars that no longer have an infrared excess, but the \citet{Dunham15} catalog still represents a very exhaustive protostar catalog. Completeness estimates for \citep{Rebull10} could not be located, but due to the similar methods used and the small distance to B18, the completeness should be as good as, if not better than what is mentioned above.  

We define as protostellar all dense cores which have peaks located less than 7.05\arcsec\ from the nearest protostar, i.e., within half the JCMT SCUBA-2 850 $\mu$m beam size.  For our analysis, we exclude dense cores associated with any class of protostar or YSO.  Especially for the more evolved protostars, we expect that most of the material associated with the YSO is on scales smaller than can be resolved with the JCMT or GBT \citep[e.g.,][]{Sadavoy18}, and hence a kinematic analysis on these larger scales could be misleading; our focus here is on the starless dense cores.
%\soedit{Should some comment be made about completeness/sensitivity here?} \rknotes{[In progress: this is proving non-trivial to find, expect an update shortly]}

\subsection{GAS Observations}\label{sec:gasObs}

The first results from GAS were recently published as ``Data Release 1" (DR1) \citep{Friesen16}, which covered L1688 in Ophiuchus, B18 in Taurus, NGC 1333 in Perseus, and the northern section of Orion A \citep[which is covered on its own in][]{Kirk17b}. \edit1{Observations were carried out using the K-Band Focal Plane Array (KFPA) on the Green Bank Telescope, using the On-The-Fly  mapping  mode with a typical noise level of 0.1 K in the main beam scale. The NH$_3$ (1,1) and (2,2) transitions (23.7 GHz) were observed with spectral resolution of 5.7 kHz (0.07 km s$^{-1}$) and spatial resolution of 32\arcsec.} Further information on the Green Bank Ammonia Survey is available in \citet{Friesen16}, including information on data reduction and a general look at the properties of the data. The final GAS data products that are vital for this analysis are the property maps for kinetic temperature (T$_{k}$) and velocity dispersion ($\sigma$), which were \edit1{both} derived \edit1{by fitting the} ammonia (1,1) and (2,2) transitions simultaneously. Since sufficiently bright lines are required to attain reliable property data, the coverage of the T$_{k}$ and $\sigma$ maps is less extensive than the total GAS fields, covering just under three quarters of identified cores in the L1688 and B18 GAS fields, and about 90\% of the cores in the NGC 1333 field.

We \edit1{use} the \edit1{full} extents \edit1{derived from the} JCMT \edit1{{\get} extraction as} pixel selection \edit1{regions for the GAS data}. While it is true that the large beam size of GBT compared to JCMT (32" vs 14.1") extends the area over which core emission appears, the same effect will make the edge of a core more contaminated by external emission, therefore suggesting that a better 
representation of the core's properties can be found by adopting the original {\get} extent, rather than a reconvolved core area. We note that we tested our analysis using both smaller and larger pixel selection radii for each core, and found similar results. 
All pixels that overlap with the core boundary are included, although core-overlapping pixels whose centers are not located within the core's extent are assigned a weighting of one quarter that of an interior pixel, to minimize the impacts of pixels tracing more turbulent gas bordering cores. This criterion for pixel selection is also employed for the column density (\S~\ref{sec:cd}) and CO maps (\S~\ref{sec:co}). 

Within this core extent, we calculate the weighted average of each kinematic property (e.g., total velocity dispersion), \edit1{weighted} by the uncertainty \edit1{of} that property.  Pixels within core extents that did not have a value determined for the given property were excluded from the average. Any cores containing no informative pixels for any single property were dropped from the analysis entirely, leaving 74 cores in L1688, 51 in NGC~1333, and nine in B18. 
Thermal broadening of the NH$_3$ emission is much lower than that of the majority H$_2$ gas that dominates these regions, and therefore we calculate the total velocity dispersion expected using the standard formulation:
\begin{equation} \label{vtoteq}
\sigma_{tot} = \sqrt{\sigma^2 - \frac{k_B T_k}{\mu_{NH3}m_H} + \frac{k_B T_k}{\mu_{mn} m_H} } ,
\end{equation}
where $k_B$ is Boltzmann's constant, $\mu_{NH3}$ is the molecular weight of NH$_3$ (i.e. 17), $m_H$ is the mass of a hydrogen atom, and $\mu_{mn}$ is the mean molecular weight of the gas, which we take to be 2.37, following \citet{Kauffmann08}. %\jknotes{I think $\sim$2.33 should be used here instead of 2.8. 2.8 is the mean molecular weight per \textit{hydrogen} molecule, but we want the mean molecular weight per free particle.} 

In theory, the total velocity dispersion within each core could have one additional component that we have not yet considered, namely a variation in the centroid velocity of NH$_3$ across the core's area, due for instance to rotation or shear.  \edit1{Using the $v_{lsr}$ maps from GAS DR1 \citep{Friesen18}, we} calculated the standard deviation in centroid velocity across each core (\edit2{${\sigma}_{v_{lsr}}$}) and found that it was on average an order of magnitude smaller than the averaged line widths. There were, however, a few cases in which the magnitude of \edit2{${\sigma}_{v_{lsr}}$} was much larger (at least 40\% of the line width), including six cores in NGC~1333, eight in L1688, and, most notably, four of the nine cores in B18. \edit1{A visual inspection of this subset of 
cores revealed that some show clear centroid velocity gradients, which are typically attributed to rotation or shear \citep{Goodman93}, and are expected to contribute to the internal support term. A similar fraction of these high-\edit2{${\sigma}_{v_{lsr}}$} cores, however,} have centroid velocity variations driven by pixels exclusively near core edges, where very different $v_{lsr}$ values often appear. The most extreme \edit2{${\sigma}_{v_{lsr}}$} value appears in a faint core in B18, where noisy edge pixels have greater influence in our uncertainty-weighted statistics. Many other cases show evidence of contamination from overlapping structures at different centroid velocities, such as two B18 cores with emission from nearby protostellar sources on their edges. Similar contamination effects are also evident in some NGC~1333 and L1688 cores. Since our visual checks reveal that most if not all cores have minimal \edit2{${\sigma}_{v_{lsr}}$} components that can be confidently attributed to potential sources of internal core support, we omit this term from all further analysis.

%If you wanted to try ds9 on the spectral cubes yourself to see what I mean, you can now easily grab the publicly available DR1 data.  Cubes for B18 are here: (commented out in case of weird latex formatting issue
%https://dataverse.harvard.edu/dataset.xhtml?persistentId=doi:10.7910/DVN/LN8YFU
%and the fitted property maps are here:
%https://dataverse.harvard.edu/dataset.xhtml?persistentId=doi:10.7910/DVN/FEW0K1
%Ah, okay.  In case it's useful in future, there is a direct link from the Friesen \& Pineda first look paper.  Or you can just go to the main Harvard DataVerse page and search for 'Green Bank Ammonia Survey' and get it all that way.

\begin{figure}
\centering
\includegraphics[width=6.8in]{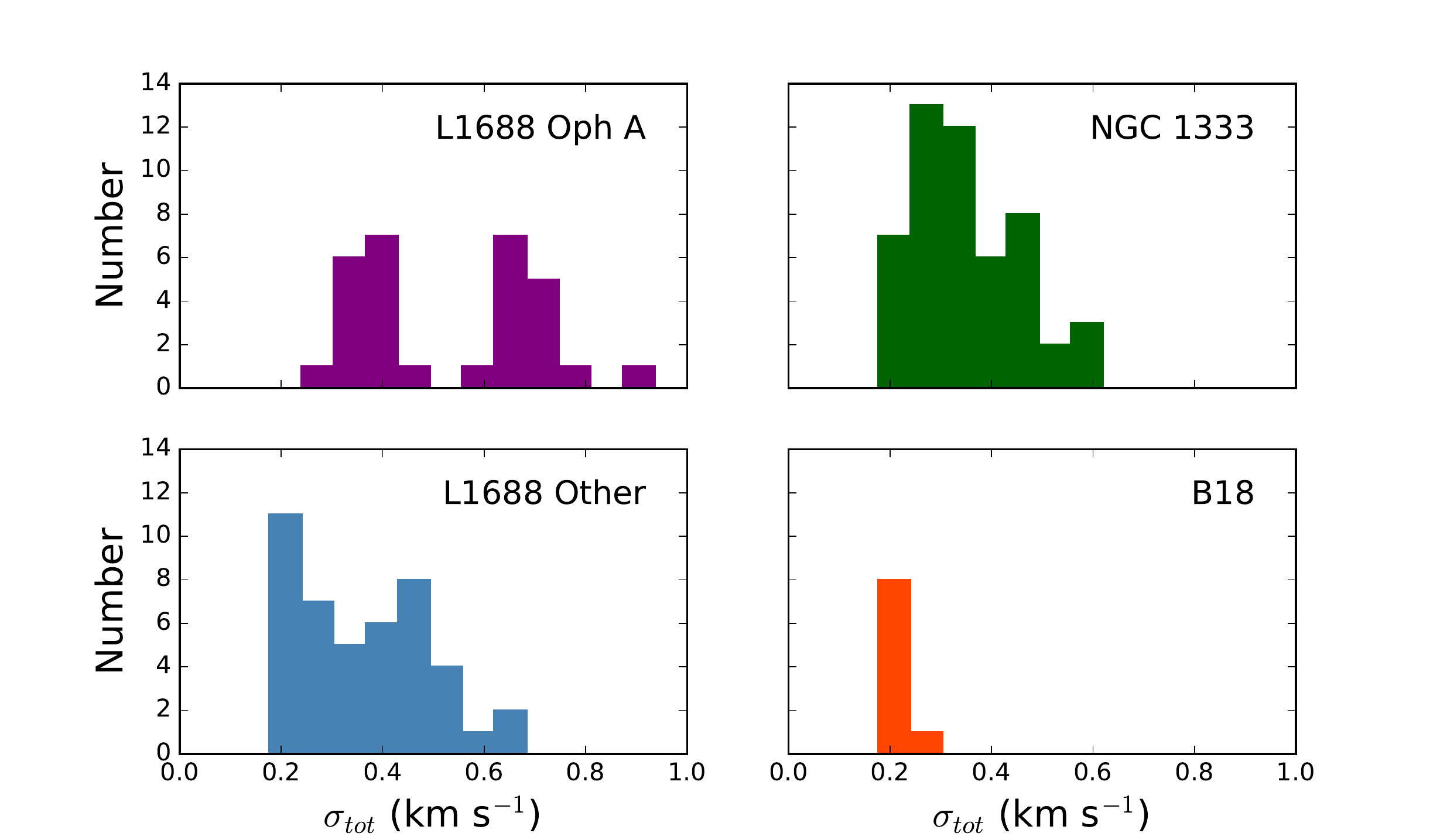}
\caption{The distribution of total line widths for starless cores found in each of the three regions.
}
\label{fig_all_sigtot}
\end{figure}

Figure~\ref{fig_all_sigtot} shows the distribution of the total velocity dispersions measured for the dense cores in each of the three clouds.  The Ophiuchus A region within L1688 (in the Ophiuchus molecular cloud) is shown separately, \edit2{as we notice very different kinematic properties there compared to the rest of the cloud. Oph A is notable for its proximity to the B star HD 147889 and other young, hot protostars \citep{Pattle15}, which are thought to raise the values of $T_k$ in the region, approaching $\sim$30-40 K in some places. Meanwhile, gas in the rest of L1688 has temperatures typically in the range of 10-20 K.
The bimodality in the velocity dispersion distribution in Oph A is caused by the structure of the region, in which a dense central filament is surrounded by less dense outlying material. The lower line widths are usually associated with the central region, whereas the higher line widths generally correspond to outlying cores. \citet{Pattle15} labelled this lower-density, high linewidth region surrounding central Oph A as Oph A', however for the sake of simplicity we present Oph A as a single region during our analysis. }
%\hknotes{HK note for Ronan -- my reading of Kate's description is that S1 is a secondary source of heating.  The main influence is HD~147889, which Oph A is closer to than the other regions.  Suggested re-word would either be to replace 'S1' with HD..., or else say 'HD... and S1'.}

\subsection{Column Density Maps} \label{sec:cd}
We also include total H$_2$ column density maps in our analysis to later estimate the pressure on the cores from the weight of surrounding cloud material. These maps are presented in \citet{Singh17}, in which a pixel-by-pixel fit to the spectral energy distribution of {\it Herschel} and {\it Planck} data is run to estimate the column density at 36{\arcsec} resolution.

\subsection{CO Maps} \label{sec:co}
Finally, we \edit1{include} 45{\arcsec} resolution $^{13}$CO (1-0) data maps from the Five College Radio Astronomy Observatory (FCRAO)\edit1{, which cover a wide region around each of the three clouds}. \edit1{$^{13}$CO typically traces gas with densities between 10$^3$ cm$^{-3}$ and 10$^4$ cm$^{-3}$, as the line becomes optically thick at higher densities \citep{Pineda08}\footnote{\edit1{Note that due to the relatively high abundance of CO, it is often subthermally excited, and hence can often be easily detected below its formal critical density.}}. This density range makes $^{13}$CO an ideal tracer for} estimating the turbulent pressure of the medium surrounding the dense cores.  The $^{13}$CO observations of NGC~1333 and L1688 were obtained by the Co-Ordinated Molecular Probe Line Extinction Thermal Emission (COMPLETE) Survey of Star Forming Regions \citep{Ridge06}, while B18 was covered by the Taurus Molecular Cloud Survey \citep{Narayanan12}. \edit1{A summary of the kinematic, virial, and {\get}-derived properties for all cores with complete data is provided in Tables \ref{tab:ophiuchus} (L1688), \ref{tab:perseus} (NGC~1333) and \ref{tab:taurus} (B18).}

\startlongtable
%\movetabledown=1.25in
\begin{deluxetable}{cccccccccccccccc}
\tablecolumns{16}
\tablewidth{0pt}
\tabletypesize{\footnotesize}
\tablecaption{Dense Core Properties in L1688 \label{tab:ophiuchus}}
\tablehead{
\colhead{ID\tablenotemark{a}} &
\colhead{R.A.\tablenotemark{a}} &
\colhead{Decl.\tablenotemark{a}} &
\colhead{M\tablenotemark{a}} &
\colhead{R$_{eff}$\tablenotemark{a}} &
%\colhead{$\sigma V$\tablenotemark{b}} &
%\colhead{$\sigma V_{err}$\tablenotemark{b}} &
%\colhead{$\sigma V_{std}$\tablenotemark{b}} &
\multicolumn{3}{c}{$\sigma_{obs}$ (km/s)\tablenotemark{b}} &
%\colhead{$T_{kin}$\tablenotemark{b}} &
%\colhead{$T_{kin,err}$\tablenotemark{b}} &
%\colhead{$T_{kin,std}$\tablenotemark{b}} &
\multicolumn{3}{c}{$T_{k}$(K)\tablenotemark{b}} &
\colhead{$\Omega_k$\tablenotemark{c}} &
\colhead{$-\Omega_g$\tablenotemark{c}}&
\colhead{$-\Omega_{p,w}$\tablenotemark{c}} &
\colhead{$-\Omega_{p,t}$\tablenotemark{c}}&
\colhead{Vir.\tablenotemark{d}}
\\
\colhead{} &
\colhead{(J2000)} &
\colhead{(J2000)} &
\colhead{(M$_{\odot}$)} &
\colhead{(pc)} &
\colhead{mn} &
\colhead{err} &
\colhead{std} &
\colhead{mn} &
\colhead{err} &
\colhead{std} &
\colhead{(erg)} &
\colhead{(erg)}&
\colhead{(erg)}&
\colhead{(erg)}&
\colhead{Rat.}
}
\startdata
2   & 246.61529 & -24.39922 & 2.60 & 0.014 & 0.263 & 0.001 & 0.016 & 17.9 & 0.1 & 0.4 & 9.5e+42 & 2.4e+43 & 1.4e+41 & 9.8e+40 & 1.3e+00 \\
3   & 246.61417 & -24.39099 & 1.95 & 0.013 & 0.242 & 0.001 & 0.027 & 17.3 & 0.1 & 0.8 & 6.4e+42 & 1.5e+43 & 9.2e+40 & 6.6e+40 & 1.2e+00 \\
9   & 246.59075 & -24.38132 & 0.08 & 0.009 & 0.674 & 0.016 & 0.051 & 26.1 & 0.7 & 1.0 & 1.2e+42 & 3.4e+40 & 3.0e+40 & 2.2e+40 & 3.5e-02 \\
12  & 246.61650 & -24.38225 & 0.38 & 0.014 & 0.264 & 0.001 & 0.052 & 17.8 & 0.1 & 0.8 & 1.4e+42 & 5.2e+41 & 1.2e+41 & 8.5e+40 & 2.6e-01 \\
13  & 246.62512 & -24.40843 & 2.71 & 0.024 & 0.244 & 0.001 & 0.050 & 18.4 & 0.1 & 1.1 & 9.3e+42 & 1.6e+43 & 6.2e+41 & 3.9e+41 & 9.2e-01 \\
16  & 246.63204 & -24.41403 & 0.13 & 0.008 & 0.211 & 0.002 & 0.038 & 17.8 & 0.2 & 0.8 & 3.7e+41 & 1.0e+41 & 2.6e+40 & 1.4e+40 & 1.9e-01 \\
21  & 246.87321 & -24.46245 & 0.14 & 0.011 & 0.329 & 0.005 & 0.077 & 15.5 & 0.2 & 0.5 & 6.5e+41 & 9.2e+40 & 6.6e+40 & 5.1e+40 & 1.6e-01 \\
25  & 246.87129 & -24.45139 & 0.15 & 0.009 & 0.269 & 0.002 & 0.065 & 13.6 & 0.1 & 0.4 & 5.2e+41 & 1.3e+41 & 3.7e+40 & 2.7e+40 & 1.9e-01 \\
26  & 246.86696 & -24.45261 & 0.43 & 0.015 & 0.367 & 0.002 & 0.056 & 14.0 & 0.1 & 0.6 & 2.3e+42 & 6.4e+41 & 1.6e+41 & 1.2e+41 & 2.0e-01 \\
27  & 246.62238 & -24.37677 & 0.17 & 0.014 & 0.320 & 0.003 & 0.054 & 19.6 & 0.2 & 1.4 & 8.3e+41 & 1.1e+41 & 1.2e+41 & 7.5e+40 & 1.8e-01 \\
28  & 246.60938 & -24.37360 & 0.10 & 0.014 & 0.327 & 0.003 & 0.078 & 21.1 & 0.2 & 1.2 & 5.3e+41 & 4.0e+40 & 1.2e+41 & 8.9e+40 & 2.4e-01 \\
38  & 246.87213 & -24.44358 & 0.21 & 0.013 & 0.327 & 0.003 & 0.061 & 13.9 & 0.2 & 0.4 & 9.4e+41 & 1.8e+41 & 1.1e+41 & 8.2e+40 & 2.0e-01 \\
39  & 246.89050 & -24.43856 & 0.91 & 0.027 & 0.286 & 0.002 & 0.078 & 14.2 & 0.1 & 0.9 & 3.4e+42 & 1.6e+42 & 8.9e+41 & 7.7e+41 & 4.8e-01 \\
40  & 246.54250 & -24.38684 & 0.04 & 0.012 & 0.571 & 0.011 & 0.018 & 24.5 & 0.5 & 1.6 & 4.9e+41 & 7.6e+39 & 6.1e+40 & 4.6e+40 & 1.2e-01 \\
41  & 246.60004 & -24.36541 & 0.03 & 0.011 & 0.308 & 0.003 & 0.064 & 21.5 & 0.2 & 1.6 & 1.3e+41 & 3.4e+39 & 5.2e+40 & 3.8e+40 & 3.7e-01 \\
44  & 246.87275 & -24.65492 & 0.01 & 0.008 & 0.630 & 0.026 & 0.009 & 21.5 & 1.1 & 0.7 & 1.5e+41 & 7.1e+38 & 2.4e+40 & 2.0e+40 & 1.5e-01 \\
45  & 246.99508 & -24.55845 & 0.58 & 0.023 & 0.125 & 0.000 & 0.014 & 11.1 & 0.1 & 0.4 & 8.5e+41 & 7.5e+41 & 5.0e+41 & 2.6e+41 & 8.9e-01 \\
46  & 246.80067 & -24.49384 & 0.49 & 0.029 & 0.420 & 0.001 & 0.033 & 15.7 & 0.1 & 0.6 & 3.3e+42 & 4.3e+41 & 1.3e+42 & 9.5e+41 & 4.1e-01 \\
48  & 246.77500 & -24.61921 & 0.02 & 0.008 & 0.577 & 0.021 & 0.075 & 15.4 & 1.0 & 1.2 & 2.4e+41 & 2.9e+39 & 2.7e+40 & 2.7e+40 & 1.2e-01 \\
50  & 246.85188 & -24.46204 & 0.27 & 0.014 & 0.427 & 0.003 & 0.029 & 13.5 & 0.1 & 0.5 & 1.8e+42 & 2.6e+41 & 1.3e+41 & 1.0e+41 & 1.4e-01 \\
51  & 246.76937 & -24.65370 & 0.55 & 0.020 & 0.134 & 0.001 & 0.032 & 12.3 & 0.1 & 0.7 & 9.0e+41 & 7.5e+41 & 3.9e+41 & 3.6e+41 & 8.4e-01 \\
52  & 246.88671 & -24.45047 & 0.14 & 0.013 & 0.188 & 0.002 & 0.035 & 12.4 & 0.1 & 0.8 & 3.1e+41 & 7.6e+40 & 1.1e+41 & 8.9e+40 & 4.6e-01 \\
53  & 246.85487 & -24.44933 & 1.28 & 0.032 & 0.413 & 0.001 & 0.052 & 13.6 & 0.1 & 0.8 & 8.1e+42 & 2.6e+42 & 1.6e+42 & 1.2e+42 & 3.4e-01 \\
57  & 246.62862 & -24.36973 & 0.09 & 0.015 & 0.364 & 0.008 & 0.050 & 22.7 & 0.6 & 1.7 & 5.5e+41 & 2.9e+40 & 1.5e+41 & 8.4e+40 & 2.4e-01 \\
58  & 246.83371 & -24.45318 & 0.11 & 0.015 & 0.405 & 0.004 & 0.052 & 14.7 & 0.2 & 0.9 & 6.9e+41 & 4.3e+40 & 1.7e+41 & 1.4e+41 & 2.5e-01 \\
59  & 246.54125 & -24.32932 & 0.12 & 0.019 & 0.608 & 0.015 & 0.078 & 24.4 & 0.7 & 2.9 & 1.5e+42 & 3.6e+40 & 2.6e+41 & 2.4e+41 & 1.7e-01 \\
60  & 246.81371 & -24.51046 & 0.44 & 0.021 & 0.308 & 0.001 & 0.042 & 14.7 & 0.1 & 1.1 & 1.8e+42 & 4.7e+41 & 5.2e+41 & 3.3e+41 & 3.6e-01 \\
61  & 246.59042 & -24.39326 & 0.02 & 0.013 & 0.605 & 0.036 & 0.057 & 31.1 & 2.1 & 1.4 & 2.7e+41 & 1.6e+39 & 9.7e+40 & 7.6e+40 & 3.2e-01 \\
62  & 246.79958 & -24.63337 & 0.11 & 0.013 & 0.485 & 0.017 & 0.097 & 11.9 & 2.5 & 0.0 & 8.7e+41 & 4.7e+40 & 1.0e+41 & 9.8e+40 & 1.4e-01 \\
63  & 246.87675 & -24.43000 & 0.04 & 0.009 & 0.380 & 0.012 & 0.097 & 14.6 & 0.9 & 0.1 & 2.1e+41 & 8.1e+39 & 3.6e+40 & 3.0e+40 & 1.7e-01 \\
65  & 246.84021 & -24.66555 & 0.38 & 0.022 & 0.450 & 0.003 & 0.163 & 15.0 & 0.1 & 1.0 & 2.8e+42 & 3.3e+41 & 4.9e+41 & 4.4e+41 & 2.2e-01 \\
66  & 246.86075 & -24.65661 & 0.01 & 0.009 & 0.615 & 0.022 & 0.057 & 20.9 & 1.3 & 0.6 & 1.3e+41 & 5.6e+38 & 2.8e+40 & 2.6e+40 & 2.1e-01 \\
69  & 246.62213 & -24.35786 & 0.01 & 0.011 & 0.183 & 0.004 & 0.034 & 18.8 & 0.6 & 0.9 & 3.2e+40 & 7.1e+38 & 5.2e+40 & 3.1e+40 & 1.3e+00 \\
70  & 246.54708 & -24.41278 & 0.02 & 0.013 & 0.836 & 0.017 & 0.075 & 27.2 & 0.6 & 3.3 & 4.8e+41 & 1.7e+39 & 8.8e+40 & 7.5e+40 & 1.7e-01 \\
71  & 246.54267 & -24.40085 & 0.01 & 0.008 & 0.555 & 0.016 & 0.040 & 26.0 & 0.9 & 2.1 & 8.9e+40 & 3.7e+38 & 2.3e+40 & 1.9e+40 & 2.4e-01 \\
72  & 246.53204 & -24.34111 & 0.09 & 0.019 & 0.501 & 0.009 & 0.064 & 21.3 & 0.4 & 2.5 & 8.1e+41 & 2.0e+40 & 2.5e+41 & 2.1e+41 & 3.0e-01 \\
73  & 247.11954 & -24.32155 & 1.18 & 0.022 & 0.102 & 0.001 & 0.006 & 8.4  & 0.7 & 0.6 & 1.3e+42 & 3.3e+42 & 2.1e+41 & 1.4e+41 & 1.4e+00 \\
75  & 246.63654 & -24.43725 & 0.02 & 0.012 & 0.570 & 0.014 & 0.099 & 25.5 & 0.7 & 4.3 & 2.8e+41 & 2.3e+39 & 8.4e+40 & 3.9e+40 & 2.3e-01 \\
76  & 246.75813 & -24.57900 & 0.30 & 0.013 & 0.144 & 0.001 & 0.012 & 10.5 & 0.1 & 0.2 & 4.6e+41 & 3.5e+41 & 1.1e+41 & 1.4e+41 & 6.5e-01 \\
77  & 246.67837 & -24.43469 & 0.09 & 0.030 & 0.290 & 0.004 & 0.092 & 18.7 & 0.2 & 1.7 & 4.0e+41 & 1.5e+40 & 1.5e+42 & 6.9e+41 & 2.7e+00 \\
78  & 246.74692 & -24.57280 & 5.87 & 0.042 & 0.133 & 0.000 & 0.026 & 10.7 & 0.0 & 0.7 & 8.7e+42 & 4.2e+43 & 4.0e+42 & 4.2e+42 & 2.9e+00 \\
79  & 246.82092 & -24.46278 & 0.05 & 0.013 & 0.405 & 0.003 & 0.040 & 14.5 & 0.1 & 0.7 & 3.2e+41 & 1.1e+40 & 1.1e+41 & 8.8e+40 & 3.3e-01 \\
83  & 246.57800 & -24.42014 & 0.02 & 0.011 & 0.663 & 0.016 & 0.070 & 26.4 & 0.8 & 1.3 & 2.9e+41 & 1.7e+39 & 5.8e+40 & 5.4e+40 & 1.9e-01 \\
84  & 246.59396 & -24.41963 & 0.01 & 0.010 & 0.630 & 0.027 & 0.033 & 25.2 & 1.4 & 2.8 & 1.5e+41 & 6.0e+38 & 4.8e+40 & 4.2e+40 & 3.0e-01 \\
87  & 246.79425 & -24.65705 & 0.20 & 0.023 & 0.483 & 0.006 & 0.226 & 15.9 & 0.3 & 1.1 & 1.7e+42 & 9.3e+40 & 5.6e+41 & 5.3e+41 & 3.5e-01 \\
88  & 246.56358 & -24.43950 & 0.02 & 0.012 & 0.638 & 0.051 & 0.022 & 25.0 & 2.7 & 0.0 & 2.3e+41 & 1.1e+39 & 6.6e+40 & 4.9e+40 & 2.6e-01 \\
90  & 246.85067 & -24.67744 & 0.17 & 0.019 & 0.130 & 0.001 & 0.048 & 12.5 & 0.1 & 1.0 & 2.8e+41 & 7.7e+40 & 3.1e+41 & 2.5e+41 & 1.2e+00 \\
91  & 246.50921 & -24.53763 & 0.03 & 0.017 & 0.333 & 0.008 & 0.039 & 23.1 & 0.5 & 1.9 & 1.6e+41 & 2.8e+39 & 1.6e+41 & 9.3e+40 & 7.7e-01 \\
92  & 246.68467 & -24.55352 & 0.09 & 0.022 & 0.148 & 0.001 & 0.030 & 12.1 & 0.1 & 1.1 & 1.7e+41 & 2.1e+40 & 5.7e+41 & 4.1e+41 & 3.0e+00 \\
93  & 246.61654 & -24.44259 & 0.03 & 0.023 & 0.750 & 0.046 & 0.075 & 30.1 & 2.1 & 2.8 & 5.2e+41 & 1.6e+39 & 5.6e+41 & 3.2e+41 & 8.4e-01 \\
94  & 246.84054 & -24.41269 & 0.06 & 0.014 & 0.123 & 0.002 & 0.007 & 12.4 & 0.3 & 0.5 & 8.6e+40 & 1.1e+40 & 1.3e+41 & 1.3e+41 & 1.6e+00 \\
95  & 247.24146 & -24.34838 & 1.80 & 0.035 & 0.120 & 0.001 & 0.009 & 10.2 & 0.1 & 0.6 & 2.4e+42 & 4.8e+42 & 6.0e+41 & 3.4e+41 & 1.2e+00 \\
96  & 246.72017 & -24.61907 & 0.02 & 0.011 & 0.186 & 0.004 & 0.065 & 12.2 & 0.6 & 1.1 & 3.8e+40 & 1.5e+39 & 7.0e+40 & 5.7e+40 & 1.7e+00 \\
99  & 246.90942 & -24.71049 & 0.09 & 0.024 & 0.142 & 0.001 & 0.026 & 13.1 & 0.1 & 1.0 & 1.7e+41 & 1.9e+40 & 5.2e+41 & 2.5e+41 & 2.4e+00 \\
101 & 246.80104 & -24.43306 & 0.04 & 0.021 & 0.426 & 0.009 & 0.052 & 18.2 & 0.6 & 1.4 & 2.9e+41 & 4.2e+39 & 5.0e+41 & 4.6e+41 & 1.6e+00 \\
102 & 246.56946 & -24.38237 & 0.01 & 0.012 & 0.692 & 0.016 & 0.090 & 22.5 & 0.7 & 2.8 & 1.5e+41 & 3.4e+38 & 7.2e+40 & 5.8e+40 & 4.5e-01 \\
104 & 246.70058 & -24.39840 & 0.03 & 0.022 & 0.333 & 0.014 & 0.071 & 22.7 & 1.2 & 2.6 & 1.5e+41 & 1.8e+39 & 5.5e+41 & 2.8e+41 & 2.8e+00 \\
105 & 246.93021 & -24.71032 & 0.05 & 0.016 & 0.235 & 0.003 & 0.043 & 14.2 & 0.2 & 0.9 & 1.3e+41 & 6.7e+39 & 1.5e+41 & 7.8e+40 & 8.9e-01 \\
107 & 246.70508 & -24.44994 & 0.12 & 0.025 & 0.346 & 0.005 & 0.096 & 17.2 & 0.4 & 2.5 & 6.2e+41 & 3.1e+40 & 8.2e+41 & 4.7e+41 & 1.1e+00 \\
108 & 246.86646 & -24.41034 & 0.22 & 0.034 & 0.134 & 0.005 & 0.046 & 11.7 & 2.2 & 0.0 & 3.4e+41 & 7.1e+40 & 1.8e+42 & 1.6e+42 & 5.1e+00 \\
109 & 246.73371 & -24.61132 & 0.06 & 0.022 & 0.181 & 0.002 & 0.104 & 14.3 & 0.2 & 1.7 & 1.4e+41 & 8.5e+39 & 5.1e+41 & 5.0e+41 & 3.7e+00 \\
110 & 246.73967 & -24.52867 & 0.71 & 0.028 & 0.123 & 0.000 & 0.023 & 10.5 & 0.1 & 0.8 & 9.9e+41 & 9.2e+41 & 1.2e+42 & 1.0e+42 & 1.6e+00 \\
112 & 246.72262 & -24.42202 & 0.02 & 0.018 & 0.336 & 0.010 & 0.050 & 19.8 & 1.3 & 2.1 & 1.0e+41 & 1.2e+39 & 3.0e+41 & 1.8e+41 & 2.3e+00 \\
113 & 246.56167 & -24.38003 & 0.02 & 0.020 & 0.613 & 0.008 & 0.082 & 22.8 & 0.4 & 2.6 & 2.6e+41 & 1.0e+39 & 3.3e+41 & 2.8e+41 & 1.2e+00 \\
115 & 246.63904 & -24.41661 & 0.06 & 0.011 & 0.204 & 0.002 & 0.029 & 17.7 & 0.2 & 0.9 & 1.8e+41 & 1.9e+40 & 6.3e+40 & 3.3e+40 & 3.2e-01 \\
116 & 246.69417 & -24.48982 & 0.64 & 0.046 & 0.263 & 0.002 & 0.114 & 18.1 & 0.1 & 2.9 & 2.3e+42 & 4.5e+41 & 5.3e+42 & 3.0e+42 & 1.9e+00 \\
117 & 247.13250 & -24.30284 & 0.50 & 0.030 & 0.094 & 0.000 & 0.007 & 10.0 & 0.1 & 0.7 & 5.8e+41 & 4.3e+41 & 5.3e+41 & 4.5e+41 & 1.2e+00 \\
118 & 246.81575 & -24.40339 & 0.07 & 0.022 & 0.201 & 0.004 & 0.051 & 15.3 & 0.4 & 1.1 & 1.7e+41 & 1.1e+40 & 5.4e+41 & 4.4e+41 & 2.8e+00 \\
123 & 246.71762 & -24.58159 & 0.13 & 0.026 & 0.354 & 0.003 & 0.056 & 14.8 & 0.2 & 0.9 & 6.7e+41 & 3.4e+40 & 9.4e+41 & 7.9e+41 & 1.3e+00 \\
124 & 246.59242 & -24.37514 & 0.03 & 0.014 & 0.575 & 0.007 & 0.100 & 23.5 & 0.3 & 1.3 & 3.1e+41 & 2.5e+39 & 1.1e+41 & 8.1e+40 & 3.1e-01 \\
126 & 246.91358 & -24.68730 & 0.07 & 0.024 & 0.376 & 0.005 & 0.047 & 15.6 & 0.3 & 2.2 & 4.1e+41 & 1.2e+40 & 5.5e+41 & 2.4e+41 & 9.7e-01 \\
127 & 246.74225 & -24.38574 & 0.50 & 0.062 & 0.231 & 0.003 & 0.056 & 16.1 & 0.3 & 3.1 & 1.5e+42 & 2.1e+41 & 1.2e+43 & 8.0e+42 & 6.5e+00 \\
129 & 246.74675 & -24.55165 & 0.24 & 0.023 & 0.163 & 0.001 & 0.023 & 12.0 & 0.1 & 0.5 & 4.5e+41 & 1.3e+41 & 6.7e+41 & 6.9e+41 & 1.7e+00 \\
133 & 246.93429 & -24.61873 & 0.25 & 0.058 & 0.438 & 0.003 & 0.049 & 17.3 & 0.1 & 1.8 & 1.8e+42 & 5.7e+40 & 8.3e+42 & 4.9e+42 & 3.6e+00 \\
\enddata
%\vspace*{0.1in}
\tablenotetext{a}{These core properties are based on our {\get} source extraction, which was performed on the JCMT GBS continuum maps. Only starless cores with coverage from GAS %, FCRAO, and the column density maps \citep{Singh17} 
are listed. These columns display the core ID, location of the core centre, mass (estimated using Equation \ref{eqn:mass}), and effective radius (geometric mean of the major and minor axes reported by {\get})}
\tablenotetext{b}{Observed velocity dispersion ($\sigma$) and kinetic temperature ($T_{k}$) are estimated
using a weighted average of GAS kinematic data. We also report the error in the weighted mean and the weighted standard deviation, the latter being more reflective of the variation in GAS values across extent of the core.}
\tablenotetext{c}{These virial parameters are derived according to the methods presented in Section \ref{sec:fullvirial}.}
\tablenotetext{d}{The virial ratio, which indicates whether or not the core is bound by the combined forces of gravity and pressure, is given by $-(\Omega_g+\Omega_{p,w}+\Omega_{p,t})/2\Omega_k$.}

\end{deluxetable}
\startlongtable
\movetabledown=1.25in
\begin{deluxetable}{cccccccccccccccc}
\tablecolumns{16}
\tablewidth{0pt}
\tabletypesize{\footnotesize}
\tablecaption{Dense Core Properties in NGC~1333 \label{tab:perseus}}
\tablehead{
\colhead{ID\tablenotemark{a}} &
\colhead{R.A.\tablenotemark{a}} &
\colhead{Decl.\tablenotemark{a}} &
\colhead{M\tablenotemark{a}} &
\colhead{R$_{eff}$\tablenotemark{a}} &
%\colhead{$\sigma V$\tablenotemark{b}} &
%\colhead{$\sigma V_{err}$\tablenotemark{b}} &
%\colhead{$\sigma V_{std}$\tablenotemark{b}} &
\multicolumn{3}{c}{$\sigma_{obs}$ (km/s)\tablenotemark{b}} &
%\colhead{$T_{kin}$\tablenotemark{b}} &
%\colhead{$T_{kin,err}$\tablenotemark{b}} &
%\colhead{$T_{kin,std}$\tablenotemark{b}} &
\multicolumn{3}{c}{$T_{k}$(K)\tablenotemark{b}} &
\colhead{$\Omega_k$\tablenotemark{c}} &
\colhead{$-\Omega_g$\tablenotemark{d}}&
\colhead{$-\Omega_{p,w}$\tablenotemark{d}} &
\colhead{$-\Omega_{p,t}$\tablenotemark{d}}&
\colhead{Vir.\tablenotemark{d}}
\\
\colhead{} &
\colhead{(J2000)} &
\colhead{(J2000)} &
\colhead{(M$_{\odot}$)} &
\colhead{(pc)} &
\colhead{mn} &
\colhead{err} &
\colhead{std} &
\colhead{mn} &
\colhead{err} &
\colhead{std} &
\colhead{(erg)} &
\colhead{(erg)}&
\colhead{(erg)}&
\colhead{(erg)}&
\colhead{Rat.}
}
\startdata
5  & 52.26300 & 31.26464 & 1.86 & 0.016 & 0.310 & 0.001 & 0.035 & 17.0 & 0.1 & 0.4 & 8.1e+42 & 1.1e+43 & 1.5e+41 & 3.6e+41 & 7.1e-01 \\
7  & 52.25846 & 31.26069 & 0.92 & 0.016 & 0.384 & 0.002 & 0.073 & 16.9 & 0.1 & 0.6 & 5.5e+42 & 2.7e+42 & 1.5e+41 & 3.7e+41 & 3.0e-01 \\
11 & 52.25717 & 31.34161 & 1.98 & 0.028 & 0.261 & 0.001 & 0.015 & 15.8 & 0.1 & 0.8 & 6.8e+42 & 7.1e+42 & 7.5e+41 & 1.4e+42 & 6.8e-01 \\
15 & 52.25900 & 31.33671 & 0.47 & 0.016 & 0.234 & 0.002 & 0.011 & 15.5 & 0.1 & 0.2 & 1.4e+42 & 7.2e+41 & 1.4e+41 & 2.7e+41 & 3.9e-01 \\
16 & 52.32925 & 31.38748 & 0.34 & 0.017 & 0.176 & 0.006 & 0.034 & 15.0 & 0.7 & 0.5 & 7.8e+41 & 3.5e+41 & 1.1e+41 & 1.9e+41 & 4.2e-01 \\
17 & 52.24992 & 31.35978 & 1.24 & 0.021 & 0.328 & 0.002 & 0.029 & 14.3 & 0.1 & 0.4 & 5.6e+42 & 3.8e+42 & 2.8e+41 & 4.5e+41 & 4.0e-01 \\
19 & 52.27758 & 31.26116 & 0.85 & 0.020 & 0.383 & 0.002 & 0.039 & 14.7 & 0.1 & 0.6 & 4.8e+42 & 1.9e+42 & 2.7e+41 & 5.8e+41 & 2.8e-01 \\
22 & 52.26150 & 31.25393 & 0.32 & 0.021 & 0.547 & 0.002 & 0.064 & 17.0 & 0.1 & 0.4 & 3.3e+42 & 2.4e+41 & 3.4e+41 & 7.8e+41 & 2.1e-01 \\
23 & 52.26713 & 31.34976 & 0.63 & 0.021 & 0.257 & 0.002 & 0.009 & 14.4 & 0.1 & 0.3 & 2.1e+42 & 9.6e+41 & 3.0e+41 & 4.7e+41 & 4.2e-01 \\
25 & 52.27925 & 31.29047 & 0.74 & 0.020 & 0.224 & 0.002 & 0.018 & 12.9 & 0.1 & 0.4 & 2.0e+42 & 1.4e+42 & 2.8e+41 & 8.5e+41 & 6.5e-01 \\
26 & 52.22392 & 31.24904 & 0.77 & 0.024 & 0.477 & 0.002 & 0.023 & 14.1 & 0.1 & 0.4 & 6.2e+42 & 1.3e+42 & 4.8e+41 & 1.0e+42 & 2.2e-01 \\
27 & 52.27150 & 31.35081 & 0.69 & 0.021 & 0.250 & 0.002 & 0.012 & 14.7 & 0.1 & 0.5 & 2.2e+42 & 1.1e+42 & 3.0e+41 & 4.4e+41 & 4.3e-01 \\
28 & 52.28754 & 31.25473 & 2.03 & 0.031 & 0.427 & 0.001 & 0.053 & 12.7 & 0.0 & 0.6 & 1.3e+43 & 6.7e+42 & 1.1e+42 & 2.2e+42 & 3.8e-01 \\
29 & 52.32633 & 31.41929 & 0.70 & 0.020 & 0.246 & 0.005 & 0.051 & 12.8 & 0.4 & 0.6 & 2.1e+42 & 1.3e+42 & 1.5e+41 & 2.1e+41 & 4.0e-01 \\
30 & 52.35658 & 31.47147 & 0.75 & 0.018 & 0.209 & 0.002 & 0.023 & 10.6 & 0.2 & 0.3 & 1.7e+42 & 1.6e+42 & 8.1e+40 & 3.9e+41 & 6.2e-01 \\
32 & 52.29475 & 31.36245 & 1.33 & 0.034 & 0.328 & 0.019 & 0.245 & 18.5 & 2.4 & 7.5 & 6.5e+42 & 2.7e+42 & 1.1e+42 & 1.5e+42 & 4.1e-01 \\
33 & 52.30683 & 31.31146 & 0.98 & 0.032 & 0.235 & 0.001 & 0.021 & 13.4 & 0.1 & 0.5 & 2.8e+42 & 1.5e+42 & 1.1e+42 & 3.2e+42 & 1.0e+00 \\
34 & 52.31754 & 31.21038 & 1.51 & 0.038 & 0.352 & 0.002 & 0.036 & 11.2 & 0.1 & 0.6 & 7.1e+42 & 3.1e+42 & 1.5e+42 & 3.2e+42 & 5.5e-01 \\
35 & 52.30712 & 31.30276 & 0.96 & 0.028 & 0.224 & 0.002 & 0.028 & 13.0 & 0.1 & 0.3 & 2.6e+42 & 1.7e+42 & 7.1e+41 & 2.1e+42 & 8.7e-01 \\
37 & 52.28554 & 31.28593 & 0.88 & 0.029 & 0.247 & 0.002 & 0.033 & 13.7 & 0.1 & 1.0 & 2.7e+42 & 1.4e+42 & 8.8e+41 & 2.4e+42 & 8.7e-01 \\
39 & 52.34996 & 31.55555 & 1.20 & 0.028 & 0.129 & 0.001 & 0.003 & 10.6 & 0.1 & 0.3 & 1.7e+42 & 2.7e+42 & 2.4e+41 & 2.3e+41 & 9.0e-01 \\
41 & 52.27354 & 31.27033 & 0.21 & 0.019 & 0.280 & 0.002 & 0.040 & 16.6 & 0.1 & 0.5 & 7.9e+41 & 1.1e+41 & 2.5e+41 & 6.0e+41 & 6.1e-01 \\
42 & 52.27150 & 31.37014 & 0.15 & 0.026 & 0.539 & 0.020 & 0.152 & 21.8 & 1.9 & 1.7 & 1.6e+42 & 4.3e+40 & 5.0e+41 & 8.7e+41 & 4.5e-01 \\
43 & 52.33325 & 31.39885 & 0.39 & 0.029 & 0.170 & 0.004 & 0.027 & 15.0 & 0.4 & 1.0 & 8.5e+41 & 2.7e+41 & 5.0e+41 & 9.2e+41 & 9.9e-01 \\
44 & 52.29983 & 31.24761 & 0.19 & 0.020 & 0.243 & 0.002 & 0.034 & 12.5 & 0.1 & 0.3 & 5.4e+41 & 8.9e+40 & 2.8e+41 & 5.1e+41 & 8.2e-01 \\
45 & 52.16458 & 31.30774 & 2.39 & 0.034 & 0.208 & 0.001 & 0.022 & 11.8 & 0.0 & 0.3 & 5.6e+42 & 8.6e+42 & 1.1e+42 & 1.5e+42 & 1.0e+00 \\
46 & 52.26054 & 31.19716 & 0.23 & 0.020 & 0.187 & 0.002 & 0.024 & 12.5 & 0.2 & 0.3 & 4.9e+41 & 1.3e+41 & 2.2e+41 & 5.3e+41 & 9.0e-01 \\
47 & 52.25308 & 31.23985 & 0.12 & 0.021 & 0.512 & 0.009 & 0.128 & 17.6 & 0.4 & 1.5 & 1.1e+42 & 3.3e+40 & 3.2e+41 & 6.9e+41 & 4.7e-01 \\
48 & 52.27067 & 31.31327 & 0.52 & 0.032 & 0.234 & 0.002 & 0.024 & 13.8 & 0.1 & 0.6 & 1.5e+42 & 4.4e+41 & 1.1e+42 & 3.8e+42 & 1.8e+00 \\
50 & 52.28117 & 31.24738 & 0.46 & 0.023 & 0.421 & 0.001 & 0.096 & 12.2 & 0.1 & 0.2 & 3.0e+42 & 4.9e+41 & 4.1e+41 & 8.1e+41 & 2.9e-01 \\
52 & 52.21137 & 31.24105 & 0.21 & 0.025 & 0.410 & 0.002 & 0.016 & 13.7 & 0.1 & 0.5 & 1.3e+42 & 8.8e+40 & 4.9e+41 & 9.4e+41 & 5.9e-01 \\
53 & 52.17858 & 31.10492 & 0.27 & 0.021 & 0.132 & 0.001 & 0.016 & 10.5 & 0.1 & 0.1 & 3.9e+41 & 1.8e+41 & 1.4e+41 & 3.9e+41 & 9.0e-01 \\
54 & 52.29508 & 31.29623 & 0.42 & 0.022 & 0.208 & 0.002 & 0.023 & 12.3 & 0.2 & 0.4 & 1.0e+42 & 4.3e+41 & 3.5e+41 & 1.0e+42 & 8.8e-01 \\
55 & 52.17567 & 31.29345 & 1.98 & 0.069 & 0.203 & 0.000 & 0.022 & 11.9 & 0.0 & 0.4 & 4.6e+42 & 2.9e+42 & 9.6e+42 & 1.4e+43 & 2.9e+00 \\
56 & 52.23917 & 31.36732 & 0.15 & 0.021 & 0.463 & 0.005 & 0.051 & 15.1 & 0.2 & 0.5 & 1.2e+42 & 5.9e+40 & 2.6e+41 & 3.7e+41 & 2.9e-01 \\
57 & 52.26900 & 31.22556 & 0.08 & 0.020 & 0.435 & 0.010 & 0.043 & 16.2 & 0.5 & 1.0 & 5.4e+41 & 1.5e+40 & 2.5e+41 & 5.1e+41 & 7.2e-01 \\
58 & 52.13675 & 31.08251 & 0.20 & 0.023 & 0.147 & 0.001 & 0.015 & 10.6 & 0.1 & 0.3 & 3.2e+41 & 9.1e+40 & 1.6e+41 & 7.3e+41 & 1.5e+00 \\
60 & 52.19796 & 31.25258 & 0.22 & 0.026 & 0.174 & 0.001 & 0.020 & 11.3 & 0.1 & 0.6 & 4.1e+41 & 9.2e+40 & 5.7e+41 & 9.1e+41 & 1.9e+00 \\
61 & 52.18950 & 31.09689 & 0.11 & 0.016 & 0.111 & 0.001 & 0.003 & 10.6 & 0.2 & 0.2 & 1.4e+41 & 3.5e+40 & 6.1e+40 & 1.7e+41 & 9.7e-01 \\
62 & 52.30725 & 31.34255 & 0.64 & 0.040 & 0.367 & 0.006 & 0.044 & 16.1 & 0.5 & 1.4 & 3.5e+42 & 5.1e+41 & 2.0e+42 & 4.1e+42 & 9.4e-01 \\
63 & 52.25246 & 31.35077 & 0.11 & 0.023 & 0.314 & 0.002 & 0.022 & 16.4 & 0.1 & 0.7 & 4.7e+41 & 2.6e+40 & 3.7e+41 & 6.4e+41 & 1.1e+00 \\
64 & 52.13550 & 31.25533 & 0.21 & 0.028 & 0.184 & 0.002 & 0.047 & 11.2 & 0.1 & 0.7 & 4.3e+41 & 8.3e+40 & 4.8e+41 & 8.1e+41 & 1.6e+00 \\
65 & 52.12367 & 31.16582 & 0.26 & 0.022 & 0.099 & 0.001 & 0.003 & 9.6  & 0.2 & 0.4 & 3.0e+41 & 1.5e+41 & 1.8e+41 & 1.0e+42 & 2.3e+00 \\
66 & 52.13075 & 31.23895 & 0.34 & 0.030 & 0.166 & 0.001 & 0.023 & 10.5 & 0.2 & 0.3 & 5.9e+41 & 1.9e+41 & 5.8e+41 & 1.2e+42 & 1.7e+00 \\
67 & 52.34829 & 31.60201 & 0.63 & 0.031 & 0.115 & 0.001 & 0.010 & 9.4  & 0.1 & 0.6 & 7.8e+41 & 6.4e+41 & 3.0e+41 & 3.4e+41 & 8.3e-01 \\
68 & 52.33888 & 31.22258 & 0.14 & 0.031 & 0.389 & 0.010 & 0.047 & 13.9 & 0.8 & 2.5 & 8.0e+41 & 3.1e+40 & 7.9e+41 & 1.7e+42 & 1.6e+00 \\
69 & 52.23613 & 31.34918 & 0.05 & 0.020 & 0.381 & 0.005 & 0.024 & 18.2 & 0.3 & 0.9 & 3.0e+41 & 6.2e+39 & 2.6e+41 & 4.2e+41 & 1.2e+00 \\
70 & 52.34258 & 31.42884 & 0.34 & 0.041 & 0.103 & 0.001 & 0.017 & 12.1 & 0.1 & 0.3 & 4.7e+41 & 1.4e+41 & 1.2e+42 & 2.4e+42 & 3.9e+00 \\
73 & 52.32963 & 31.19548 & 1.81 & 0.049 & 0.310 & 0.002 & 0.067 & 10.8 & 0.1 & 0.8 & 6.9e+42 & 3.4e+42 & 2.8e+42 & 6.0e+42 & 8.8e-01 \\
75 & 52.23483 & 31.32137 & 0.18 & 0.034 & 0.199 & 0.001 & 0.026 & 15.9 & 0.1 & 1.5 & 4.7e+41 & 4.8e+40 & 1.4e+42 & 3.9e+42 & 5.7e+00 \\
76 & 52.16329 & 31.31961 & 0.70 & 0.040 & 0.261 & 0.001 & 0.039 & 13.1 & 0.1 & 0.8 & 2.2e+42 & 6.3e+41 & 1.6e+42 & 3.1e+42 & 1.2e+00
\enddata
\vspace*{0.1in}
\tablenotetext{a}{These core properties are based on our {\get} source extraction, which was performed on the JCMT GBS continuum maps. Only starless cores with coverage from GAS, FCRAO, and the column density maps \citep{Singh17} are listed. These columns display the core ID, location of the core centre, mass (estimated using Equation \ref{eqn:mass}), and effective radius (geometric mean of the major and minor axes reported by {\get})}
\tablenotetext{b}{Observed velocity dispersion ($\sigma$) and kinetic temperature ($T_{k}$) are estimated using a weighted average of GAS kinematic data. We also report the error in the weighted mean and the weighted standard deviation, the latter being more reflective of the variation in GAS values across extent of the core.}
\tablenotetext{c}{These virial parameters are derived according to the methods presented in Section \ref{sec:fullvirial}}
\tablenotetext{d}{The virial ratio, which indicates whether or not the core is bound by the combined forces of gravity and pressure, is given by $-(\Omega_g+\Omega_{p,w}+\Omega_{p,t})/2\Omega_k$.}
\end{deluxetable}
\startlongtable
\movetabledown=1.25in
\begin{deluxetable}{cccccccccccccccc}
\tablecolumns{16}
\tablewidth{0pt}
\tabletypesize{\footnotesize}
\tablecaption{Dense Core Properties in B18 \label{tab:taurus}}
\tablehead{
\colhead{ID\tablenotemark{a}} &
\colhead{R.A.\tablenotemark{a}} &
\colhead{Decl.\tablenotemark{a}} &
\colhead{M\tablenotemark{a}} &
\colhead{R$_{eff}$\tablenotemark{a}} &
%\colhead{$\sigma V$\tablenotemark{b}} &
%\colhead{$\sigma V_{err}$\tablenotemark{b}} &
%\colhead{$\sigma V_{std}$\tablenotemark{b}} &
\multicolumn{3}{c}{$\sigma_{obs}$ (km/s)\tablenotemark{b}} &
%\colhead{$T_{kin}$\tablenotemark{b}} &
%\colhead{$T_{kin,err}$\tablenotemark{b}} &
%\colhead{$T_{kin,std}$\tablenotemark{b}} &
\multicolumn{3}{c}{$T_{k}$(K)\tablenotemark{b}} &
\colhead{$\Omega_k$\tablenotemark{c}} &
\colhead{$-\Omega_g$\tablenotemark{d}}&
\colhead{$-\Omega_{p,w}$\tablenotemark{d}} &
\colhead{$-\Omega_{p,t}$\tablenotemark{d}}&
\colhead{Vir.\tablenotemark{d}}
\\
\colhead{} &
\colhead{(J2000)} &
\colhead{(J2000)} &
\colhead{(M$_{\odot}$)} &
\colhead{(pc)} &
\colhead{mn} &
\colhead{err} &
\colhead{std} &
\colhead{mn} &
\colhead{err} &
\colhead{std} &
\colhead{(erg)} &
\colhead{(erg)}&
\colhead{(erg)}&
\colhead{(erg)}&
\colhead{Rat.}
}
\startdata
8  & 67.34042 & 24.55539 & 0.13 & 0.019 & 0.118 & 0.002 & 0.013 & 12.8 & 0.4 & 1.1 & 2.0e+41 & 4.6e+40 & 6.3e+40 & 1.3e+41 & 5.8e-01 \\
9 & 68.90600 & 24.15507 & 4.85 & 0.044 & 0.133 & 0.000 & 0.025 & 9.9  & 0.0 & 0.3 & 6.9e+42 & 2.8e+43 & 4.6e+41 & 8.1e+41 & 2.1e+00 \\
11 & 67.98225 & 24.54840 & 1.43 & 0.031 & 0.103 & 0.000 & 0.009 & 8.8  & 0.0 & 0.3 & 1.6e+42 & 3.4e+42 & 2.1e+41 & 4.7e+41 & 1.3e+00 \\
12 & 66.64562 & 24.69573 & 0.14 & 0.029 & 0.116 & 0.003 & 0.029 & 14.7 & 0.9 & 3.8 & 2.4e+41 & 3.5e+40 & 1.3e+41 & 2.4e+42 & 5.3e+00 \\
13 & 68.18908 & 24.36316 & 0.49 & 0.028 & 0.151 & 0.001 & 0.022 & 9.8  & 0.2 & 0.7 & 7.6e+41 & 4.4e+41 & 1.7e+41 & 3.2e+41 & 6.1e-01 \\
14 & 67.34754 & 24.57879 & 0.66 & 0.043 & 0.111 & 0.000 & 0.007 & 10.0 & 0.0 & 0.5 & 8.3e+41 & 5.2e+41 & 7.2e+41 & 1.4e+42 & 1.6e+00 \\
15 & 68.19529 & 24.42679 & 0.81 & 0.036 & 0.097 & 0.000 & 0.010 & 9.3  & 0.0 & 0.5 & 9.0e+41 & 9.3e+41 & 3.7e+41 & 5.7e+41 & 1.0e+00 \\
16 & 68.18333 & 24.38401 & 0.66 & 0.032 & 0.122 & 0.000 & 0.009 & 9.3  & 0.1 & 0.5 & 8.4e+41 & 6.9e+41 & 2.7e+41 & 4.7e+41 & 8.5e-01 \\
17 & 68.96121 & 24.16152 & 0.58 & 0.037 & 0.118 & 0.001 & 0.023 & 10.6 & 0.3 & 1.6 & 8.0e+41 & 4.7e+41 & 2.7e+41 & 5.9e+41 & 8.3e-01
\enddata
\vspace*{0.1in}
\tablenotetext{a}{These core properties are based on our {\get} source extraction, which was performed on the JCMT GBS continuum maps. Only starless cores with coverage from GAS, FCRAO, and the column density maps \citep{Singh17} are listed. These columns display the core ID, location of the core centre, mass (estimated using Equation \ref{eqn:mass}), and effective radius (geometric mean of the major and minor axes reported by {\get})}
\tablenotetext{b}{Observed velocity dispersion ($\sigma$) and kinetic temperature ($T_{k}$) are estimated using a weighted average of GAS kinematic data. We also report the error in the weighted mean and the weighted standard deviation, the latter being more reflective of the variation in GAS values across extent of the core.}
\tablenotetext{c}{These virial parameters are derived according to the methods presented in Section \ref{sec:fullvirial}}
\tablenotetext{d}{The virial ratio, which indicates whether or not the core is bound by the combined forces of gravity and pressure, is given by $-(\Omega_g+\Omega_{p,w}+\Omega_{p,t})/2\Omega_k$.}

\end{deluxetable}

\section{Basic Jeans Analysis}\label{sec:Jeans}

\begin{figure}
%\centering
\includegraphics[width=3.5in]{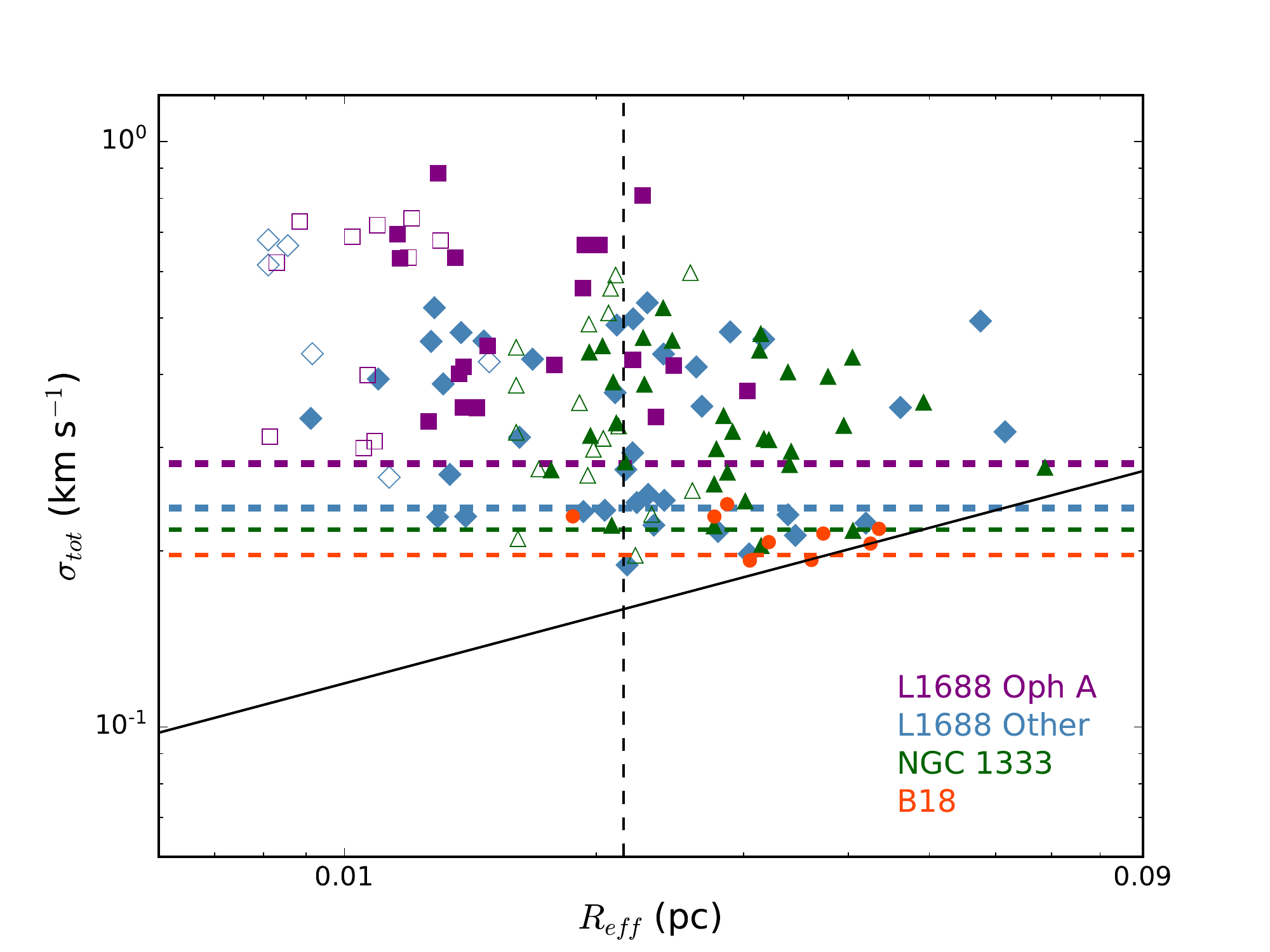}
\includegraphics[width=3.5in]{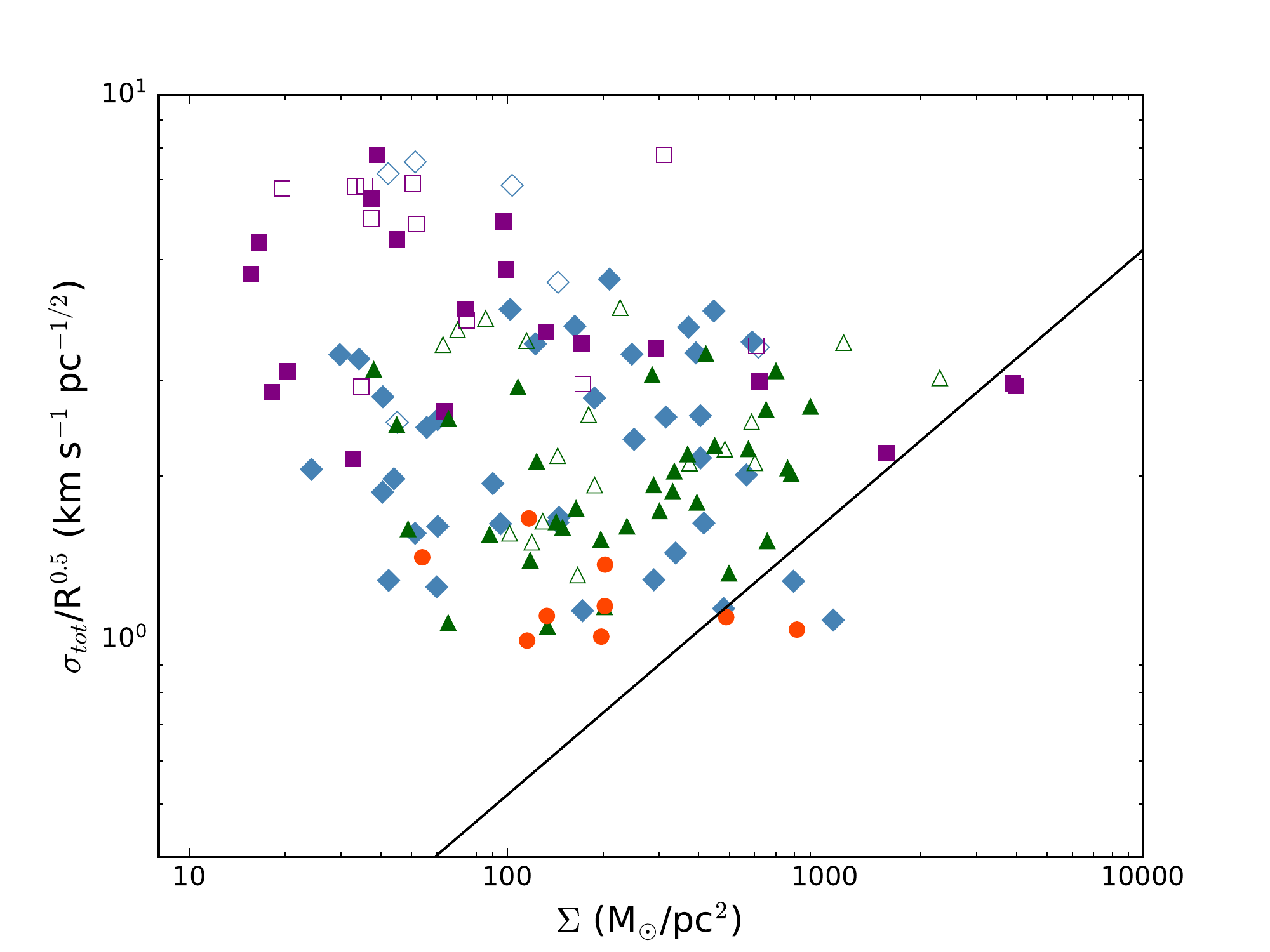}
\caption{\edit1{The first panel shows t}otal velocity dispersion versus size for all cores analyzed.  Dashed horizontal lines indicate the thermal velocity dispersion for the mean core kinetic temperature in each cloud, color coded according to the same convention as the markers for cores. The solid black diagonal line shows Larson's Law. \edit1{The second panel is analagous to Figure 7 in \citet{Heyer09}, with their proposed size-line width relation shown as a solid diagonal line. In both panels, purple squares show cores in the Oph A region of L1688 (Ophiuchus), blue diamonds show the rest of L1688, green triangles show NGC 1333 (Perseus), and red circles show B18 (Taurus). Open symbols represent unresolved cores (i.e., where $R_{eff}$ is an upper limit).}}
\label{fig_larson}
\end{figure}

For the first portion of our analysis, we combine the information from the JCMT GBS on core masses and sizes with GAS data on core temperatures and velocity dispersions.  \edit1{Our results are shown in relation to previously-suggested size-line width relations in Figure~\ref{fig_larson}. Like many previous studies of cores \citep[e.g.,][]{Goodman98}, we do not see the characteristic `Larson's Law' \citep{Larson81} where $\sigma_{tot}$ varies with $R_{eff}$ to some power between $1/3$ and $1/2$ \citep{Larson81,Solomon87}.} % Our results also do not agree with more recent results suggesting an additional dependence on the square root of column density \citep{BallesterosParedes11, Heyer09}, as at densities of 10$^{21}$--10$^{22}$cm$^{-2}$ which are typical of our cores, these trends differ minimally from \citet{Larson81} \citep{BallesterosParedes11}. }
%\edit1{A review of papeyrs investigating trends between core size and line width is provided in \citet{BallesterosParedes11}, which finds that the scaling relation of $\sigma_{tot} \propto \Sigma^{1/2}R^{1/2}$ from \citet{Heyer09} best describes the combined results from the earlier studies. In regions with typical cloud column densities of 10$^{21}$-10$^{22}$cm$^{-2}$, which encompasses the lower end of the column density range seen in our data, the results of \citet{Heyer09} and \citet{Larson81}, agree with each other quite closely.
%\hknotes{[HK note Jan9/19: I suggest leaving out the new text here.  This type of detailed response is more appropriate for the referee, and comes a bit out of the blue in the body of the text.  If you wish to add something additional here, it could be just a single short sentence or footnote saying something like : 'Note that for the typical column densities of these clouds, the updated Heyer09 relationship is largely the same as Larson's Law.']}
Using a least squares fitting algorithm, we find a weak \textit{negative} trend that is approximated by $\sigma_{tot} \propto R^{-0.34}$. \edit1{A similar trend is seen between column density and $\sigma_{tot}$/R$^{0.5}$, which does not match the relation presented in \citet{Heyer09}}. Often, the lack of correlation between $\sigma_{tot}$ and $R_{eff}$ is interpreted as an indication of structures that have decoupled from the larger-scale turbulent cascade.  In our case, however, the total velocity dispersions that we measure are often non-thermally dominated, with slightly over half of the cores having total line widths that are at least 50\% larger than the thermal line width. In Section~\ref{sec_high_veldisps}, we investigate these large line widths further. There, we find evidence that the GAS data may be biased to larger line widths than should be attributed to dense core gas, in part due to crowding.  We find typically much narrower line widths in B18, which has many more isolated dense cores. %\edit1{Similar systematic effects may impact other studies investigating trends in line width, and subsequently future investigations should be careful to consider whether these effects are present.}
%\hknotes{[while this sentiment has some merit, it would need to be expressed with a lot more care - e.g., what properties of this dataset lend it to being so vulnerable to the overestimates, and to what other studies might this be applicable?  This is all reasonably well addressed in the appendix, and any reader who is concerned about possible overestimates in other studies will already jump straight to the appendix to get all of the details, so it's best not to disrupt the flow of the text here to add more details.}

\begin{figure}
\centering
\includegraphics[width=6in]{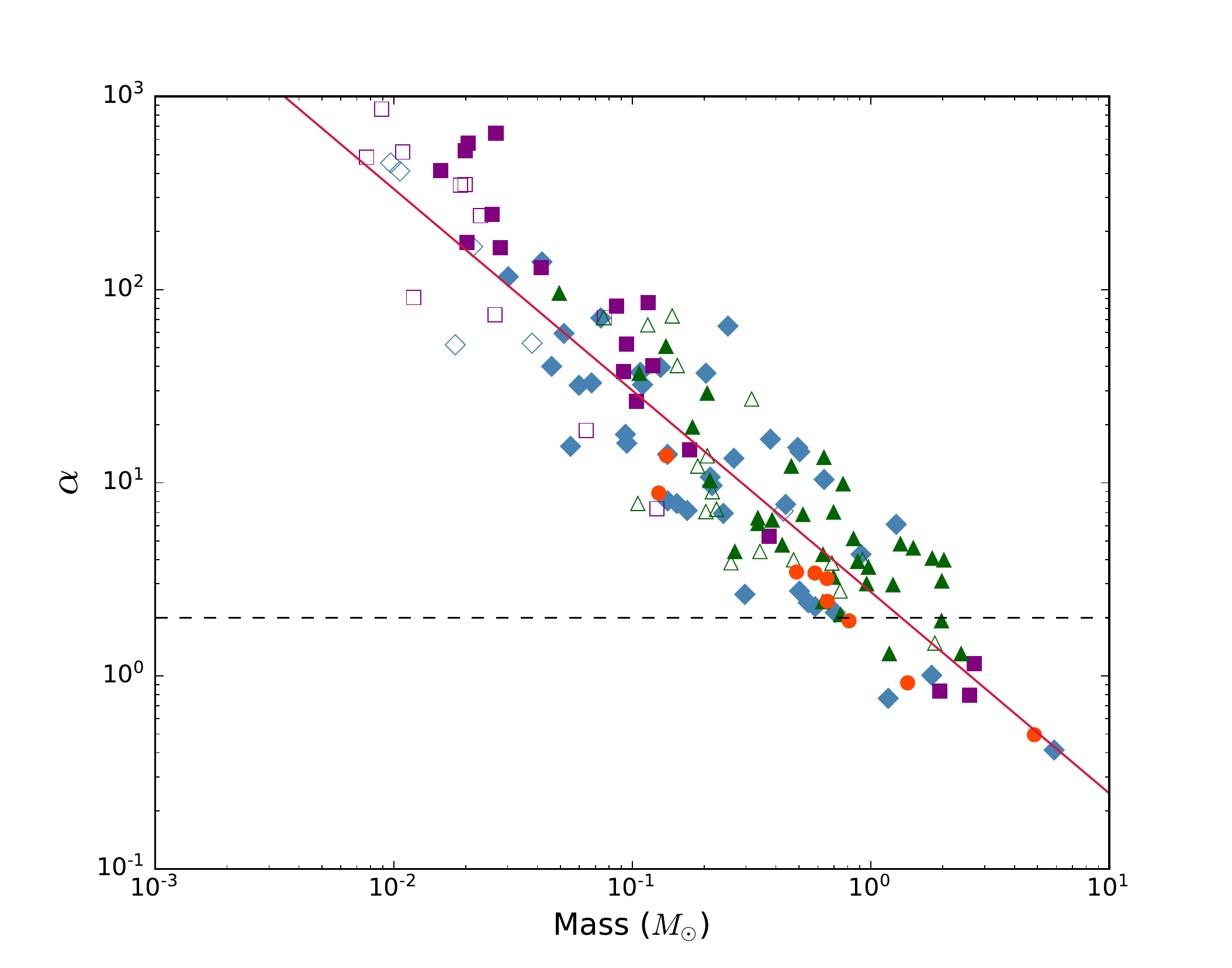}
\caption{The virial parameter, $\alpha$, versus core mass for all regions studied. Cores lying above the horizontal dashed line are unbound when considering only the balance between self-gravity and thermal plus non-thermal velocity dispersion. See Figure~\ref{fig_larson} for the plotting conventions used. \edit1{The red diagonal line shows the least-squares best fit to the power law relation between M and $\alpha$.}}
\label{fig_alpha_gs}
\end{figure}

We next compute $\alpha$, the virial parameter, following the standard method outlined in \citet{Bertoldi92}. It is valid for uniform, ellipsoidal clouds, and includes only internal turbulent motions and self-gravity:
\begin{equation}
\alpha = \frac{5\sigma_{tot}^2R}{GM} ,
\end{equation}
where $\sigma_{tot}$ is the total velocity dispersion, $R$ is the core radius, $G$ is the gravitational constant, and $M$ is the core mass.  In Figure~\ref{fig_alpha_gs}, we show the virial parameter as a function of mass for each of the three regions. Like many existing studies, \citep[e.g.,][]{Kirk17b, Keown17, Kauffmann08, Friesen16b}, our plot of M and $\alpha$ shows a clear linear negative trend in log-log space, which corresponds roughly to a power law relationship of $\alpha \propto$ M$^{-1.04 \pm 0.03}$.  A qualitatively similar slope is seen in the Cepheus-L1251 analysis in \citet{Keown17}, but the Orion A analysis in \citet{Kirk17b} shows a slightly steeper trend. We also find that the mass at which cores transition from bound to unbound in B18 from this paper is very similar to that found in Cepheus-L1251 \citep{Keown17}, while the other two regions transition at a slightly higher mass, although all are near 1 M$_{\odot}$. \edit1{At around 2 M$_{\odot}$,} the transition mass in Orion A \citep{Kirk17b} is slightly higher than all three regions studied in this paper\edit1{, a difference primarily caused by the larger core velocity dispersions there.}

\section{Virial Analysis Including External Pressure}\label{sec:fullvirial}

While it is perhaps surprising that such a large fraction of the dense cores surveyed appear to be gravitationally unbound in Section~\ref{sec:Jeans}, we note that the full virial equation includes more than just terms for self-gravity and internal motions.  A number of previous studies \citep[e.g.,][]{Pattle15, Kirk17b} have suggested that cores may be bound at least in part due to `external pressure'.  We add that to our analysis in the following sections, considering several sources of pressure.

First, we follow \citet{Pattle15} and introduce the formalism of energy density to represent better the full virial equation. The energy density from internal motions, self-gravity, and external pressure can be expressed as
\begin{equation}\label{eqn_kin}
\Omega_k=\frac{3}{2}M{\sigma}_{tot}^2 ,
\end{equation}
\begin{equation}\label{eqn_grav}
\Omega_g=-\frac{3}{5}\frac{GM^2}{R} ,
\end{equation}
\begin{equation} \label{eqn_press}
\Omega_{p}=-4\pi{PR^3} ,
\end{equation}
\noindent respectively. Note that this formulation alone does not provide a complete virial analysis, as it excludes the magnetic field term. This point is discussed in Section \ref{sec:mag}. Also note that we have adopted a slightly different constant factor in Equation~\ref{eqn_grav} than was used in \citet{Pattle15} and \citet{Kirk17b}.  The constant factor used in both of those analyses, $\frac{-1}{2\sqrt{\pi}}$, is appropriate for sources with a Gaussian density distribution, while a factor of $-\frac{3}{5}$ is appropriate for sources of constant density.  While neither of these density distributions exactly describes the cores, the latter factor provides a closer approximation to sources with density distributions following the typical $\rho \propto R^{-1}$ profile of dense cores.  We note that \citet{Sipila11} has run a similar calculation for the Bonnor-Ebert sphere, another popular model used to characterize dense cores.  Figure~B1 of \citet{Sipila11} shows that a factor of -0.6 to -0.73 is appropriate for Bonnor-Ebert sphere models ranging from the uniform sphere to the critical Bonnor-Ebert sphere.

\subsection{Cloud Weight Pressure}
\label{sec:cloudpress}

\citet{Kirk17b} found that dense cores in Orion~A appear to be primarily confined due to the pressure from the weight of overlying molecular cloud material. 
Following this work, we estimate pressure from the weight of the surrounding cloud using \citep{McKee89,Kirk06}:

\begin{equation} \label{eqn:cwp}
P_{w}={}{\pi}G{\Sigma}{\bar{\Sigma}}
\end{equation}
where P$_{w}$ is the pressure from cloud weight, $\Sigma$ is the surface mass density (column density multiplied by mean particle mass) along the line of sight of the core, and $\bar{\Sigma}$ is the mean surface mass density across the entire cloud. This equation assumes that the large-scale structure of the molecular cloud is well approximated by a sphere, and that the density profile is similar to $\rho \propto r^{-k}$ with $k=1$. Somewhat more centrally concentrated density profiles in the range of $k=1-3$ introduce a second positive term to these relationships, so in that case Equation \ref{eqn:cwp} would provide a lower limit \citep{Kirk17b}. Additional information on the derivation of this cloud weight pressure equation and relevant geometry can be found in \citet{Kirk17b}.

As in \citet{Kirk17b}, we apply an \textit{\'{a} Trous} wavelet decomposition to the column density maps to estimate the contributions of cloud material versus core material. In our source extraction, cores are generally less than 0.1 pc across \citep[comparable to the typical core sizes of 0.03-0.2 pc given in][]{Bergin07, DiFrancesco07}. For our main analysis, we conservatively assign scales of approximately 0.5~pc and above as belonging to the cloud. The \textit{\'{a} Trous} algorithm decomposes structures on scales of $2^N$ pixels, while the column density maps have a pixel scale of 14\arcsec\ per pixel.  We therefore adopted a scale of 64 pixels for L1688 and B18, and a scale of 32 pixels for NGC 1333, corresponding to physical scales of approximately 0.6~pc in all three regions.  
The effects of choosing a different scale are discussed further explored in Appendix \ref{app:filscale}. Aside from changing the magnitude of the cloud weight pressure term, the filter scale choice has a relatively small impact on the qualitative results of this paper, largely due to the relatively modest strength of cloud weight pressure in the regions studied. %This can be seen through the statistics displayed in Table \ref{tab:systematics}.

For all three regions that we analyze, we find that the pressure attributed to cloud weight produces a relatively small virial term ($\Omega_{p,w}$) relative to the findings of \citet{Kirk17b} in Orion~A.  This difference is likely attributable to the typical column density associated with each of these regions: Orion~A is a Giant Molecular Cloud, with a mean column density of $\bar{N} = 3.9 \times 10^{22}$~{cm}$^{-2}$ over the area analyzed in \citet{Kirk17b}.  Meanwhile, the mean column densities estimated for the three clouds that we analyze here are $\bar{N}=4.6 \times 10^{21}$~{cm}$^{-2}$, $2.7 \times 10^{21}$~{cm}$^{-2}$, and $2.0 \times 10^{21}$~{cm}$^{-2}$ for L1688, NGC~1333, and B18, respectively, i.e., roughly an order of magnitude smaller than Orion~A. Since the pressure attributable to cloud weight is relatively small, it does not dominate the forces contributing to core confinement as seen in Orion A \citep{Kirk17b}. We therefore estimate additional pressure sources before presenting the results of our virial analysis.

\subsection{Turbulent Pressure} \label{sec:turpress}

A separate form of pressure is exerted on each core by turbulent motions of the surrounding cloud. These motions can be traced by transitions from molecules that are common at lower densities, such as $^{13}$CO, which will be used in this paper. Line widths are found by fitting a Gaussian distribution to the $^{13}$CO lines, and the result is put into the following equation, which follows from the ideal gas law:

\begin{equation} \label{eqn:turpress}
{P_t}={}n_{^{13}\mathrm{CO}}{\mu_{mn}}{m_{H}}{(\sigma_{tot, ^{13}\mathrm{CO}}^2)}\\
\end{equation}
where $n_{^{13}\mathrm{CO}}$ is the density of the gas that $^{13}$CO is probing. Since $^{13}$CO freezes out at densities above $10^5$ $\x{cm}^{-3}$ \citep{DiFrancesco07}, we conservatively select a %somewhat 
less dense value of $3\times10^3$ $\x{cm}^{-3}$ to represent the typical 
density of the ambient cloud. This value is somewhat lower than that of the $5\times10^3$ $\x{cm}^{-3}$ employed in \citet{Kirk17b}, which reflects the lower mean densities in these clouds compared to Orion A, and the lower density gas traced by $^{13}$CO compared to the C$^{18}$O used in \citet{Kirk17b}. The resulting value for $P_t$ can be used in Equation \ref{eqn_press} to generate a value for an external turbulent pressure term, $\Omega_{p,t}$. When this value is added to the cloud weight pressure term, a total external pressure term is generated.  We discuss the implications of changing some \edit1{of} the assumptions made in computing the turbulent pressure term in Section \ref{sec:systematics}.

\begin{figure}
\includegraphics[width=7in]{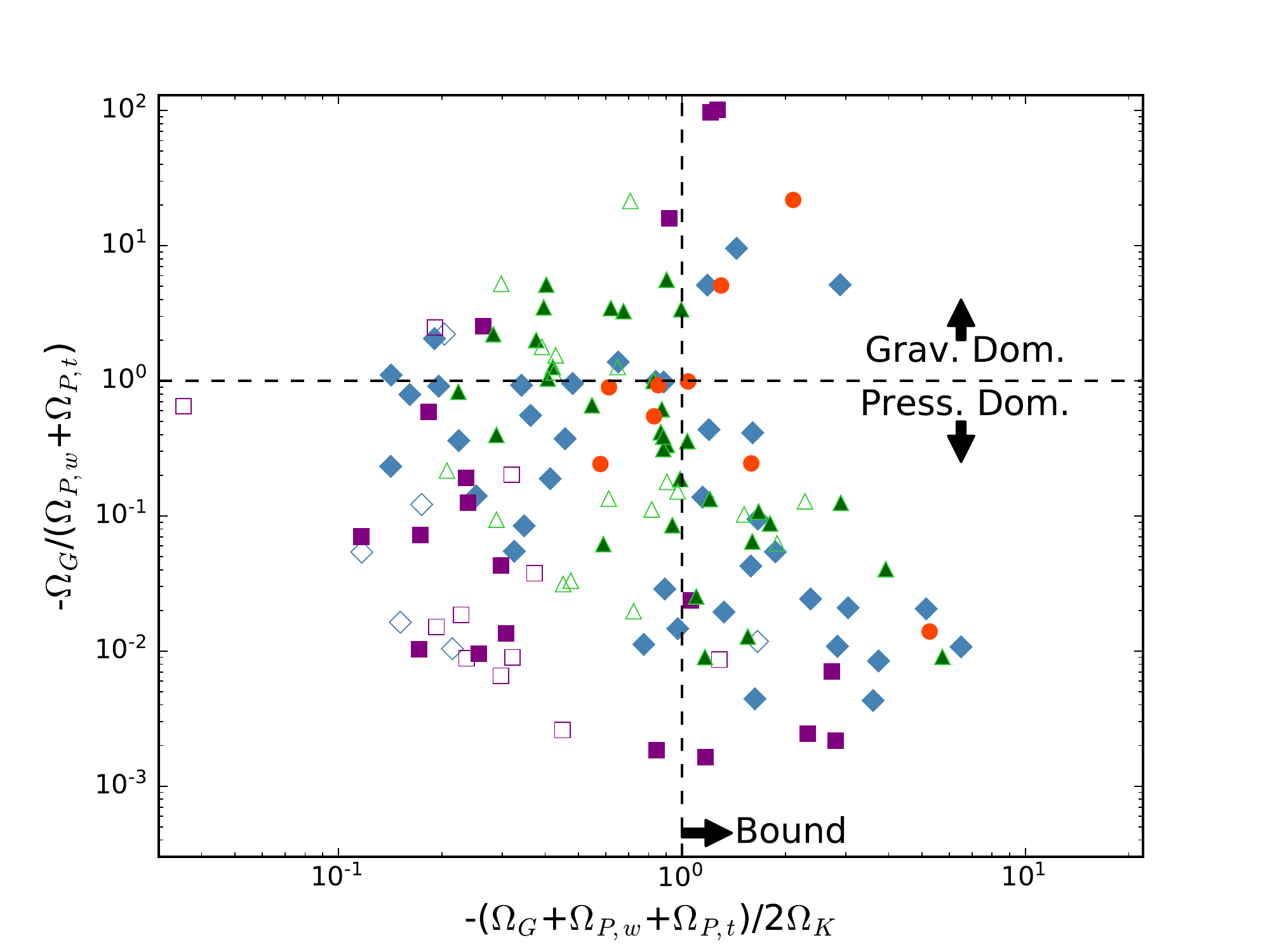}
\caption{Confinement ratio versus total energy density for cores in all regions surveyed. Cores lying above the horizontal dashed line have gravitational binding larger than the combined pressure, and cores to the right of the vertical dashed line are bound. See Figure~\ref{fig_larson} for the plotting conventions used.
}
\label{fig_oph_fullvir_gs}
\end{figure}

The result of a more complete virial analysis, including both pressure terms, is shown in Figure \ref{fig_oph_fullvir_gs}. \edit1{The vertical axis shows the ratio between the gravitational and pressure terms, which describes which of the two terms dominates the binding of cores. This ratio is plotted against the virial ratio, $-(\Omega_g+\Omega_{p,t}+\Omega_{p,w})/2\Omega_k$. A virial ratio of one indicates virial equilibrium, while a ratio greater than one indicates that the core is bound. }With the addition of turbulent and cloud weight pressure, 36\% of cores in L1688, 27\% of cores in NGC~1333, and 56\% of B18's cores qualify as bound. Furthermore, over 75\% of cores across all regions have a pressure term, $\Omega_p$, that is larger than the corresponding gravity term, $\Omega_g$, implying that external pressure dominates the binding of the dense cores.
We also find that this pressure is dominated by turbulent pressure in both NGC~1333 and B18, with a median turbulent pressure term of roughly double the cloud weight pressure in both regions. L1688 has more equal contributions from the two pressure sources, with turbulent pressure contributing roughly 0.75 times that of the cloud weight to the total pressure term in L1688. This balance of pressures is in stark contrast to Orion A \citep{Kirk17b}, which has a median cloud weight pressure term that is 15 times higher than the corresponding turbulent pressure term. Comparisons to the \citet{Kirk17b} results in Orion A are discussed more extensively in Section \ref{sec:largecontext}. More information comparing the relative strengths of forces binding cores can be found in Table \ref{tab_pressure}, which includes Orion A. 

%We also provide a separate virial analysis using {\fw} rather than {\get} for the generation of our source catalogue, which is discussed in Appendix \ref{app:fellwalker}. The results are qualitatively quite similar to those of our {\get} extraction. See Table \ref{tab:systematics} for some statistics related to this {\fw}-based analysis.   

In Appendix \ref{sec:uncvir}, we examine the sources of uncertainty in our virial parameter estimates, including the dust-based core catalogue, the NH$_3$-based core line widths, the large-scale cloud weight pressure, and the ambient turbulent pressure. Of particular note is our examination of the core line widths in Appendix~\ref{sec_high_veldisps}, where we find evidence that the NH$_3$ observations may be biased to larger values than should be attributed to the dense core gas. \edit1{We discuss the implications of this conclusion in Section \ref{sec:rvr}. }
%Table~\ref{tab:systematics} in Appendix~\ref{sec:uncvir} summarizes the influence that varying certain assumptions and techniques has our final virial analysis.  Under this full range of assumptions, pressure binding is always found to play an important role in core confinement.

\begin{deluxetable}{ccccc}
\tablecolumns{5}
\tablewidth{0pt}
\tablecaption{A summary of the virial terms.\label{tab_pressure}}
\tablehead{
\colhead{} &
\colhead{L1688} &
\colhead{NGC~1333} &
\colhead{B18} &
\colhead{Orion A \tablenotemark{c}}
}
\startdata
$\rm{log}|\Omega_k|$\tablenotemark{a}\tablenotemark{b}& 41.58$^{+0.38}_{-0.37}$ & 42.19$^{+0.41}_{-0.30}$ & 41.90$^{+0.04}_{-0.38}$ &  $42.46_{-0.36}^{+0.46} $\\
$\rm{log}|\Omega_{g}|$\tablenotemark{a}\tablenotemark{b}& 40.30$^{+0.91}_{-0.89}$ & 41.86$^{+0.43}_{-0.72}$ & 41.67$^{+0.24}_{-0.46}$ & $41.35_{-0.44}^{+0.65} $\\ 
$\rm{log}|\Omega_{p,w}|$\tablenotemark{a}\tablenotemark{b}& 41.05$^{+0.66}_{-0.47}$ & 41.45$^{+0.36}_{-0.27}$ & 41.31$^{+0.19}_{-0.34}$ & $43.16_{-0.34}^{+0.31}$\\
$\rm{log}|\Omega_{p,t}|$\tablenotemark{a}\tablenotemark{b}& 40.92$^{+0.60}_{-0.42}$ & 41.81$^{+0.38}_{-0.29}$ & 41.67$^{+0.17}_{-0.33}$ & $41.94_{-0.31}^{+0.34}$\\
\hline
%\hline
$\rm{log}|\Omega_g/\Omega_k|$\tablenotemark{a}   & -1.21$^{+0.62}_{-0.53}$ &-0.41$^{+0.13}_{-0.38}$ & -0.20$^{+0.22}_{-0.03}$ & $-0.96_{-0.38}^{+0.33}$\\
$\rm{log}|\Omega_{p,w}/\Omega_k|$\tablenotemark{a}& -0.47$^{+0.59}_{-0.29}$ &-0.48$^{+0.24}_{-0.37}$ &-0.49$^{+0.11}_{-0.16}$ & $0.69_{-0.45}^{+0.40}$\\ 
$\rm{log}|\Omega_{p,t}/\Omega_k|$\tablenotemark{a}& -0.56$^{+0.54}_{-0.29}$ & -0.12$^{+0.24}_{-0.52}$ & -0.21$^{+0.08}_{-0.17}$ & $-0.46_{-0.44}^{+0.47}$\\
$\rm{log}(\Omega_{p,t}/\Omega_{p,w})$\tablenotemark{a}& -0.12$^{+0.05}_{-0.09}$ & 0.31$^{+0.07}_{-0.08}$ & 0.28$^{+0.04}_{-0.02}$ & $-1.18_{-0.12}^{+0.14}$\\
$\rm{log}(\Omega_{g}/(\Omega_{p,t}+\Omega_{p,w}))$\tablenotemark{a}& -1.20$^{+1.00}_{-0.74}$ & -0.51$^{+0.61}_{-0.54}$ & -0.05$^{+0.04}_{-0.56}$ & $-1.68_{-0.80}^{+0.58}$\\ 
$\rm{log}|(\Omega_{p,t}+\Omega_{p,w}+\Omega_{g})/\Omega_k|$\tablenotemark{a}& -0.35$^{+0.46}_{-0.29}$ & -0.08$^{+0.11}_{-0.29}$ & -0.02$^{+0.18}_{-0.10}$ & $0.74_{-0.39}^{+0.39}$\\ 
\enddata

\tablenotetext{a}{For each virial term listed, we provide the median value and the range spanned by the upper and lower quartiles in the distribution.}
\tablenotetext{b}{Virial terms are given in units of ergs.}
\tablenotetext{c}{Values calculated from \citet{Kirk17b} with their protostellar cores excluded for consistency with our analysis.}
\end{deluxetable}

%\soedit{If you assume the NH3 is smeared out and limit the calculation to the center of the identified core, does this give values more similar to the N2H+ and N2D+?} %\rknotes{[Generally no. There are cases where the core is enclosed with higher-velocity dispersion material, but this surrounding turbulent material is normally fairly noisy, and sharp contrasts are only ever seen near bright cores, so there is little contribution from them. Aside from that, most regions have sufficiently blended ammonia data that using a different radius definition has little impact. The problem appears to be that a lot of cores are contaminated by additional emission along their line of sight, as inferred in a few places throughout 5.1]}
%\hknotes{[HK adds - might be helpful to add one or two sentences directly to the paper with this information in case other readers wonder something similar.  i.e., something like 'In most cases, we find a similar velocity dispersion in the central core pixel as in the core outskirts....']}

\section{Discussion}\label{sec:discussion}

\subsection{Magnetic Fields} \label{sec:mag}

Magnetic fields constitute a potentially important additional factor in the virial state of cores, working internally to support cores against collapse from gravity and external pressure. Subsequently, a strong magnetic term could change our conclusions about the virial state of cores. The importance of magnetic fields in the binding of cores, however, is not clear. In L1688, \citet{Pattle15} used magnetic field strength data from \citet{Crutcher93} and \citet{Troland96} to argue that the magnetic support term was small, with a ratio of $\Omega_{M}$ to $\Omega_{k,NT}$ (i.e., the ratio of the magnetic field to non-thermal kinetic energy terms) of only 0.11 on average.

The recent JCMT BISTRO survey \citep{WardThompson17} provides polarization maps for many JCMT GBS clouds, including NGC~1333 and L1688, but not B18. These data allow updated magnetic field strength estimates to be made for various star forming regions.
Using the Chandrasekhar-Fermi method \citep{ChandrasekharFermi53}, \citet{Soam18} estimated the mean plane-of-the-sky magnetic field strength in Oph B2 to be 630$\pm$410 $\mu$G, while a similar study in Oph A \citep{Kwon18} found a slightly smaller plane-of-the-sky magnetic field strength of between 200 $\mu$G and 500 $\mu$G. Measurements were also recorded for Oph B1, but the magnetic field there was found to be disorganized on the cloud scale, so an overall estimate could not be produced \citep{Soam18}.
Assuming that the magnetic field strengths are uniform across each separate region, these values would correspond to dramatically higher magnetic terms than those used in \citet{Pattle15}, with a $\Omega_{M}$ to $\Omega_{k,NT}$ ratio of 1.9 on average in Oph A and 2.0 in Oph B2, which would certainly have a significant impact on the virial state of many cores in the sample.

BISTRO-based estimates of magnetic field strength have only been made for a handful of regions \citep{Pattle17b, Soam18,Kwon18} so far. Additional research will be necessary to provide sufficiently detailed estimates in the regions of interest, which must also take regional variations into account in clouds as diverse as L1688. Therefore, since high-resolution estimates
of the magnetic field are not available in all regions explored, we exclude the term for the sake of maintaining a self-consistent analysis across all regions. Our results for the fraction of bound cores are subsequently best interpreted as providing an upper limit to the bound fraction of cores, as any non-zero magnetic term would make cores appear less bound. Once more extensive and detailed studies on the magnetic field strength become available for many regions, future virial analyses should include this term, as the early BISTRO magnetic field strength estimates \citep[e.g.,][]{Soam18} suggest
that the impact of magnetic fields may be significant. 

%\subsection{Revised Virial Results} 
\subsection{Influence of Potential Overestimates to Core Velocity Dispersions}

\begin{figure}
\includegraphics[width=7in]{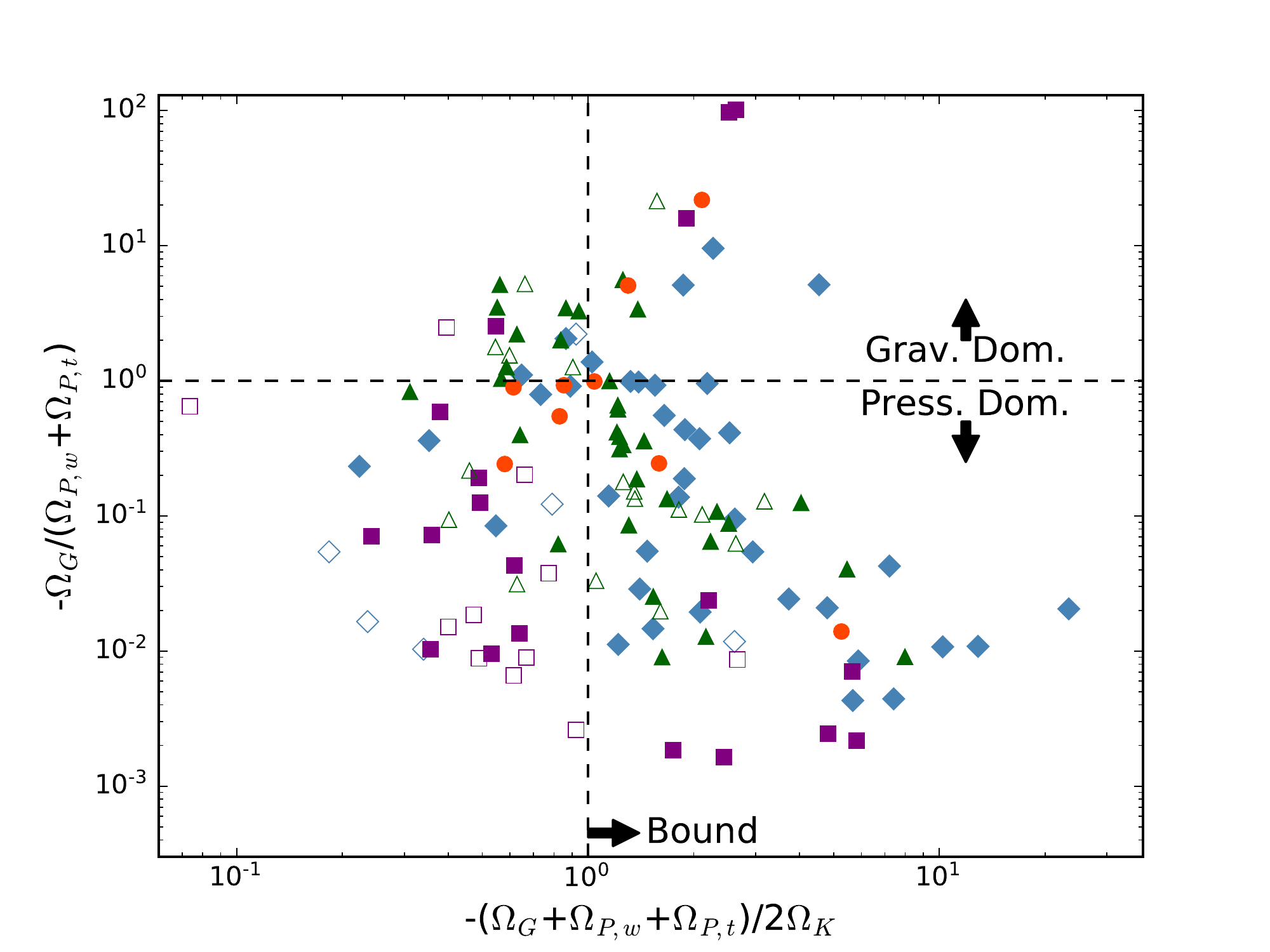}
\caption{Confinement ratio versus total energy density for cores in all regions surveyed, with average ratios between our observations and {\ndp} and {\nhp} observations applied. See Figure~\ref{fig_larson} for the plotting conventions used.}
\label{fig_all_correctvir_gs}
\end{figure}

\edit1{In Appendix~\ref{sec_high_veldisps}, we investigate the unusually large dense core velocity dispersions noted in Section~\ref{sec:Jeans}.  We find evidence suggesting that the observed NH$_3$ line widths may be larger than should be attributed solely to the dense core material.  While corrections are not possible for each individual core, in Appendix~\ref{sec_high_veldisps}, we derive representative correction factors for the velocity dispersion of cores in each of the regions.  Figure~\ref{fig_all_correctvir_gs} shows the virial ratios after these correction factors have been applied.  Using these approximate correction factors, the number of unbound cores is reduced to 55 of the total 134 cores.  Higher resolution observations of a higher-density gas tracer for each core in the sample are needed to verify the true dense core velocity dispersions.}

\subsection{Larger Context, Role of Pressure} \label{sec:largecontext}

Our results present a primarily pressure-dominated view of cores in L1688, NGC~1333 and B18, a conclusion shared by many previous analyses of dense cores
\citep[e.g.,][]{Maruta10, Barnes11}. While virial analyses of dense cores have traditionally focused on comparisons between gravitational confinement and internal support, the importance of pressure has recently become a much more consistent theme. \citet{Bertoldi92} was among the first to consider this term, concluding that of their sample of cores in Ophiuchus, Orion, the Rosette, and Cepheus OB3, pressure, primarily rooted in cloud turbulence, is required to bind a significant population of mainly smaller cores. Later studies such as \citet{Maruta10} built on this research by demonstrating that external pressure is consistently quite large relative to other forces in these systems, making it potentially critical to the virial state of cores. \citet{Maruta10} also provided a virial analysis, confirming that in their sample of Ophiuchus cores, the external pressure term was more important than the gravitational term in most cases.

Many more recent papers have supported a similar narrative. An analysis of the CHaMP survey region \citep{Barnes11}, which focused on a diverse set of more distant clouds between 1.1 kpc and 6.8 kpc away, also showed generally small gravitational terms, with evidence suggesting that external pressure, mainly due to the turbulence of the surrounding cloud, may bind these cores in most cases. Studies making use of the JCMT GBS continuum data in both Ophiuchus and the Cepheus Flare region \citep{Pattle15,Pattle17} also concluded that pressure is the dominant force in many cores, even more so in the Cepheus Flare region than Ophiuchus. Papers adopting newer
core identification methods have reached similar conclusions. \citet{HChen19} concluded that most objects in their unique set of droplets were pressure dominated, while \citet{Seo15}'s dendrogram analysis found that almost half of the gravitationally unbound cores identified by their survey can be bound with the consideration of this pressure term.

While pressure does appear to be important in many virial analyses, some studies identify cores whose binding is dominated by self-gravity.
In the analysis of dendrogram leaves in Cepheus-L1251 presented by \citet{Keown17}, and the B59 virial analysis by \citet{Redaelli17}, for example, many of the cores appear to be dominated by self-gravity. Furthermore, in dendrogram analyses of Perseus-L1448 \citep{Rosolowsky08} and Serpens South \citep{Friesen16b}, it is suggested that larger-scale structures may be more likely to be self-gravitating. Similarly, in our present analysis, we identified 13 cores bound by gravity alone, and these cores tended to be the most massive of the group. %\rknotes{ER: Just thinking here, is there a link between fraction of self-gravitationally bound cores and i.e. present star formation efficiency of a cloud? My thought is that if the present dense, starless cores population of a molecular cloud is representative of the past one, one should find that higher SFE are linked with higher fraction of self-gravitating objects (assuming that you have to have gravity to win over other terms to have a gravitational collapse..).}

We further conclude from our analysis that external turbulent pressure is dominant over cloud weight pressure in NGC~1333 and B18, while the two pressure terms share a similar level of importance in L1688. An interesting contrast can be made between our conclusions and those of \citet{Kirk17b}, which observed the much more massive Orion A cloud, using essentially the same data sets as this paper: {\get} catalogs derived from JCMT SCUBA-2 data, GAS DR1 data for internal kinetic properties, and cloud weight pressure estimates from column density maps derived using \textit{Planck} and \textit{Herschel} data \citep{Lombardi14}. The paper concluded that cloud weight pressure is the dominant force in binding Orion A cores, contributing a binding term 15 times greater than the corresponding turbulent pressure term on average. This situation is primarily due to the high column density of the region. In our analysis, cloud weight pressure was either roughly the same size as or smaller than the turbulent pressure, ranging from a cloud weight pressure to turbulent pressure ratio of 1.33 in L1688 to roughly 0.5 in NGC~1333 and B18. Like in L1688, the turbulent pressure term derived in Orion A is also smaller relative to the internal kinetic energy term than it is in NGC~1333 and B18, suggesting that the strength of turbulent pressure may correlate loosely with the level of clustering in the parent cloud. 

This small sample of observations provides evidence that cloud weight pressure may tend to dominate the pressure budget of cores in larger, more clustered systems, rather than turbulent pressure, which may dominate in sparser regions. It is difficult, however, to make definitive statements about these trends given the current sample size of clouds that have been analyzed in an identical manner.
While some virial analyses do argue that cloud pressure \citep[e.g.,][]{Lada08, Kirk17b} or turbulent pressure \citep[e.g.,][]{Barnes11, Pattle15} dominates, many analyses treat the different forms of pressure as a single combined force, and therefore our ability to compare these different pressure terms across a diverse array of regions is extremely limited. Virial analyses considering both sources of pressure on different star-forming regions will be critical to the development of our knowledge of the role that cloud weight and turbulent pressure have in a diverse range of star-forming environments.  Future GAS observations, which cover numerous additional regions, will represent an important contribution to the study of these forces. 

\section{Conclusion}\label{sec:conclusion}

Through new surveys such as the Green Bank Ammonia Survey (GAS), large studies that provide uniform, high quality {\it maps} of dense gas emission across a range of star-forming environments are now a reality. This advancement enables us to compare systematically star formation properties on a cloud-by-cloud basis.  Our analysis focuses on a simple core virial analysis using GAS DR1 data on Ophiuchus L1688, Perseus NGC~1333, and Taurus B18.  
We combined data to identify dense cores and estimate their self-gravity (using dust emission from the JCMT Gould Belt Survey), measure their thermal and non-thermal motions (using NH$_3$ data from GAS), and estimate their external binding pressures, including cloud weight (using column density estimates based on {\it Herschel} data) and cloud turbulent pressure (using FCRAO $^{13}$CO data). We restricted our analysis to starless dense cores using various {\it Spitzer} protostar catalogues.

Traditionally, virial analyses of dense cores often neglect terms beyond self-gravity and thermal/non-thermal support.  Following this simplified approach, we find that only a small fraction of the dense cores are bound.  When binding pressure from the surrounding cloud material is included, our analysis shows that a significant fraction are bound: 36\% in L1688, 27\% in NGC~1333, and 55\% in B18.  Comparisons with other kinematic datasets suggest that at least some of our GAS NH$_3$ spectra may be biased to larger line widths than should be attributable solely to the dense core gas, e.g., due to crowding along the line of sight and / or the sensitivity of NH$_3$(1,1) to lower density gas.  We attempt to account for this potential bias, and find that almost 60\% of cores across all three star-forming regions qualify as bound.  With or without this correction, pressure dominates gravity in confining 
the dense cores. 
Unlike an earlier study in Orion~A by \citet{Kirk17b} using similar methods, however, most of the pressure binding the cores in NGC~1333 and B18 appears to be induced by the turbulence of the surrounding medium, while L1688 has more equal quantities of turbulent and cloud weight pressure. 

One complication that is becoming increasingly apparent in all virial analyses is the importance of how dense cores are defined.  Our analysis uses the fairly simple and traditional method of defining dense cores based on their dust emission.  Other methods recently explored by the GAS team include using dendrograms directly on the NH$_3$ map \citep{Keown17} as well as searching for coherent and quiescent zones within the NH$_3$ emission cube \citep{HChen19}.  The detailed virial parameters of an individual core necessarily changes under each of these schemes, and a careful examination of the best method of identifying cores in the context of a virial analysis is clearly needed.  Nonetheless, our over-arching conclusion that the pressure term is important appears to hold across all core identification methods.  Furthermore, conclusions based on comparisons between clouds, or between a cloud and numerical simulations, should be less prone to variation than detailed conclusions about a single core identified under a variety of methods, as long as exact methodologies are replicated across the full analysis sample.

\acknowledgments

RK, HK, and JDF thank the co-op student program at the National Research Council of Canada, which enabled this research to be conducted.
The authors thank the JCMT GBS team for access to their catalogues prior to their publication, which were used to make the core catalogues for this analysis, and the GBT staff for their help with the many hours of observing needed for GAS. 
RK thanks Alvaro Hacar for a helpful discussion about {\nhp} data sets in Taurus. \edit1{The authors thank the anonymous referee, whose comments greatly improved the final paper.} The JCMT has historically been operated by the Joint Astronomy Centre on behalf of the Science and Technology Facilities Council of the United Kingdom, the National Research Council of Canada, and the Netherlands Organisation for Scientific Research.  Additional funds for the construction of SCUBA-2 were provided by the Canada Foundation for Innovation. The Green Bank Observatory is a facility of the National Science Foundation operated under cooperative agreement by Associated Universities, Inc. The National Radio Astronomy Observatory is a facility of the National Science Foundation operated under cooperative agreement by Associated Universities, Inc. 
JEP, PC, and ACT acknowledge the financial support of the European
Research Council (ERC; project PALs 320620). JDF and EWR are supported by Discovery Grants
from the NSERC of Canada. SSRO recognizes the support of NSF grant AST-1510021. AP Acknowledges the support of the Russian Science Foundation project 18-12-00351. Y.L.S. was supported in part by the NSF Grant AST-1410190.

\vspace{5mm}
\facilities{GBT (KFPA); JCMT (SCUBA-2); Spitzer; FCRAO; 2MASS; \textit{Herschel}; \textit{Planck}}

\software{Astropy library \citep{Robitaille13};  Matplotlib \citep{Hunter07}; Starlink \citep{Currie14}; Lmfit \citep{LmFit16}}

\appendix

\section{Uncertainties in Virial Parameters}\label{sec:uncvir}

Some of the quantities used in our virial analysis are difficult to measure definitively using observations, and therefore multiple competing methods have been used in literature. In this section we summarize the many sources of uncertainty introduced by varying our assumptions, and estimate the degree to which any such change would affect our conclusions. 

\subsection{Source-Finding Algorithms} \label{app:sfa}

The method used to identify dense cores has long been a point of discussion, and many different techniques have been used in past virial analyses.
A common technique, which is applied in this paper, defines cores according to their extent in dust emission. As shown in Appendix \ref{app:sfa}, there are many ways to extract sources in this manner, though they all  conceptually work in the same way -- by identifying peaks in the emission map and assigning an emission region to that peak. 
Cores can also be identified via gas emission, however. Since the total core mass can be more difficult to estimate using only a single gas tracer, gas data are often supplemented with dust continuum observations, which can be applied at the location of the identified structures to compute masses. Finally, hybrid approaches also exist, drawing from gas, dust, and temperature data. These approaches attempt to identify cores based on their expected properties, such as lower line widths, lower temperature, and higher dust intensity  \citep[e.g.,][]{HChen19}. 

A recent study by \citet{Keown17} highlights the importance of the core identification method on virial analyses.  Using Cepheus-L1251 observations from GAS and {\it Herschel}, \citet{Keown17} performed two complementary virial analyses.  Dense cores were identified both in the GAS map using a dendrogram analysis technique \citep{Rosolowsky08}, and also in the {\it Herschel} dust map using {\get}.  Remaining portions of the analysis were very similar \edit1{to} those of this paper: \textit{Herschel} observations were used to estimate core masses, GAS NH$_3$ data was used to estimate core temperatures and non-thermal velocity dispersions, $^{13}$CO observations were used to determine external turbulent pressure, and H$_2$ column density maps were used to determine cloud weight pressure. \citet{Keown17} found that while most {\it Herschel}-identified cores appear pressure confined, similar to our results, the majority of the GAS-based dendrogram leaves were instead gravitationally bound.  This result suggests that the core identification method may strongly influence the subsequent virial analysis. Other gas-based dendrogram virial analyses, such as \citet{Friesen16b} in Serpens South and \citet{Rosolowsky08} in Perseus L1448, also tend to conclude that self-gravity plays a greater role than we find, but without a direct comparison using the same regions and datasets, it is not possible to attribute this difference to the core-identification method.

\citet{HChen19} uses a more complicated, hybrid technique for core identification to perform a virial analysis on L1688 and B18 GAS data, in which a very specific subset of dense cores, (referred to as ``droplets'' in that paper)
are identified using their observed kinematic properties. Droplets are identified by dividing up contiguous intersections of regions with 10-$\sigma$ integrated intensity emission and subsonic velocity dispersion into smaller regions with one peak each. These are then compared to \textit{Herschel} column density and dust temperature maps, to ensure that each core contains a minimum in dust temperature and a maximum in column density. If an object satisfies these criteria and is resolved by the GAS beam (32\arcsec), is it included. \citet{Hasenberger18} also uses minima in dust temperatures in their identification of cores, so there is a slowly-expanding body of literature exploring the detection of coherent cores through their kinematic properties. 
Similar to our conclusions, \citet{HChen19} concludes that turbulent pressure dominates the binding of cores.  An important difference, however, is that the techniques in \citet{HChen19} target a specific subset of dense cores, which resemble smaller variants of ``coherent cores" \citep{Goodman98}, with much lower levels of non-thermal support than we estimate in our dense core sample. Due to the focus on cores of this variety, these methods are not applicable to as wide of a variety of sources 
as most other core identification methods.   

At present, it is unclear which source identification technique or techniques is best for virial analyses and related studies.  Fortunately, surveys such as GAS are allowing progress to be made in this regard.  With a consistent dataset across a variety of clouds, GAS data allow for a multi-region analysis, highlighting similarities and differences in core populations that can be measured with any single core identification technique.  Furthermore, the application of multiple core identification techniques across one or more regions in a consistent manner will provide a way to test and understand the biases and advantages of each technique.

\begin{figure}[htb]
\centering
\includegraphics[width=7in]{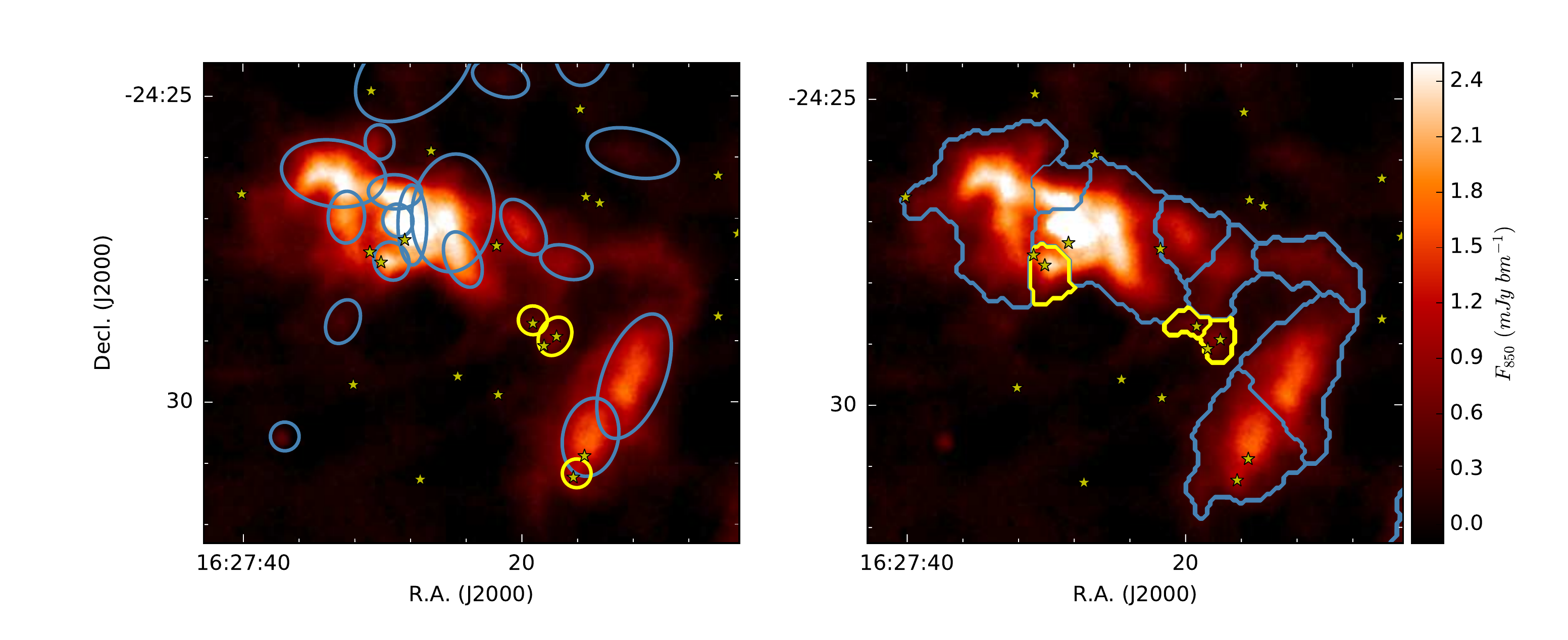}
\caption{The full core extents in Oph B, as found using {\get} (left) and {\fw} (right). The background image in both panels shows the 850~$\mu$m emission measured by the JCMT GBS. Core detection algorithms are most prone to disagreement in crowded regions like this, due to the difficulty they often have in separating dense emission. Yellow extents are labeled as protostellar, while blue ones are starless. The yellow stars icons are known protostars}
\label{fig:gsvfw}
\end{figure}

As part this paper, we run our analysis on dense cores in all three regions using two different source-finding algorithms: {\get} and {\fw}. Here, we describe the implementation of both algorithms as well as the resulting core catalogues. An example comparison of the results of both source extraction methods is shown in Figure \ref{fig:gsvfw}, in one of the more crowded regions of our sample, Oph~B. We note that well-separating emission in crowded areas is a known challenge \citep{Pineda09}, and that employing and comparing the results from multiple algorithms helps to reduce uncertainties in the potential biases associated with any single technique. We also compare these results to those from the \citet{Pattle15} L1688 dense core catalog, which uses another separate source-finding algorithm called CuTEx. 

\subsubsection{{\get}}\label{app:getsources}

{\get}, as described in \citet{Menshchikov12}, is a source extraction algorithm originally designed for {\it Herschel Space Telescope} observations, and can use multi-wavelength observations to perform a single extraction. {\get} is especially notable in that it corrects for significant background sources of flux, such as filamentary structure and larger-scale cloud structures. Using decompositions at different scales, it methodically subtracts the background structure, leaving behind just the core. Hypothetically, the subtraction of structure along the line of sight should make for more accurate flux estimates for any given core. For each of the three clouds, the full {\get} pipeline was run using the JCMT SCUBA-2 maps at 450 $\mu$m and 850 $\mu$m, using parameters recommended in the {\get} Quick Start Guide. Although the 850 $\mu$m maps typically have better signal-to-noise ratios, {\get} excels at multi-chromatic extractions, which use additional image frames to constrain better the significance of sources. The smaller beam size for the 450 $\mu$m observations also helps to separate overlapping cores in clustered regions, most notably in the centre of Oph A.  

In the initial output, B18 had 23 source detections labeled as `reliable' by {\get}, while L1688 had 190 and NGC 1333 had 103. The relative detection numbers are consistent with expectations, as B18 is known to be a quiescent region with minimal active star formation, compared to the more active NGC~1333 cloud and even more active L1688 cloud. We visually inspected the resulting core catalogues and found that some likely spurious sources did arise from the extraction, and therefore additional culling of the catalogue was needed. We required that the {\get}-defined Global Goodness be greater than one (\texttt{GOOD}\textgreater 1), the Global Significance be greater than ten (\texttt{SIG\char`_GLOB}\textgreater 10), the 850 $\mu$m detection have a significance of more than five (\texttt{SIG\char`_MONO02}\textgreater5) and a ratio of peak flux to noise greater than one (\texttt{FXP\char`_BEST02}/\texttt{FXP\char`_ERRO02}\textgreater 1). These conditions ensured that any remaining sources had both robust global characteristics and at least a 5-$\sigma$ detection significance at 850 $\mu$m. The culled catalogs had around 25\% fewer sources, leaving L1688 with 134 sources, NGC~1333 with 78, and B18 with 17. A visual inspection of the culled sources showed that most were very faint and very few (if any) were associated with significant emission. These culled {\get} catalogues were used in the analysis presented in the main body of this paper.

\begin{figure}[htb]
\centering
\includegraphics[width=7.5in]{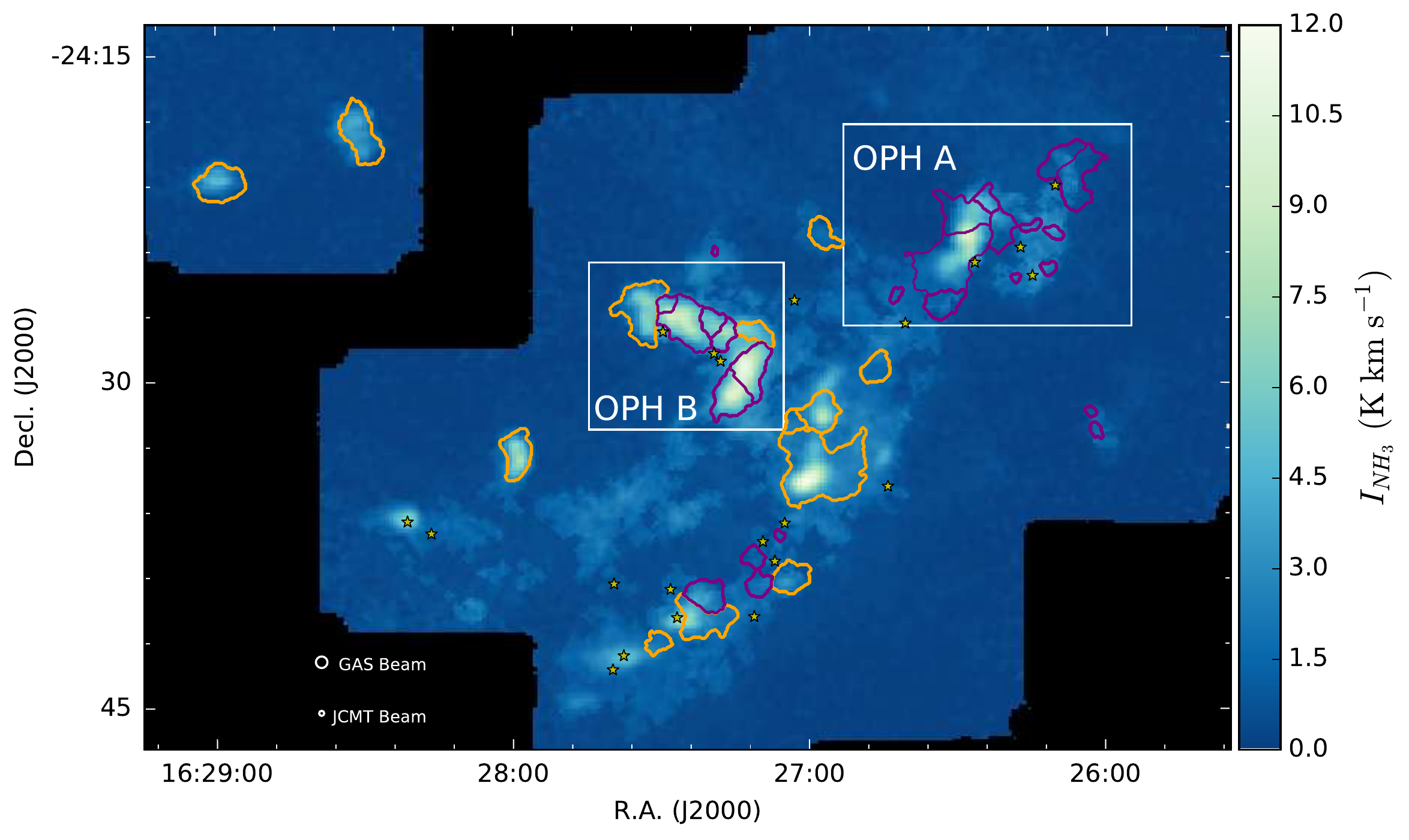}
\caption{Dense cores identified in L1688 in Ophiuchus using {\fw}, plotted on the GAS NH$_3$ (1,1) integrated intensity map. The plotting conventions are the same as in Figure~\ref{fig_oph_gs}.}
\label{fig_oph_fw}
\end{figure}

\begin{figure}[htb]
\centering
\includegraphics[width=6.2in]{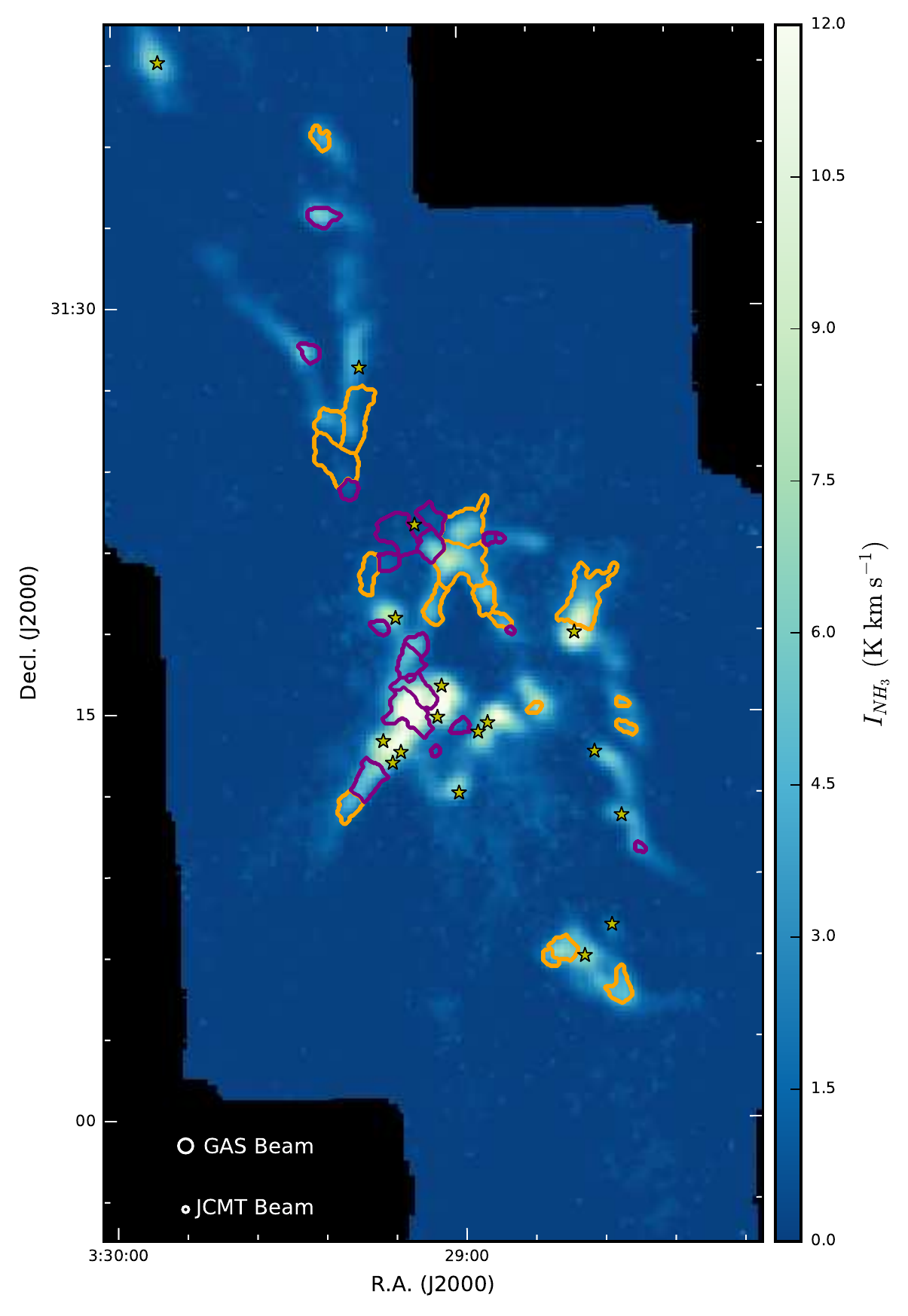}
\caption{Cores identified in NGC1333 in Perseus using {\fw}.  See Figure~\ref{fig_oph_fw} for details.}
\label{fig_pers_fw}
\end{figure}

\begin{figure}[htb]
\centering
\includegraphics[width=9in,angle=270]{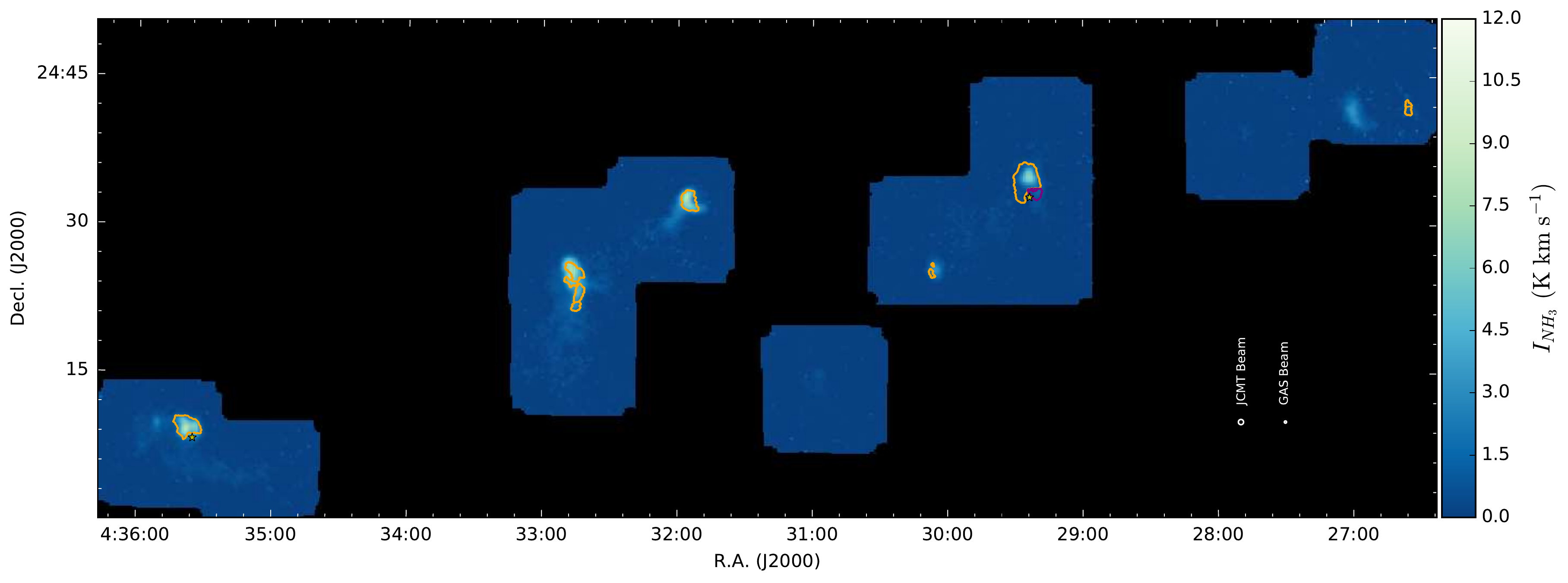}
\caption{Cores identified in B18 in Taurus using {\fw}.  See Figure~\ref{fig_oph_fw} for details.  
}
\label{fig_tau_fw}
\end{figure}

\subsubsection{{\fw}} \label{app:fellwalker}

We repeated our main analysis using instead the {\fw} algorithm to identify cores. {\fw} is an algorithm which identifies significant peaks based on the number of pixels visited on all possible paths to the peak.
{\fw} has a number of user-adjustable parameters. We kept most of these at the default recommended values, but changed several settings where testing suggested that the final core catalogue was improved, typically by providing cores that were better separated from surrounding filamentary structure. All of these non-default parameters are summarized in Table~\ref{tab_FWparams}. We applied the optional ``Findback'' function \citep{Currie14} which subtracts larger-scale background emission from the individual cores. Finally, we culled the resulting core catalogue to remove non-robust detections. 
First, we required that at least one third of the pixels in each core have values of greater than five times the local noise level. To rule out contamination by noise spikes, we also slightly smoothed the image (using a Gaussian smoothing kernel of $\sigma$=2 pixels) and required that at least one third of the pixels in the smoothed core have a value of greater than two times the local noise level. Roughly one third of cores were eliminated through the culls,
leaving 38 in L1688, 37 in NGC~1333, and 12 in B18. The cores identified by {\fw} in each region are shown in Figures~\ref{fig_oph_fw}-\ref{fig_tau_fw}.

\begin{figure}
\includegraphics[width=7.2in]{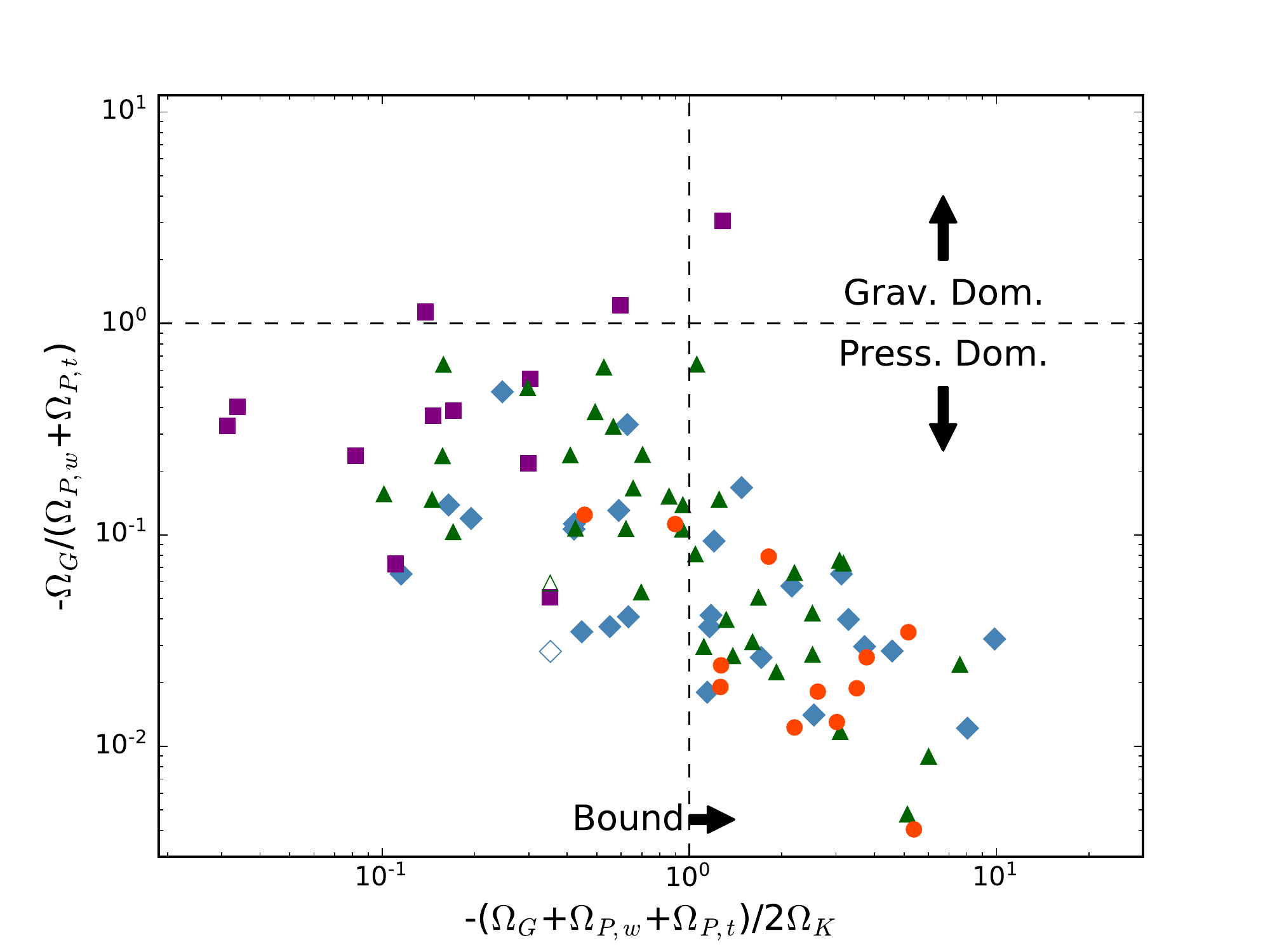}
\caption{The confinement ratio versus the total energy density for cores in all regions surveyed, using the {\fw} catalog. See Figure~\ref{fig_larson} for the plotting conventions used.}
\label{fig_all_fullvir_fw}
\end{figure}

We ran our virial analysis on the {\fw} catalogue, and show the results in Figure~\ref{fig_all_fullvir_fw}. About
40\% of the cores in L1688, 50\% in NGC~1333, and over 80\% of B18 cores are identified as bound by the combination of gravity and pressure, which are all higher than the values of 36\%, 27\%, and 56\%, respectively, that we estimated in our main {\get} analysis. Furthermore, all cores except three have their binding dominated by pressure, not gravity, which, like in the {\get} results, underlines the importance of pressure in these cores. The fact that we find a higher fraction of bound cores using {\fw}, especially in NGC~1333 and B18, 
may be due to the nature of the {\fw} cores, which 
are often larger than those found by {\get}. Larger effective radii increase the area over which the pressure acts, therefore increasing the pressure terms. 

Due to its ability to separate cores with closely spaced peaks using additional higher-resolution but noisier 450~$\mu$m data, {\get} detects significantly more cores in crowded regions compared to {\fw}. As can be seen in Figure \ref{fig:gsvfw}, 
{\fw} sometimes groups together visually distinct faint emission structures with nearby brighter features. Furthermore, since {\fw} does not model and subtract larger scale structures during core detection, it often has difficulty separating cores from bright, larger-scale structures. As such, in regions with complex emission structures, {\get} appears to perform better than {\fw} in separating nearby peaks from large-scale emission. While we conclude that {\get} is preferable to {\fw} for the purposes of our analysis, it is reassuring to see that the two algorithms provide qualitatively similar conclusions in a virial analysis. 

\begin{deluxetable}{ccc}
\tablecolumns{3}
\tablewidth{0pt}
\tablecaption{Non-Default Settings Used For {\fw}\label{tab_FWparams}}
\tablehead{
\colhead{Parameter} &
\colhead{Value} &
\colhead{Description}
}
\startdata
\texttt{FellWalker.AllowEdge} & 0 & Do not include cores that touch the image boundary\\
\texttt{FellWalker.CleanIter} & 6 & Apply smoothing to boundaries of clumps\\
\texttt{FellWalker.MaxJump} & 2 & The maximum distance (in pixels) that a path can skip \\
\texttt{FellWalker.Noise} & 3.2*RMS & The minimum value from which to start climbing\\
\texttt{FellWalker.RMS} & 0.05 & The RMS Noise level of the data \\[1ex]
\enddata
\end{deluxetable}

\subsubsection{\citet{Pattle15} Catalog (CuTEx)}

CuTEx \citep{Molinari11} is a source finding algorithm that detects sources using the second derivative of a (single, monochromatic) flux map.  The algorithm therefore excels at detecting sharp peaks in data, while it is prone to miss more diffuse, less peaked cores. \citet{Pattle15} identified dense cores in the SCUBA-2 map of L1688 by running CuTEx, and then folded in \nhp(1-0) and C$^{18}$O(2-1) observations to trace the core and cloud material respectively and subsequently run a full virial analysis. 
The characterization of the turbulent pressure term in \citet{Pattle15} therefore differs somewhat from that of this paper in their
use of C$^{18}$O (2-1), which traces denser gas than $^{13}$CO (1-0) emission \citep{Nishimura15}. Subsequently, the surrounding cloud density estimate is also higher, at 10$^5$~cm$^{-3}$ compared to our estimate of 3$\times$10$^3$~cm$^{-3}$. Therefore, although the underlying methods behind \citet{Pattle15} are quite similar to ours, the different choices for mean ambient cloud density and its corresponding dense tracer do not permit the same level of comparison as presented above for {\fw} cores.

Therefore, we instead performed a virial analysis on the \citet{Pattle15} CuTEx core sample using exactly the same calculation methods employed earlier in the paper, including the use of our lower-density $^{13}$CO observations, and typical cloud density assumption of $3\times10^3$~cm$^{-3}$. Using this method, we find that none of the cores are bound, clearly a significantly lower fraction compared to what was found in both the {\fw} (39\%) and {\get} (36\%) results\footnote{This fraction of bound cores is considerably lower than the original fraction found in \citet{Pattle15}, which is around 70\%. The significantly higher cloud density estimate drives this difference, as the ratio between the line widths from the $^{13}$CO and C$^{18}$O data is on average only 1.1.}. This lack of bound cores in the CuTEx results is heavily impacted by the selection bias of the algorithm. When we examine the subset of {\get} cores that have matches in the CuTEx catalogue, we similarly find that an extremely low fraction, only 3\%, are bound.  This result suggests that CuTEx and {\get} differ primarily in the population of cores that are identified, and much less in the properties (mass, size) ascribed to the cores.  Specifically, CuTEx excludes larger and fluffier cores, which tend to be the ones where pressure binding is strongest due to their larger surface area.  We therefore conclude that the use of {\get} is more appropriate for our work because it identifies a more diverse range of cores.

\subsection{Core Line Width Estimates} \label{sec_high_veldisps}

As noted in Section~\ref{sec:Jeans}, the line widths measured by GAS for the dense cores are often significantly non-thermal, particularly in L1688 and NGC~1333.  This basic observation runs contrary to most previously published studies, where dense cores are typically found to be thermally-dominated structures \citep[e.g.,][]{Jijina99,Kirk07,Foster09}. The consequence of the large line widths in our subsequent analysis is to imply that the dense cores are mostly gravitationally unbound, and therefore significant pressure binding is required to prevent the cores from expanding.  In this section, we use comparisons with other datasets to determine whether or not the GAS NH$_3$ data is biased towards systematically higher line widths than should be attributed to the dense cores.  Specifically, there are two issues that we wish to test.  First, due to the moderately low resolution of GAS (32\arcsec), there may be cases where GAS measures emission from more than one dense structure along the line of sight, thus increasing the observed line width.  Second, NH$_3$ is sensitive to gas at both high \textit{and} moderate densities, and it may therefore be tracing not only the dense core but also the surrounding lower-density envelope, which is typically more turbulent. Outflows may also contaminate the NH$_3$ data, as outflows have been shown to host significantly more NH$_3$ emission compared to denser gas tracers \citep[e.g.,][]{Tafalla95}. For each of the three regions analyzed, we compare the GAS data to available complementary data to establish whether either of these biases might be present, and estimate the degree to which they affect our results. 

\subsubsection{NGC~1333}
\label{sec:per_lwc}
\begin{figure}
\includegraphics[width=3.6in]{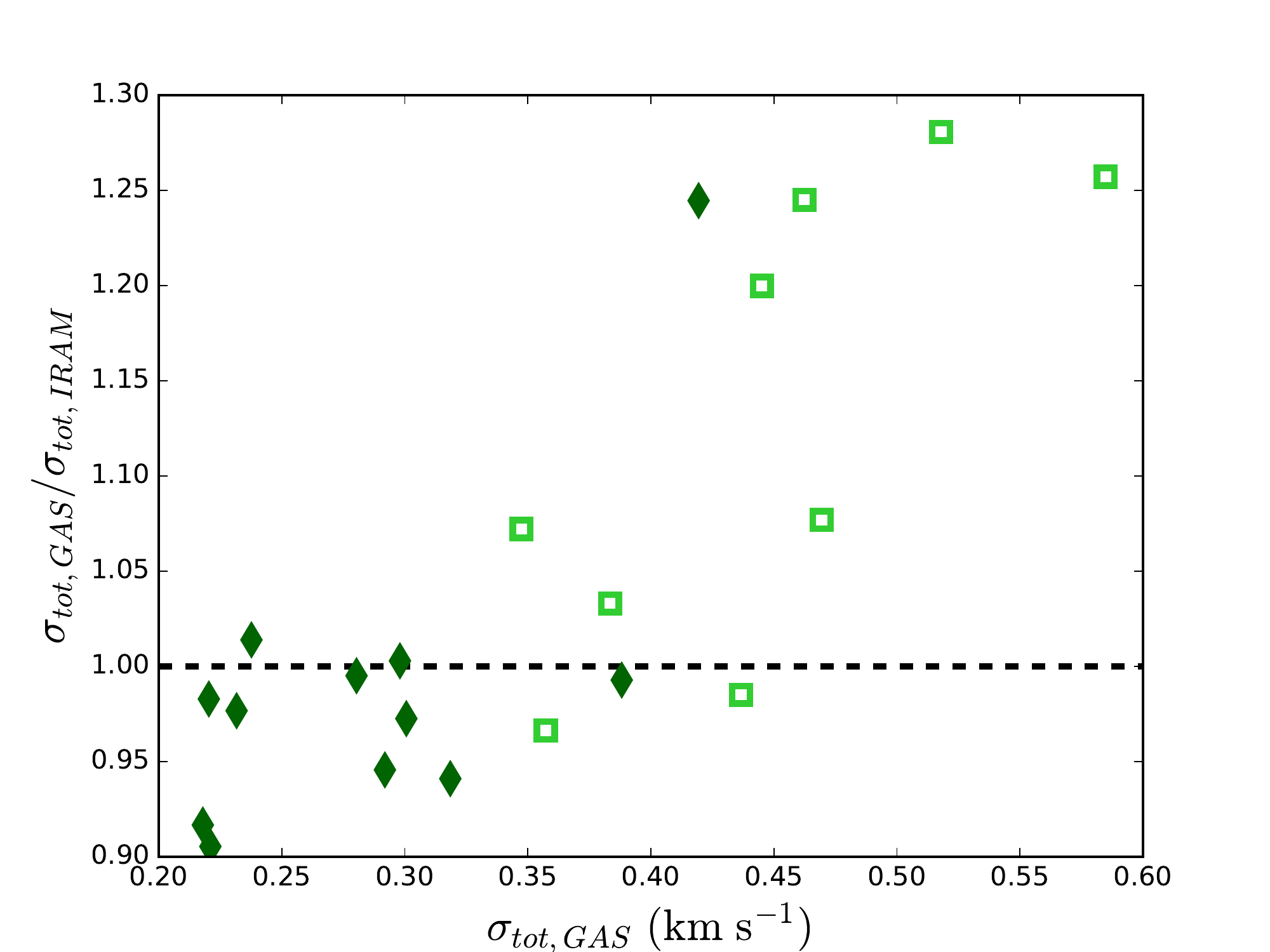}
\includegraphics[width=3.6in]{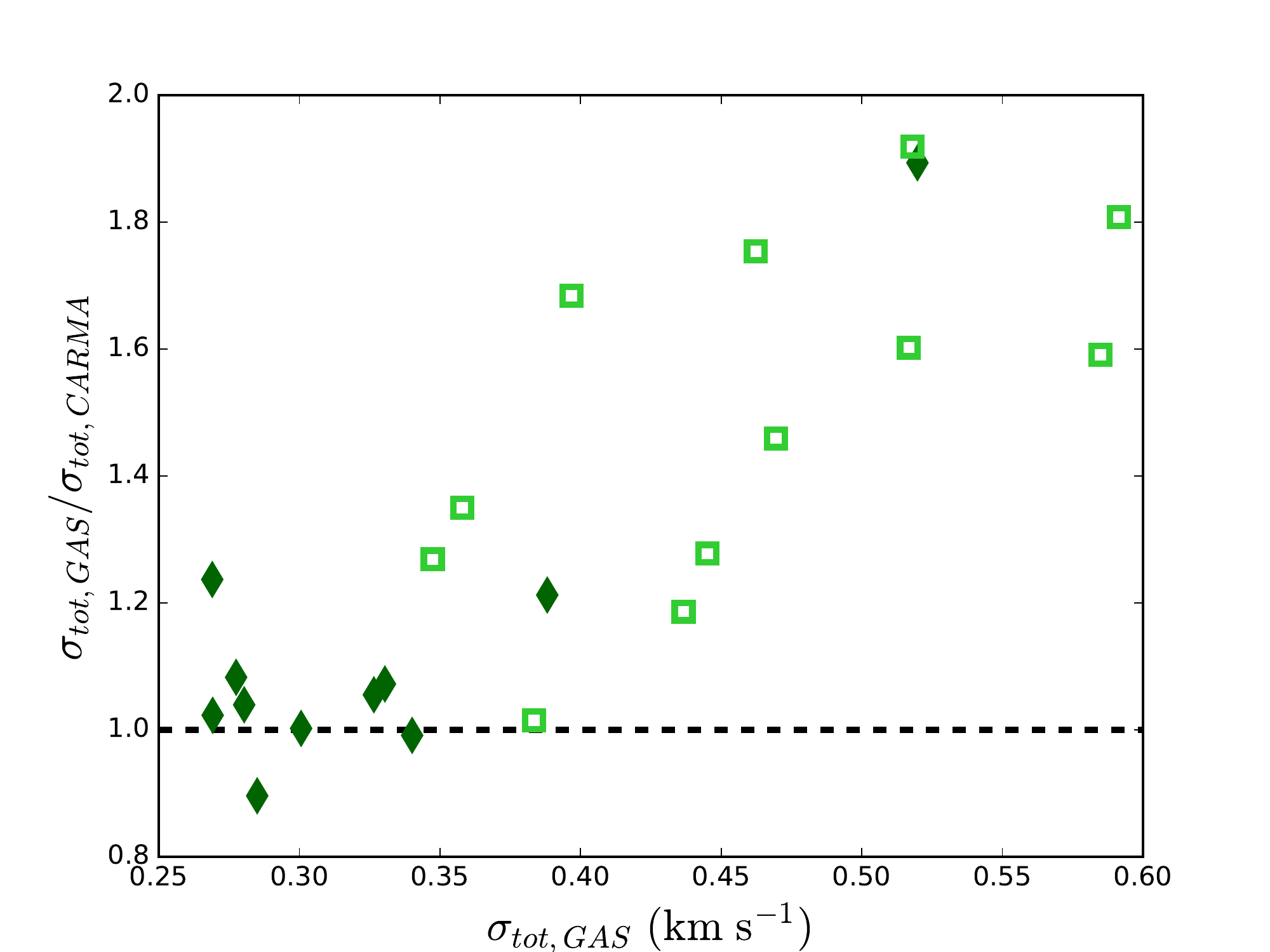}
\caption{A comparison of line widths for the dense cores in NGC~1333 as measured by GAS in NH$_3$(1,1) and previously published values for \nhp(1,0) from the IRAM~30m \citep{Kirk07} (left panel) and CARMA \citep{Walsh07} (right panel).  In both panels, the vertical axis shows the ratio of the GAS line width to the corresponding \nhp\ measurement.
Open squares correspond to cores in the clustered centre of NGC~1333, while closed diamonds represent cores in outlying regions. The dashed line represents where the two compared quantities would be equal.
}
\label{fig:N1333lwc}
\end{figure}

%\soedit{In terms of the similarity between N2H+ and NH3 Gaches et al. 2015 is relevant to cite here. This paper showed that the emission of these two tracers, including how the line profiles changed with scale were statistially similar. This work is particularly applicable especially because the chemistry was computed for a 3D, turbulent star-forming (simulated) cloud - rather than 1D/symmetric conditions used by many astrochem studies.} 

%\ernotes{ER: From the observational point of view, I think that Johnstone+(2010) also investigated this point.}

\nhp\ is a frequently observed molecular tracer of dense cores, and subsequently, diverse N$_2$H$^+$ data exist across the GAS DR1 regions. The \nhp(1-0) transition is not strictly a tracer of denser gas than NH$_3$(1,1); the latter likely traces a larger range of densities than the former \citep[see, e.g., the discussion in][]{Kirk17b}. Subsequently, NH$_3$ (1,1) and {\nhp} (1-0) often have very similar emission profiles \citep{Gaches15,Johnstone10}. Nonetheless, the effective critical density probed by \nhp(1-0) is higher than that of NH$_3$(1,1) \citep{Shirley15}, 
and therefore \nhp(1-0) is likely a tracer of higher-density gas in some conditions.

\citet{Kirk07} used the IRAM 30~m telescope to observe \nhp(1-0) in single pointings at the peaks of 157 dense core candidates across the Perseus molecular cloud.  The angular resolution of 25\arcsec\ is similar to that of GAS (32\arcsec). We identified 21 observations where the IRAM \nhp\-observed position was within one IRAM beam (25\arcsec) of one of our cores. For our comparison of line widths, we separately examine cores located in the most clustered part of NGC~1333 and the remainder of the sample. Here, we define the clustered area as a 4{\arcmin}-wide square centered on J2000 (03:29:07, 31:14:42), that roughly encapsulates the filament surrounding the protostar IRAS 4. Figure~\ref{fig:N1333lwc} shows a comparison of the total line widths measured with GAS NH$_3$ and IRAM {\nhp}, where the latter
were calculated assuming the same gas temperature as observed with GAS. 

We find that the ratio between the GAS NH$_3$ and IRAM {\nhp} line widths in the clustered centre of NGC~1333 is 1.12 $\pm$ 0.12 (mean and standard deviation), i.e., the NH$_3$ line width is typically 12\% higher than the \nhp\ line width. The results outside of the central cluster on average agree quite closely between the two studies, with a ratio of 0.99 $\pm$ 0.08. Excluding one outlier, outside of the central cluster, no NH$_3$ line widths exceed the {\nhp} results by more than 2\%. Using GAS line widths and kinetic temperatures taken directly at the location of the {\nhp} pointings rather than averaged over the core returns similar ratios, albeit with a notably wider spread in values, reaching ratios of nearly 1.5 at most, despite having a somewhat lower average. These relatively subtle differences under most circumstances are consistent with the results in \citet{Johnstone10}, which also compared pointed GBT NH$_3$ to pointed IRAM {\nhp} in the Perseus Molecular Cloud. The similarity between the pointed IRAM {\nhp} and current GAS core-averaged NH$_3$ also illustrates the similarity between pointed and core-averaged line width measurements. %\rfnotes{It might be worthwhile to check the GAS line widths *at* the exact locations of the \nhp\ pointings. In clustered regions, the impact of averaging over the core might be more important than in the outer regions.}

We next compare our measurements with CARMA observations of N$_2$H$^+$(1-0) at 10\arcsec\ resolution from \citet{Walsh07}.  We require matched cores to have peak positions separated by less than the JCMT beam size (14\farcs1), finding a total of 23 matches, of which roughly half (12) lie within the central dense portion of NGC~1333. In the outskirts of NGC~1333, we find relatively small variations in the line width; the ratio of GAS to CARMA line widths is 1.14~$\pm$ 0.26, and only three cores matched here have GAS line widths that are more than 10\% larger than their CARMA counterparts.
In the central part of NGC~1333, the ratio between the GAS and CARMA line widths is much larger, at 1.49$\pm$0.27. Several of the cores have ratios of close to 2.

We additionally run a test within the GAS data itself. \citet{Friesen16} ran a visual check of all GAS DR1 spectra, and found no more than \edit1{10\% to 15\%}
of the spectra \edit1{in zones with more complex emission} appeared to contain more than one velocity component.  The hyperfine structure of NH$_3$(1,1) emission can, nonetheless, make the identification of multiple velocity components difficult, particularly if the second component is much fainter.  \citet{Chen19} recently pioneered a new technique for extracting such additional velocity components.  
%\jepnotes{Is this 40\% of hte whole cube? Because that is not what we get, we see a very low value.} \rknotes{yes it is, Mike's paper found many more locations where multiple components matter than in the DR1 paper} \mcnotes{[MC: it's of the `fitted pixels,' see my email for details.]}
In NGC~1333, \citet{Chen19} find that about 25\% of the pixels which are bright enough to enable reliable line-fitting are better fit with two velocity components, where the second velocity component is typically associated with faint and broad emission. Dense cores identified using {\get}, which are generally found in more crowded regions, 
are much more likely than the rest of the cloud to have multiple components, with 68\% of the dense cores containing pixels with a second velocity component within their extent. Assuming that the brightest of the two velocity components corresponds with the true dense core emission, we compare the line width measured by \citet{Chen19} with the single-component DR1 line width. For 50\% of these comparisons, the two line width measurements agree within 10\%, i.e., the second velocity component identified by \citet{Chen19} is so faint that it did not influence the single component fit used in DR1.  Yet 27\% of the sample has significant discrepancies of 25\% or more (corresponding to a change in $\Omega_k$ of at least 50\%) between the original GAS DR1 and two-component line width results. In the most extreme case, the DR1 line width is almost double the brighter two-component velocity fit from \citet{Chen19}.  These cores with discrepant line widths represent about 18\% of the total population in NGC~1333, i.e., of the 68\% of all cores best fit by multiple velocity components, 27\% have significant discrepancies.
This result suggests that, in a non-negligible number of cases, the NH$_3$ line widths derived for the cores based on the GAS DR1 fits are significantly influenced by the presence of multiple %substantial 
reservoirs of dense gas along the line of sight.

The \citet{Chen19} two-component line widths in the central filament of NGC~1333 are on average larger than those in the \citet{Walsh07} CARMA data, which may be due to the presence of additional, unfit components. Slightly narrower line widths are found in the two-component GAS line widths compared to CARMA in outlying regions, which is somewhat unexpected, as these regions are not expected to contain much crowded emission. This difference, however, is small relative to the standard deviation of the line width distribution, so it can likely be explained by the significantly larger sample size available for the two-component comparison, which contains more cores in the marginally more crowded outer filaments of the cloud.

Through our two comparisons to other catalogs and additional check within GAS, we find that the line widths of cores in the outskirts of NGC~1333 are minimally affected by both the angular resolution and the dense gas tracer, while the most clustered cores often have noticeably elevated line widths in GAS. We find that the smallest line widths are found in the \citet{Walsh07} CARMA \nhp\ observations for the central filament, and the \citet{Chen19} two-component fits for outlying regions. The average ratios between GAS DR1 line widths and these lowest-line width data sets (1.18 in outlying regions, 1.49 in the clustered centre) correspond to an $\Omega_k$ term that is on average a factor of 1.4 larger in outlying regions and 2.2 in the clustered centre - significant effects. We \edit1{applied} these two ratios in line widths to the full population of isolated and clustered cores in NGC~1333 when re-calculating $\Omega_k$ 
in Section \ref{sec:rvr}, thereby providing a rough correction to the biases that may be present in the GAS DR1 single-component fits used for our main analysis.

\subsubsection{L1688}

\citet{Punanova16} used the IRAM~30 m telescope to observe {\ndp}(1-0) emission in L1688 with 32\farcs1 resolution. As the deuterated form of {\nhp}, {\ndp} distributions are regulated by enhanced deuterium fractionation towards the central regions of dense cores, where freeze-out happens more rapidly \citep{Caselli02}. This effect is reflected in the higher {\ndp} to {\nhp} ratios found in regions with higher H$_2$ densities \citep{Pagani07}. Therefore,  {\ndp}(1-0) can be used to trace gas at typically higher densities than using NH$_3$(1,1) emission.
%\rknotes{[Note to other collaborators: it would be nice to make a statement on the density ranges here, does anyone have a good citation?]}
%while the range of densities that it traces does still overlap somewhat with NH$_3$ \rkflag \rknotes{I think that the discussion in your Orion A paper and the pagani papers lead to this conclusion, but I would like something more direct if possible}, {\ndp} emission is, on average, more concentrated in dense regions than NH$_3$ or {\nhp}, making it a very reliable dense gas tracer. 
The \citet{Punanova16} {\ndp}(1-0) data have a similar angular resolution to our GAS observations, allowing us to compare the two tracers on the same scale. We identify 24 cases where the GAS and IRAM core locations agree within 32{\arcsec} (i.e., the GAS resolution), and perform line width comparisons for those cores. \edit2{Note that in Oph A, all cores that have \ndp observations lie in the central region, and therefore the higher-line width outlying cores are not included in any statistics discussed here}
%\hknotes{HK question -- 'all cores with full coverage' is confusing.  Do you mean 'all cores that have \ndp observations'?}

We find that the ratio of GAS NH$_3$ to the IRAM {\ndp}(1-0) line widths is 1.13 $\pm$ 0.09 in Oph A, 1.20 $\pm$ 0.37 in Oph B, and 1.02 $\pm$ 0.39 in outlying regions. These are very similar ratios to those seen in NGC~1333 for crowded versus isolated emission regions (see Section \ref{sec:per_lwc}). The ratio of line widths is larger in Oph~B than in Oph~A, which was unexpected given that Oph~A appears to have more clustered emission, as evidenced by the numerous tightly-packed, often overlapping {\get}-identified cores found in the GBS maps for the region (see the {\get} contours in the upper right of Figure \ref{fig_oph_gs}).  It is possible that the mean density in Oph~A is sufficiently high that lower-density contaminating features along the line of sight continue to contribute even in {\ndp}(1-0) emission.

We also compare the GAS line widths to those measured in higher-resolution (16\farcs3 beam) \ndp(2-1) observations also published in \citet{Punanova16}. Similar line width ratios to those listed above are found in both Oph~A and the sparser regions of the cloud, while the ratio in Oph~B increases significantly to 1.46$\pm$0.38. A visual comparison to the line widths found in slightly higher-resolution (13{\arcsec}) {\nhp} observations of Oph A in \citet{DiFrancesco04}, indicate similar ratios to what is shown above. While the results in Oph B and outlying regions continue to reflect the results in NGC~1333, the Oph A ratios remain unexpectedly low. This may imply that even better angular resolution is required to resolve individual cores fully \citep[as suggested in][]{Punanova16}, or it may simply reflect Oph A's warmer, more active nature. Recent studies using ALMA have also shown evidence of additional protostellar cores in Oph A \citep{Kawabe18,Friesen14,Friesen18}, so our sample may be skewed towards larger line widths by the unintentional inclusion of protostellar cores and gas from any associated outflows. 

Finally, we present an additional test to check for multiple velocity components, using the same two-component line-fitting techniques \citep{Chen19} discussed in Section \ref{sec:per_lwc}. Comparing the GAS DR1 NH$_3$ line width to the narrower component in the fit by \citet{Chen19}, we find ratios of 1.20 $\pm$ 0.20 in Oph~A, 1.24 $\pm$ 0.16 in Oph~B, and 1.12 $\pm$ 0.20 in the less-clustered regions.  In Oph~B, these line width ratios are less than those measured using the comparison with \ndp(2-1) emission, similar to our results in the NGC~1333 central filament.  
Oph~A shows the opposite trend, with line width ratios {\it larger} when compared to the {\ndp} to NH$_3$ ratios. This result further suggests that even the \citet{Punanova16} \ndp (2-1) observations are unable to disentangle tightly clustered gas pockets along the line of sight, especially in Oph A. The two-component fitting also derives slightly thinner line widths (compared to the N$_2$D$^+$ observations) in outlying regions. Like in NGC~1333, this is likely due to selection effects, since more of the outlying cores with data in the GAS two-component fits are in slightly more crowded regions compared to the \citet{Punanova16} cores. %\rfnotes{I still think mentioning James' 2004 paper with high resolution \nhp\ here would be helpful. He still finds large linewidths in some regions with higher resolution than in Punanova et al.}\rknotes{[James' paper uses FWHM rather than sigma, so when you convert the line widths basically match with the comparison to Punanova et al. I made a note of this in the third paragraph]}

For each of Oph~A, Oph~B, and the outlying regions, we conservatively \edit1{applied} the largest ratio in line widths found above as a possible correction factor to the GAS measurements \edit1{in Section \ref{sec:rvr}}.  These ratios are
1.20 in Oph~A, 1.46 in Oph~B, and 1.12 in the outlying regions.
%The ratios that will be applied to the virial results in Section \ref{sec:rvr} are therefore 1.51 in Oph B, 1.22 in Oph A, and 1.13 in outlying regions. 

\subsubsection{B18}

Unlike NGC~1333 and L1688, GAS observations show that B18 tends to be a thermally dominated environment; none of the cores have line widths exceeding 30\% above the thermal value.  With such relatively small line widths measured, any intrinsic bias toward larger widths must be necessarily much less than we found in the other regions.  Nonetheless, we attempt to quantify the likely magnitude of such effects. We were unable to locate any large-scale N$_2$H$^+$ or N$_2$D$^+$ surveys of B18, so we cannot perform the same depth of comparisons as in the other regions. There are, however, studies employing NH$_3$(1,1) and N$_2$H$^+$(1-0) in the adjacent L1495-B218 complex, which we will use as a proxy for B18. \citet{Seo15} observed NH$_3$(1,1) in L1495-B218 using the Green Bank Telescope under a similar set up to GAS. \citet{Seo15} find that the line widths are thermally dominated, with an average dense core $\sigma_{tot}$ of 0.209 km/s, which is about 7\% less than the average core line width that we measure in B18.

We compare the \citet{Seo15} NH$_3$(1,1) data to combined FCRAO and IRAM~30 m observations of \nhp(1-0) in L1495-B218 by \citet{Hacar13}. All cores were mapped with FCRAO at 57{\arcsec} resolution, while regions where emission appeared more complex received IRAM follow-up at 25{\arcsec} resolution. For each NH$_3$ dense core in \citet{Seo15}, we search for a matching \nhp\ line width fit within 31\arcsec, which were derived by \citet{Hacar13} at 375 locations in L1495-B218. For the 46 matched cases identified, we find very similar line widths, with an average line width ratio (NH$_3$  to \nhp) of 1.02 $\pm$ 0.08. Only one ratio was greater than 1.2. We therefore conclude that if B18 is indeed similar to L1495-B218, NH$_3$(1,1) and \nhp(1-0) observations made at comparable resolutions are likely to have nearly identical line widths in B18. We therefore \edit1{did} not apply a corrective ratio to the revised B18 results in Section \ref{sec:rvr}. 

We were unable to locate any higher-resolution surveys of B18 or L1495-B218 to test the influence of angular resolution on the line widths measured.  We note, however, that with its sparse emission at higher densities, B18 is unlikely to have dense core measurements that are influenced by crowding along either the line of sight or in the plane of the sky.

\begin{deluxetable}{cccccc}
\tablecolumns{6}
\tablewidth{0pt}
\tablecaption{Line width ratios between GAS and other dense gas tracers\label{tab_lwch}.}%The line width ratios between GAS and alternative data presented in this section, with standard deviations. B18 is excluded, since no direct comparisons were possible there. The boldface entries indicate the corrective ratios used in Section \ref{sec:rvr}\label{tab_lwch}}
\tablehead{
\colhead{} &
\colhead{L1688 Oph~A\tablenotemark{a}} &
\colhead{L1688 Oph~B\tablenotemark{a}} &
\colhead{L1688 Out\tablenotemark{a}} &
\colhead{NGC~1333-Cen\tablenotemark{a}} &
\colhead{NGC~1333-Out\tablenotemark{a}}
}
\startdata
Denser Tracer Only \tablenotemark{b}& 1.13 $\pm$ 0.09 & 1.20 $\pm$ 0.37 & 1.02 $\pm$ 0.39 &1.12 $\pm$ {0.12} &0.99$\pm$ 0.08\\
Denser Tracer \& Higher Resolution \tablenotemark{c}& 1.11 $\pm$ 0.05 & \textbf{1.46 $\pm$ 0.38}  & 1.04 $\pm$ 0.45 & \textbf{1.49 $\pm$ 0.27}  & 1.14 $\pm$ 0.26\\ 
Multi-Component Fitting on GAS\tablenotemark{d} & \textbf{1.20 $\pm$ 0.20} & 1.24 $\pm$ 0.16 & \textbf{1.12 $\pm$ 0.20}  &1.35 $\pm$ 0.25  &\textbf{1.18 $\pm$ 0.27}\\
\enddata
\tablenotetext{a}{The line width ratios between GAS and other dense gas observations.  'Out' indicates the outlying portions of each region, while 'Cen' indicates the central portion.  B18 is excluded, as no direct comparisons were possible there.  The boldface entries indicate the corrective ratios applied in Section~\ref{sec:rvr}.}
\tablenotetext{b}{Comparison with a denser gas tracer at a similar resolution to GAS.  For L1688, the denser tracer is {\ndp} \citep[from][]{Punanova16}, and for NGC~1333 it is {\nhp} \citep[from][]{Walsh07,Kirk07}.}
\tablenotetext{c}{The ratios use the same dense gas tracer as above, but at 16{\farcs}3 in L1688 and 10{\arcsec} in NGC~1333.}
\tablenotetext{d}{A comparison with the multiple velocity component fitting by \citet{Chen19}.}
\end{deluxetable}

\subsubsection{Revised Virial Estimates}
\label{sec:rvr}
We next estimate the influence of the possible NH$_3$-based core line width estimates discussed in the previous sections on our virial analysis. To do so, we multiply $\sigma_{tot}$ for each core by the ratio found between GAS data and reference data in the core's parent region as stated in the preceding subsections (see Table \ref{tab_lwch}). We note that we did not use any core-specific correction ratios, as not all cores in our GAS analysis had other kinematic data available for a direct comparison. The resulting virial plot is shown in Figure \ref{fig_all_correctvir_gs}, and some important statistics are shown in Table \ref{tab:systematics}.

After applying these ratios, we find that only about 40\% of cores in both L1688 and NGC~1333 are identified as unbound, corresponding to a total of 55 of 134 cores across all three regions. Oph B and the clustered centre of NGC~1333 both show a majority of bound cores under this correction, despite containing all unbound cores in the original uncorrected result. A total of 19 cores remain identified as unbound Oph~A, however. The relatively large fraction of apparently unbound cores in Oph~A compared to the other clustered regions merits additional research, especially at higher resolution, to determine whether this difference is physical, perhaps relating to the somewhat warmer, active nature of the region \edit1{\citep[e.g.,][]{Friesen16, Stemellos07}}, or just a side-effect of observational limitations. We emphasize that the above calculation provides a statistical correction to the potential over-estimate in the core line widths, and the results are not expected to be exactly correct for any individual core.  Nonetheless, this estimated correction illustrates two important points.  First, it suggests that many of the dense cores that we identify as being unbound in our analysis may in fact be classified as bounded by the combination of pressure and gravity, if we were instead using a higher density tracer observed at higher angular resolution.  Second, our main conclusion, that pressure is the dominant factor in binding the cores, remains true even accounting for potential biases in the reported core line widths; in Figure~\ref{fig_all_correctvir_gs}, the vast majority of the cores lie below the horizontal dotted line.

\subsection{Cloud Weight Pressure and Filtering Scale} \label{app:filscale}

\begin{deluxetable}{cccc}
\tablecolumns{4}
\tablewidth{0pt}
\tablecaption{Variation in cloud pressure with filter scale\label{tab_filter}}
\tablehead{
\colhead{Filter Scale (pc)\tablenotemark{a}} &
\colhead{L1688} &
\colhead{NGC~1333} &
\colhead{B18} \\
\hline
\colhead{} &
\multicolumn{3}{c}{$\overline{{\rm log} (P_w/k_B)}~~({\rm log}~{\rm cm}^{-3}{\rm K})$}
}
\startdata
%\cutinhead{ \hspace{1.2in} $\overline{{\rm log} (P_w/k_B)}~~({\rm log}~{\rm cm}^{-3}{\rm K})$} 
0 & 6.47 & 6.23 & 5.94 \\
0.3& 6.10& 5.90 & 5.39 \\
0.6 & 5.95 & 5.74 & 5.16 \\
1.2 & 5.71 & 5.50   & 4.94 \\
\cutinhead{ \hspace{1.2in} $\overline{\rm{log}(-(\Omega_g + \Omega_{p,w}+\Omega_{p,t})/2\Omega_k)}$}
0  &-0.04& 0.03 & 0.73  \\   
0.3 & -0.39 & -0.18 & 0.23 \\   
0.6  & -0.55 & -0.26 & 0.20 \\   
1.2 &  -0.77 & -0.35  & 0.13 \\
\enddata
\tablenotetext{a}{Approximate filtering scale applied.  Exact values vary slightly by region due to the differing distances to each cloud.}
\end{deluxetable}

As discussed in Section~\ref{sec:cloudpress}, the binding of dense cores due to the weight of overlying cloud material is another factor in our virial calculation for the cores. The calculation of this parameter relies on several assumptions, including approximate spherical symmetry of the cloud, which are discussed in more detail in \citet{Kirk17b}.  One assumption which is poorly constrained and might vary from cloud to cloud is the size scale of core versus cloud material.  In our simple model, we associate the column density at any location with an effective cloud pressure, and must therefore take care to remove the contribution of the column density associated with the dense core itself beforehand. We separated the core and cloud material using a simple spatial filtering algorithm, the \textit{\'{a} Trous} wavelet decomposition.  In this scheme, the size scale dividing core and cloud must be represented as $2^N$ pixels.  In our main analysis, we conservatively selected a scale of approximately 0.6~pc (64 pixels in L1688 and B18 and 32 pixels in NGC~1333) to represent the division between cores and the larger structures that they inhabit.  Here, we examine the influence of that selection by re-running our analysis using a suite of different filtering scales.

Table~\ref{tab_filter} shows the typical (mean) estimated cloud weight pressure for different assumptions about the filter scale.  Increasing or decreasing the filter scale by a factor of two results in pressure estimates which vary by a similar factor, roughly two or less ($<$0.3 in logarithmic units), while not applying any filtering (i.e. attributing {\it all} material with the cloud) increases the cloud pressure by a factor of about 3 in L1688 and NGC~1333 and a factor of 6 in B18.  The case of no filtering represents an extreme unphysical case, since the dense cores must have some material associated with them. Still, only L1688 seems to show significant changes in the virial ratio as a result of applying a somewhat higher or lower filtering scale choice, with a virial ratio that varies by up to 60\% (0.2 in logarithmic units) under a doubling or halving of the filter scale. Both NGC~1333 and B18 show much more modest variations under these filter scale changes, with typical variations in derived core virial parameters of less than 25\% ($<$0.1 in logarithmic units). As noted in Section \ref{sec:turpress}, cloud weight pressure provides a minor contribution to the binding of the dense cores in both B18 and NGC~1333, hence its lack of impact on the virial ratios. Subsequently, since L1688 has a cloud weight pressure term with a magnitude that is more equal to the contribution from turbulent pressure, changes to the filter scale have a more significant impact. 

\begin{deluxetable}{ccccc}
\tablecolumns{5}
\tablewidth{0pt}
\tablecaption{Virial parameters estimated for different analysis methods. \label{tab:systematics}}
\tablehead{
\colhead{} &
\colhead{$\overline{(\Omega_{p,t}/\Omega_{p,w})}$} &
\colhead{$\overline{\rm{log}(\Omega_{g}/(\Omega_{p,t}+\Omega_{p,w}))}$} &
\colhead{$\overline{\rm{log}(-(\Omega_g + \Omega_{p,w}+\Omega_{p,t})/2\Omega_k)}$}&
\colhead{\% Bound Cores}
}
\startdata
%%L1688
\cutinhead{ \hspace{1.2in} L1688}
Original {\get} Results   & 0.75 &-2.34 & -0.55 & 36\\
{\fw} Core Catalogue \tablenotemark{a} & 0.75 &-2.28 &-0.55&39\\ 
Corrected GAS Core Line Widths \tablenotemark{b}& 0.75 & -2.34 & 0.26 &57\\ 
$P_t$ Uses GAS NH$_3$ Properties \tablenotemark{c} & 0.19 & -1.96 & -0.84 &27\\
20\% Lower $^{13}$CO Line Widths \tablenotemark{d} & 0.48 & -2.18 & -0.68 &32\\
$P_t$ Uses $n_{^{13}\mathrm{CO}} = 1\times10^3$ \tablenotemark{e} & 0.25 & -2.01 & -0.81 & 28\\
$P_t$ Uses $n_{^{13}\mathrm{CO}} = 1\times10^4$ \tablenotemark{e} & 2.50 & -3.03 & 0.00 & 46\\
Filter Scale = 0.3 pc \tablenotemark{f} & 0.54 & -2.54 & -0.39 & 41\\
Filter Scale = 1.2 pc \tablenotemark{f} & 1.32 & -2.07 & -0.77& 30\\
%%NGC1333
\cutinhead{ \hspace{1.2in} NGC~1333}
Original {\get} Results   &2.16 &-1.17 &-0.26 & 27\\
{\fw} Core Catalogue \tablenotemark{a} &1.78 & -2.45& -0.09&49\\ 
Corrected GAS Core Line Widths \tablenotemark{b}&2.16 &-1.17 & 0.17&63\\ 
$P_t$ Uses GAS NH$_3$ Properties \tablenotemark{c} &0.25 & -0.24& -0.83&6\\
20\% Lower $^{13}$CO Line Widths \tablenotemark{d} &1.38 &-0.89 & -0.44&20\\
$P_t$ Uses $n_{^{13}\mathrm{CO}} = 1\times10^3$ \tablenotemark{e} & 0.72& -0.57& -0.64&10\\
$P_t$ Uses $n_{^{13}\mathrm{CO}} = 1\times10^4$ \tablenotemark{e} &7.19 & -2.11& 0.44& 67\\
Filter Scale = 0.3 pc \tablenotemark{f} & 1.58 & -1.29 & -0.18 & 35\\
Filter Scale = 1.2 pc \tablenotemark{f} & 3.71 & -1.03 & -0.35 & 22\\
%B18
\cutinhead{ \hspace{1.2in} B18}
Original {\get} Results   & 3.47 & -0.35& 0.20 & 56\\
{\fw} Core Catalogue \tablenotemark{a} &4.26 & -3.64& 0.74&83\\ 
Corrected GAS Core Line Widths \tablenotemark{b}& 3.47 & -0.35& 0.20&56\\ 
$P_t$ Uses GAS NH$_3$ Properties \tablenotemark{c} &0.40 & 0.59 & -0.36&22\\
20\% Lower $^{13}$CO Line Widths \tablenotemark{d} & 2.22& -0.07& 0.04&44\\
$P_t$ Uses $n_{^{13}\mathrm{CO}} = 1\times10^3$ \tablenotemark{e} &1.15 &0.27 & -0.14&44\\
$P_t$ Uses $n_{^{13}\mathrm{CO}} = 1\times10^4$ \tablenotemark{e} & 11.5  &-1.31 & 0.81& 100\\
Filter Scale = 0.3 pc \tablenotemark{f} &2.06 &-0.55 & 0.31&56 \\
Filter Scale = 1.2 pc \tablenotemark{f} &6.16 &-0.22 & 0.13&44 \\ 
\enddata
\tablenotetext{a}{These results use the Fellwalker core catalog described in Appendix \ref{app:fellwalker}}
\tablenotetext{b}{Employs the corrections described in Section \ref{sec:rvr}. No corrective ratio is employed for B18.}
\tablenotetext{c}{This provides an extreme lower limit to the estimate of external turbulent pressure by applying the mean GAS NH$_3$ line widths outside of identified cores to trace the kinematics of the ambient cloud. See Section \ref{sec:systematics} for details.}
\tablenotetext{d}{This provides a rough approximation to possible overestimates resulting from our use of $^{13}$CO rather than denser gas tracers in our estimate for $P_t$. See Section \ref{sec:systematics} for details.}
\tablenotetext{e}{The value used for $n_{13CO}$ in the main analysis is $3\times10^3$ cm$^{-3}$. See Section \ref{sec:systematics} for more information.}
\tablenotetext{f}{The scale used in the main analysis is 0.6 pc. See Appendix \ref{app:filscale} for details.}
\end{deluxetable}

\subsection{Turbulent Pressure Estimates} \label{sec:systematics}

The turbulent pressure estimate required several assumptions. Here we examine the likely range of uncertainty introduced into our final $\Omega_{p,t}$ values as a result. In our main analysis in Section~\ref{sec:turpress}, we estimated the turbulent pressure using $^{13}$CO(1-0) to measure the random motions in the medium surrounding each core, and assumed a representative density of $3\times10^3$~cm$^{-3}$. Both of these assumptions may cause us to over-estimate the true turbulent pressure. Indeed, the $^{13}$CO emission may be tracing a much larger volume along the line of sight, and the mean density that this emission represents may be different than assumed.

We first present a likely \textit{under}estimate of the turbulent pressure, to illustrate the maximum range of potential values.  Previous research \citep[e.g.,][]{Pineda10,HChen19} suggests that the transition between quiescent dense cores and their turbulent surroundings can be observed directly in NH$_3$(1,1) observations.  As an initial, coarse approximation, we take the median value of all line widths in the GAS maps that are \edit1{outside of all core
boundaries}, which yields a value of 0.37~km~s$^{-1}$ in L1688, 0.31~km~s$^{-1}$ in NGC~1333, and 0.18~km~s$^{-1}$ in B18. This result would suggest significantly lower turbulent pressure terms in all three regions. These small line widths, however, likely
underestimate the true inter-core line width, as there are likely additional quiescent gas structures which have not been adequately captured by the GBS-based core catalogue. 

For a more accurate estimate of the inter-core line width, we visually examine the GAS line width values surrounding a subset of more isolated cores, where the transition to coherence is clearer.  We find line widths that are often upwards of 0.8~km~s$^{-1}$ in both NGC~1333 and L1688, significantly larger than the median line width estimates listed above, and critically, only about 20\% smaller than the typical line widths found using $^{13}$CO. This comparison suggests that our original estimate using the $^{13}$CO line widths may not be substantially over-estimated. This suggestion is also in line with a comparison of the $^{13}$CO line widths measured in L1688 and the C$^{18}$O line widths presented in \citet{Pattle15}. C$^{18}$O also traces denser material than $^{13}$CO, making it less susceptible to overestimating the turbulence in the local core environment. We find, however, that the C$^{18}$O line widths are typically only 10\% smaller than the $^{13}$CO line widths. While these somewhat smaller line widths would imply
a smaller effect from turbulent pressure, no major qualitative conclusions would be changed by a line width reduction of 20\%. Even the importance of turbulent pressure relative to cloud weight pressure in NGC~1333 and B18 remains evident, as shown in Table~\ref{tab:systematics}.

Finally, we examine the influence of the mean gas density that we assumed.  We re-compute the turbulent pressure assuming instead mean densities of $10^3$~cm$^{-3}$ and $10^4$~cm$^{-3}$, which are a factor of $\sim$3 lower and higher than the value used in our main analysis respectively. Assuming $n_{^{13}\mathrm{CO}}\sim 10^3$~cm$^{-3}$ decreases the turbulent pressure on each core by a factor of 3, which then increases the relative importance of the cloud weight pressure in binding the cores, with cloud weight pressure becoming the dominant source of binding in both L1688 and NGC~1333.

Table~\ref{tab:systematics} shows that for the most plausible range of assumptions in calculating the turbulent pressure and other quantities, our final conclusions are qualitatively unchanged: pressure plays a dominant role in binding the majority of dense cores in all three clouds.  We emphasize that the {\it relative} importance of each virial term in the three clouds is even more robust, since the same datasets and analysis methods were applied to each of them.

\bibliographystyle{apj}
\bibliography{HKmod_Ronan.bib}{}



\end{document}